\documentclass{pasj00}
\draft

\def\tsutsui{$E_p$--$T_L$--$L_p$ }
\def\yonetoku{$E_p$--$L_p$ }
\def\amati{$E_p$--$E_{\rm iso}$ }
\def\firmani{$E_p$--$T_{\rm 0.45}$--$L_p$}

\def\swift{{$\it Swift$} }
\def\bepposax{{$\it BeppoSAX$} }
\def\fermi{{$\it Fermi$} }
\def\svom{{$\it SVOM$} }

\begin{document}
\SetRunningHead{R. Tsutsui et al.}{Improved Ep-TL-Lp diagram}

\title{The Improved Ep-TL-Lp Diagram and\\ a Robust Regression Method}

\author{Ryo \textsc{Tsutsui},\altaffilmark{1}\email{tsutsui@tap.scphys.kyoto-u.ac.jp}
Takashi \textsc{Nakamura},\altaffilmark{1}
Daisuke \textsc{Yonetoku},\altaffilmark{2}
Toshio \textsc{Murakami},\altaffilmark{2}
Yoshiyuki \textsc{Morihara}\altaffilmark{2}
and
Keitaro \textsc{Takahashi},\altaffilmark{3}
}
\altaffiltext{1}{Department of Physics, Kyoto University,
Kyoto 606-8502, Japan}
\altaffiltext{2}{Department of Physics, Kanazawa University, Kakuma, Kanazawa,
Ishikawa 920-1192, Japan}
\altaffiltext{3}{Department of Physics and Astrophysics,
Nagoya University, Fro-cho, Chikusa-ku, Nagoya, 464-8602, Japan}

%

\KeyWords{gamma rays: bursts ---  gamma rays: observations
--- gamma rays: cosmology} 

\maketitle

\begin{abstract}
The accuracy and reliability of gamma-ray bursts (GRBs)
as distance indicators are strongly restricted by their systematic
errors which are larger than statistical errors.
These systematic errors might come from
either intrinsic variations of GRBs, or systematic errors in observations.
In this paper, we consider the possible origins of systematic errors
in the following observables,
(i) the spectral peak energies ($E_p$) estimated by Cut-off power
law (CPL) function, 
(ii) the peak luminosities ($L_p$) estimated by 1 second in observer time.
Removing or correcting them, we reveal the true intrinsic variation
of the \tsutsui relation of GRBs. Here $T_L$ is the third parameter
of GRBs defined  as $T_L \equiv E_{\rm iso} / L_p$.
Not only the time resolution of $L_p$ is converted from observer time
to GRB rest frame time, the time resolution with the largest likelihood
is sought for. After removing obvious origin of systematic errors
in observation mentioned above, there seems to be still remain
some outliers. For this reason, we take account another origin of
the systematic error as below, 
(iii) the contamination of short GRBs or other populations.
To estimate the best fit parameters of the \tsutsui relations
from data including outliers, we develop a new method which combine
robust regression and an outlier identification technique.
Using our new method for 18 GRBs with $\sigma_{E_p}/E_p < 0.1$,
we detect 6 outliers and find the \tsutsui relation become
the tightest around 3 second.

\end{abstract}

\section{Introduction}

There are some correlations between the spectral peak energies ($E_p$)
and the brightness of gamma ray bursts (GRBs), as well as correlations
between temporal properties of bursts (variability, spectral lag, etc)
and their brightness
\citep{Fenimore:2000p1151,Reichart:2001p1424,Norris:2000p753}.
The first correlation found is the $E_p$-isotropic equivalent energy
($E_{\rm iso}$) correlation found by \citet{Amati:2002p454}.
This correlation is confirmed and extended to both higher and
lower energy by many satellite teams 
\citep{Amati:2006p1577,Sakamoto:2004p1731,Amati:2009p761,Krimm:2009p1283}.
Although there are some paper against the \amati relation, all of them
use poor data without redshift and/or spectropic $E_p$
\citep{Butler:2007p1782, Ghirlanda:2008p432, Nava:2008p433}.
It is difficult to know whether these results come from the intrinsic
property of GRBs or the systematic effect by using poor data.

Another correlation is the $E_p$-jet collimated energy ($E_{\gamma}$)
correlation found by \citep{Ghirlanda:2004p266}. This is one of
the tightest correlation and expected to be used as distance
indicator toward high redshift universe \citep{Ghirlanda:2006p318}, 
but the need for jet break time make difficult to increase their sample.
More importantly, the complexity of early afterglow observed
by \swift makes it difficult to identify the jet break time.
Thus, it seems to be not so effective to use the $E_p$--$E_{\gamma}$
correlation and also $E_p$--$T_{\rm break}$--$E_{\rm iso}$ correlation
\citep{Liang:2005p2354} for cosmology.

A correlation between $E_p$ and their 1-second peak luminosity
($L_p$) was also found by \citep{Yonetoku:2004p196}. 
\citet{Firmani:2006p302} proposed that taking the high signal time
($T_{0.45}$) as third parameter improves the \yonetoku correlation.
Further, \cite{Rossi:2008p765} showed that the \firmani correlation
becomes as tight as \amati correlation, using 41 GRBs observed by
\bepposax and \swift. Recently, \citet{Tsutsui:2009p158} found similar
but different correlation between $E_p$, luminosity time
($T_L \equiv E_{\rm iso}/L_p$) and $L_p$. Because these relations
use only prompt emission parameters, they might become useful
distance indicators for high redshift universe.
However, there are some researches against these correlations.
\cite{Collazzi:2008p1106} studied whether adding any timescale
improves the \yonetoku correlation
(not only $T_{0.45}$, but also $T_{90}$, $T_L$, and so on.), but they
found that there is no timescale which improves the relation.
From the result they insist that the \firmani (\tsutsui) relation
is equivalent to the \yonetoku relation. However there seems to be
some room for further investigation.

First of all, we must point out that there are many reasons
which cause the scatter around the relation in addition to
the intrinsic dispersion and the measurement uncertainties.
As discussed in some previous works
\citep{Kaneko:2006p1899,Krimm:2009p1283,Shahmoradi:2010p2946},
the peak energies estimated by the Cut-off power law (CPL) model
become systematically higher than the peak energies estimated
by the Band model. Besides, the timescales of GRBs must be
defined in GRB rest frame, because fixed observed timescales
correspond to different rest-frame timescales for GRBs at
different redshifts. Then the parameters defined by observer
time also would cause extra scatter and redshift dependence of
these relations \citep{Tsutsui:2008p162,Yonetoku:2010p2942}.
For these reasons, we use only the GRBs whose peak energies
are estimated by the Band model, and peak luminosities
computed using a fixed timescale in GRB rest frame in this paper.

Even if we take these factors into account, another problem still
remains: how many populations of GRBs exist?
Clearly there are some GRBs which are far from most of GRBs
in the relations, e.g. GRB980425, and some short GRBs.
Furthermore, there might be other unknown populations.
Therefore, when we derive correlations, applying ordinary regression
to the data including different populations would result
in misleading results. Thus, we must separate different populations
of GRBs in our regression analysis, or must use robust statistics.
As far as we know, there are no reference to deal with data
which potentially contains both intrinsic dispersion and/or 
multiple populations. In this paper, we develop a new method
to do this.
 
The structure of this paper is as follows. First we describe
our database of 86 GRBs with known redshift, public light curve
and their spectrum parameters (\S-\ref{sec:data}).
In \S-\ref{sec:method}, we describe the new method we developed.
Using our new database and method, we estimate the best-fit
relation and true intrinsic dispersion in the \tsutsui plane
(\S-\ref{sec:plane}). In \S-\ref{sec:discussion}, we present and
discuss our results.

\section{Data Description}
\label{sec:data}
In \citet{Yonetoku:2010p2942}, we constructed a database
selecting 109 GRBs from GCN Circular Archive \citep{gcn}
and GRBlog \citep{Quimby:2004p2271}. In this section,
we briefly describe our database.

In many cases, the prompt gamma-ray spectrum is well fitted
with the spectral model of the exponentially-connected
broken power-law function suggested by \citet{Band:1993p1901}.
This Band function has four parameters, the low-energy photon
index $\alpha$, the high-energy photon index $\beta$,
the spectral break energy $E_0$ and the normalization.
The peak energy ($E_p$), at which the flux is maximum
in the $\nu F_{\nu}$ spectrum, can be calculated as
$E_p = (2 + \alpha) E_0$.

However, for some GRBs, the photon index (mostly $\beta$)
cannot be determined due to the limited energy range of
the detector and/or the lack of the number of photons
\citep{Pendleton:1997p2118}. When the observation of
high-energy range is not enough, the spectrum is sometimes
fitted with the cut-off power-law (CPL) function.
This CPL function is very similar to the Band function,
but the high energy end is exponentially cut-off.
This function is composed by three parameters,
the low-energy photon index $\alpha$,
the spectral break energy $E_0$ and the normalization.
The peak energy can be also expressed as
$E_p = (2 + \alpha) E_0$ similar to the Band function.
Note that, for an observed GRB spectrum, even if
the reduced $\chi^2$ value of the CPL model is smaller
than that of the Band function, it is difficult to say
whether this model reflects the intrinsic property of
the GRB or it is just due to the poor statistics in
the high-energy range.

In \citet{Yonetoku:2010p2942}, we calculated the bolometric
energy and the peak luminosity in the energy range
1-10,000~keV in the rest frame of each GRB by extending
the observed spectrum. Here, it should be noted that
the integration was performed assuming the Band function
even for GRBs whose spectra were not fitted by
the Band function and the high energy photon index were not reported.
In these cases we assumed the typical values
$\alpha = -1$ and $\beta = -2.25$ to calculate
the bolometric fluence ($S_{\rm bol}$)
and the bolometric peak flux $F_{\rm p,bol}$.
These values are suggested by BATSE observations
\citep{Preece:2000p2353} and also supported by Fermi
observations of GRBs up to 100~GeV energy range. 
\citet{Zhang:2010} confirmed that the time-resolved spectra of 14 out of 17 GRBs 
are best modeled with the Band function over the entire
 Fermi spectral range.
Then the bolometric 
isotropic energy ($E_{\rm iso}$) and the 1-second
peak luminosity ($L_p$) can be simply calculated as
$
E_{\rm iso} = 4 \pi d_L^2 S_{\rm bol}/(1+z)~{\rm (erg)},
$
and
$
L_p = 4 \pi d_L^2 F_{\rm p,bol}~{\rm (erg~s^{-1})}.
$
Here, $d_L$ is the luminosity distance calculated for 
the flat $\Lambda$-CDM universe with the cosmological parameters of
$(\Omega_{\rm m}, \Omega_{\Lambda}) = (0.3, 0.7)$ and
the Hubble parameter of $H_0 = 70~{\rm km~s^{-1} Mpc^{-1}}$.
Further we define the luminosity time as the third parameter
of GRB prompt emission as
$
T_L \equiv E_{\rm iso}/L_p.
$
The error of the luminosity time is estimated by using error
propagation equation. We can neglect the crossterm between
$L_p$ and $E_{\rm iso}$ because of the independence of
the \amati and \yonetoku relation shown in \citep{Tsutsui:2009p158}.

Thus, for GRB whose observed photon number is small,
there are two possible systematic effects. One comes from
the fact that the peak energy $E_p$ is determined by fitting
the spectrum with either the Band function or CPL function.
As \citet{Kaneko:2006p1899} pointed out that the CPL function
tends to overestimate $E_p$ compared to the Band function.
This would induce a systematic error in the correlations
related to $E_p$. On the other hand, although $L_p$ and $E_{\rm iso}$
are determined in a single straightforward way,
the photon indices are set to the typical values
if the number of detected photons is small.
This would also cause a systematic error.

To estimate the impact of systematic errors, we separate GRBs into
three group. One is the platinum data set which consists of GRBs whose
spectrum is well observed and fitted by the Band function
with small $E_p$ error, $\sigma_{E_p}/E_p < 0.1$. The second, 
gold data set, is defined by GRBs fitted by the Band function
with relatively large error, $\sigma_{E_p}/E_p \geq 0.1$.
Finally bronze data set consists of all other GRBs.

In the previous works, the peak luminosity was culculated with 
1-second peak flux in observer frame. This means that the
time scale of the peak luminosity is different from
event by event in GRB frame because of the different redshift.
So the time scale should be defined in GRB rest frame.
Then we must convert the peak luminosity reported in
\citet{Yonetoku:2010p2942} to the $\tau$-second peak luminosity
($L_p(\tau)$) in GRB rest frame. Let us explain the conversion method.
The archived lightcurves of CGRO-BATSE and Konus-Wind are
provided with 64~msec time resolution, and we can also create
the lightcurve of Swift-BAT with the same time resolution.
We used only 86 GRBs observed by these three satellites in this paper,
although there are 109 GRBs in the database of \citet{Yonetoku:2010p2942}.
This is because there are no lightcurves with 64~msec resolution
 for the remaining 23 GRBs mainly observed by BeppoSAX or HETE-II.
When we consider the $\tau$-second peak luminosity in GRB frame,
the time scale of observed peak flux is equivalent to
$\tau (1+z)$-second because of the cosmological time dilation.
Then we performed the re-binning with the number of
$N(\tau) = \tau (1+z)/0.064$ bins (round off to the nearest
whole number), and estimate the observed peak photon flux.
Here, we have $N(\tau)$ degrees of freedom to choose
the start point of re-binning. Therefore we searched all
patterns of re-binning to find the brightest peak photon
flux $P(\tau(1+z))$. We convert the bolometric peak flux
$F_{p,bol}(\tau(1+z))$ with
\begin{eqnarray}
F_{p,bol}(\tau(1+z)) = F_{p,bol}(1.024) \times 
\frac{P(\tau(1+z))}{P(1.024)}.
\end{eqnarray}
Here we refer to the 1-second bolometric peak flux
$F_{p,bol}(1.024)$ for each event summarized in 
\citet{Yonetoku:2010p2942}. The peak luminosity calculated by
$L_p(\tau) = 4 \pi d_L^{2} F_{p,bol}(\tau(1+z))$
becomes the same time interval of $\tau$-second in GRB frame.
Therefore $L_p(\tau)$ may be more appropriate than the
previous definition by \citet{Yonetoku:2010p2942} to discuss
the $E_p$--brightness correlations of GRBs.
In the following sections, we use newly estimated
$L_p(\tau)$ as the bolometric peak luminosity $L_p$.

In the whole analysis of this paper,
we do not use GRB980425 which is a famous outlier for both
\yonetoku and \amati relations. 
The data are summarized in table.~\ref{tab:data1}-\ref{tab:data3}.

\section{Correlation analysis with outliers and multiple populations}
\label{sec:method}

Ordinary regression assumes that the scatter of data around
the relation follows a Gaussian distribution. However, in many cases of
astrophysics, there are some obstacles which lead to misleading results.
For example, as in the case of Cepheid variables, there might be
several populations whose relations are very different
from each other. Also there might be some experimental mistakes
which result in outliers. In these cases, we must separate each
population or identify and remove outliers, because they would dominate
the chi square value and bias the result of regression.
In the history of astronomy, such kind of mistakes was frequently seen.
For example, it is very famous that Hubble combined different types
of Cepheid variables, without knowing it, to estimate the distance
to galaxies. Consequently he overestimated the expansion rate of
the universe by a factor of two. Likewise, there are some clear
outliers in the correlations of GRBs, such as GRB980425 and
some short GRBs. Thus, we should remove the outliers to obtain
the true correlations and estimate the intrinsic dispersion reasonably.

If we knew a criterion to distinguish different populations and
the value of the intrinsic dispersion of the relation in advance, 
there are some reliable ways to eliminate outliers
\citep{Kowalski:2008p1580}. However, we cannot find any reference
to do this without any prior knowledge. Because the number of
GRB events with small observational errors is currently very small,
it is important to eliminate outliers in a systematic manner
rather than in an {\it ad hoc} manner as is often seen.
Here we develop a statistically reliable way to derive
a correlation and estimate the intrinsic dispersion.

The basic idea is to combine a robust regression with outlier
detection \citep{Hampel2005robust}. To identify outliers,
we have to evaluate the residuals of samples from the best-fit
relation. However, if we perform ordinary chi square regression,
outliers can influence the result of regression itself so much
that we cannot measure the residuals correctly. Thus, we should
adopt a more robust regression in the sense that the result is
not so affected by the presence of outliers. To do this,
following {\it Numerical Recipes} \citep{Press2007numerical},
we first perform a regression based on an assumption that
the residuals follow a Lorentzian distribution rather than
a Gaussian distribution. Then we derive a tentative
relation and evaluate the intrinsic dispersion, which can be
used to identify outliers. Eventually, removing the outliers,
we perform ordinary chi square regression to derive
a final relation with confidence intervals of fitting parameters.
Here it should be noted that the robust regression does not
give the confidence intervals, which is why we have to remove
outliers and perform ordinary chi square regression.

Our method can be summarized as the following five steps.
\begin{enumerate}
\item Choose samples with small measurement uncertainties.
\item Fit a relation using a robust regression based
on a Lorentzian distribution of the scatter (\S-\ref{sec:robust}).
\item Calculate the robust standard deviation of the residuals
and estimate the intrinsic dispersion of the relation
(\S-\ref{sec:intrinsic_dispersion}).
\item Identify outliers (\S-\ref{sec:outlier}).
\item Remove the outliers and perform ordinary chi square
regression on the remaining samples.
\end{enumerate}
We find that the first step is important for the data analysis
of GRB whose measurement uncertainties are not uniform.
The largest uncertainty about $E_p$ among our gold samples is
$\sigma_{E_p} / E_p \approx 1$, while the smallest one is
$\sigma_{E_p} / E_p \approx 0.01$. Even if there would be
different populations in the diagram of GRBs, large observational
uncertainties would make it difficult to distinguish them.
 
The goal of this section is to describe each step of our method
to derive the \tsutsui relation and estimate the intrinsic
dispersion. We will show a demonstration of our method by
performing Monte Carlo simulations in Appendix \ref{sec:MC}.

\subsection{Robust regression}
\label{sec:robust}

Before explaining robust regression, let us review ordinary
least-square regression. We assume a linear correlation
between $E_p$, $T_L$ and $L_p$ in logarithmic scale as,
\begin{equation}
\log L_p (E_p,T_L)
= A + B \log \left(E_p / {\bar E_p}\right)
  + C \log \left(T_L / {\bar T_L}\right),
\end{equation}
where $A$, $B$ and $C$ are the parameters of the model.
We adjust these parameters  to maximize the likelihood function
given by,
\begin{eqnarray}
&& P(A,B,C)
\propto \prod_{i=1}^{N} \exp{(-\frac{1}{2}z_{i}^{2})} \Delta L, \\
&& z_i
= \frac{\log L_{p,i}-\log L_p(E_{p,i},T_{L,i})}
         {\sqrt{(1 + 2 C) \sigma^2_{\log L_{p,i}}
                + B^2 \sigma^2_{\log E_{p,i}}
                + C^2 \sigma^2_{\log T_{L,i}}
                + \sigma_{\rm int}^2
               }
         },
\label{eq:z_i}
\end{eqnarray}
where $\Delta L$ is an arbitrary small constant and $\sigma_{\rm int}$
is the intrinsic dispersion of the correlation. The $2 C$ factor
in front of $\sigma_{\log L_{p,i}}$ comes from the fact that
the definition of $T_L$ includes $L_p$. Maximizing this likelihood
function is equivalent to minimizing chi square function,
\begin{equation}
\label{eq:chi square}
\chi^2(A,B,C) = \sum_{i=1}^{N} z_i^2.
\end{equation}
Because not only the numerator but also denominator of $z_i$
depend on model parameters, this chi square function is not
a linear function of model parameters. Therefore nonlinear
regression algorithm, Levenberg-Marquardt Method 
\citep{Levenberg:1944,Marquat:1968}, is used to
find the best fit parameters. If $\sigma_{\rm int}$ is not known
in advance or there are extra components of error which cause
the scatter, $\sigma_{\rm int}$ is adjusted to hold
$\chi^2_{\rm min}/{\rm d.o.f.} = 1$. However we should point
out that this procedure to estimate $\sigma_{\rm int}$ is based
on an assumption that the intrinsic dispersion around the model
follows a Gaussian distribution. Therefore, contamination of outliers
and/or different populations will make the estimation incorrect.
We thus develop a more reasonable way to estimate the intrinsic
dispersion of the correlation, which does not depend on
the existence of outliers and/or different populations
(\S-\ref{sec:intrinsic_dispersion}).

Let us move to robust regression. In this paper, we assume
the scatter of the samples around the relation follows a Lorentzian
distribution, rather than a Gaussian as suggested by
{\it Numerical Recipes} \citep{Press2007numerical}. In this case,
the likelihood function is,
\begin{equation}
P(A,B,C)
\propto \prod_{i=1}^{N} \frac{1}{1 + \frac{1}{2} z_i^2} \Delta L.
\end{equation}
This distribution has wide tails so that the existence of
outliers is common. We maximize this likelihood function,
or equivalently, minimize the sum of the negative logarithms.
Ignoring constant terms, this means minimization of the Lorentzian
merit function defined by,
\begin{equation}
M(A,B,C) = \sum_{i=1}^{N} \ln \left[ 1 + \frac{1}{2} z_i^2 \right].
\end{equation}
Using this merit function with $\sigma_{\rm int}=0$, a tentative set of model parameters
are determined. Regression based on a Lorentzian distribution is
much more robust than one based on a Gaussian distribution,
because a Lorentzian distribution has wide tails and the contribution
of outliers to the merit function is highly suppressed.

Although the robust regression can find reasonable best-fit values
of the model parameters, it doesn't provide reliable confidence
intervals for them, that is, it can't be used to compare the fit of
different sets of parameters. This is why we perform an ordinary
least-square regression in the final step.

\subsection{Estimation of intrinsic dispersion}
\label{sec:intrinsic_dispersion}

To identify outliers, we need to measure how much each sample deviates
from the relation, which is evaluated by $z_i$ in Eq. (\ref{eq:z_i}).
The value of the intrinsic dispersion $\sigma_{\rm int}$ is then necessary
and we estimate it as follows.

If the residuals follow a Gaussian distribution,
the distribution of $z_i$ follows a Gaussian distribution whose
mean is zero and standard deviation is unity. In this case,
we can expect that the standard deviation is estimated as,
\begin{equation}
\sum_{i=1}^{N}
\left\{ \log L_{p,i} - \log L_p(E_{p,i},T_{L,i}) \right\}^2
= \sum_{i=1}^{N}
  \left\{ (1+2C) \sigma_{\log L_{p,i}}^{2} + B^2 \sigma_{\log E_{p,i}}^2
         + C^2 \sigma_{\log T_{L,i}}^2 \right\}
  + N \sigma_{\rm int}^2.
\label{eq:square}
\end{equation}
However, in the current case, the l.h.s. of Eq. (\ref{eq:square})
is strongly influenced by outliers. Thus, we replace it with
a robust standard deviation of the residuals $(\sigma_{\rm RSD})$,
for which we adopt the median absolute deviation (MAD),
\begin{equation}
\sigma_{\rm RSD}
\equiv
\frac{{\rm median} \left[|\log L_{p,i}-\log L_p(E_{p,i},T_{L,i})| \right]}
     {0.6745}.
\label{eq:RSD}
\end{equation}
Here the factor $0.6745$ comes from the fact that $50\%$ of a Gaussian
distribution lies in $0.6745$ standard deviation of the mean.
Obviously this $\sigma_{\rm RSD}$ is not affected so much by
the presence of outliers. It should be noted that, if the scatter
around the relation follows a Gaussian distribution,
this $\sigma_{\rm RSD}$ is equivalent to the normal standard deviation.

If we identify the l.h.s. of Eq. (\ref{eq:square}) with $N \sigma_{\rm RSD}^{2}$,
we can estimate the $\sigma_{\rm int}^2$ as follows,
\begin{equation}
\sigma_{\rm int}^2
= \sigma_{RSD}^2
  - \frac{1}{N}
    \sum_{i}^{N}
    \left\{ (1 + 2 C) \sigma_{\log L_{p,i}}^2 + B^2 \sigma_{\log E_{p,i}}^2
            + C^2 \sigma_{\log T_{L,i}}^2 \right\}.
\label{eq:int}
\end{equation}
We use this value of $\sigma_{\rm int}$ to calculate $z_i$ and then
detect outliers. In Appendix \ref{sec:MC}, we demonstrate that
this way of calculating the intrinsic dispersion is reasonable.

\subsection{Detecting outliers}
\label{sec:outlier}

Let us explain how to detect outliers. Because we have obtained
a tentative set of model parameters and the intrinsic dispersion, 
we can compute, for each sample, $t = |z_i|$ and the two-tailed
P-value from the t distribution with $(N-K)$ degrees of freedom,
where $N$ and $K$ are the numbers of samples and model parameters,
respectively. Following \cite{Motulsky:2006p1549}, we adopt
the False Discovery Rate (FDR) method of \citep{Benjamini:1995p2956}.
We need to determine the threshold value of P-value which decides
whether a sample is an outlier or not. To do this,
\cite{Motulsky:2006p1549} ranked the P-value from
high to low and defined the threshold value of P-value
for $i$-th sample as $\alpha_i = Q (N-i+1)/N$ where $Q$ is
an arbitrary number less than unity. This means that
the $i$-th sample with $P_i < \alpha_i$ is regarded as an outlier.
In this paper, we simply define the threshold value as,
\begin{equation}
\alpha \equiv Q/N,
\end{equation}
for all $i$, which makes our criterion more conservative,
and we take $Q = 0.1$. For this choice of $Q$, we mistakenly
regard $10/N\%$ of samples as outliers on average even if there
are actually no outliers.

Here, we have estimated the intrinsic dispersion for 
data set including outliers and then it is slightly larger than the value without outliers. 
For more reasonable estimation of the intrinsic dispersion and outlier detection, 
we repeat the steps 2-4 again.

By now, we have shown a new method to derive correlation in
presence of many outliers and/or different populations.
We apply this method to estimate the best fit parameters
and the intrinsic dispersion of the \tsutsui relation.

\section{The improved \tsutsui plane}
\label{sec:plane}

In this section we apply the method which we have developed in
the previous section to the data described in section~\ref{sec:data}.
Here, we take a time resolution of $L_p$ as a free parameter of
the relation, and find the most favored time resolution of the relation.

\begin{figure}
\begin{center}
\begin{tabular}{cc}
 \FigureFile(80mm,50mm){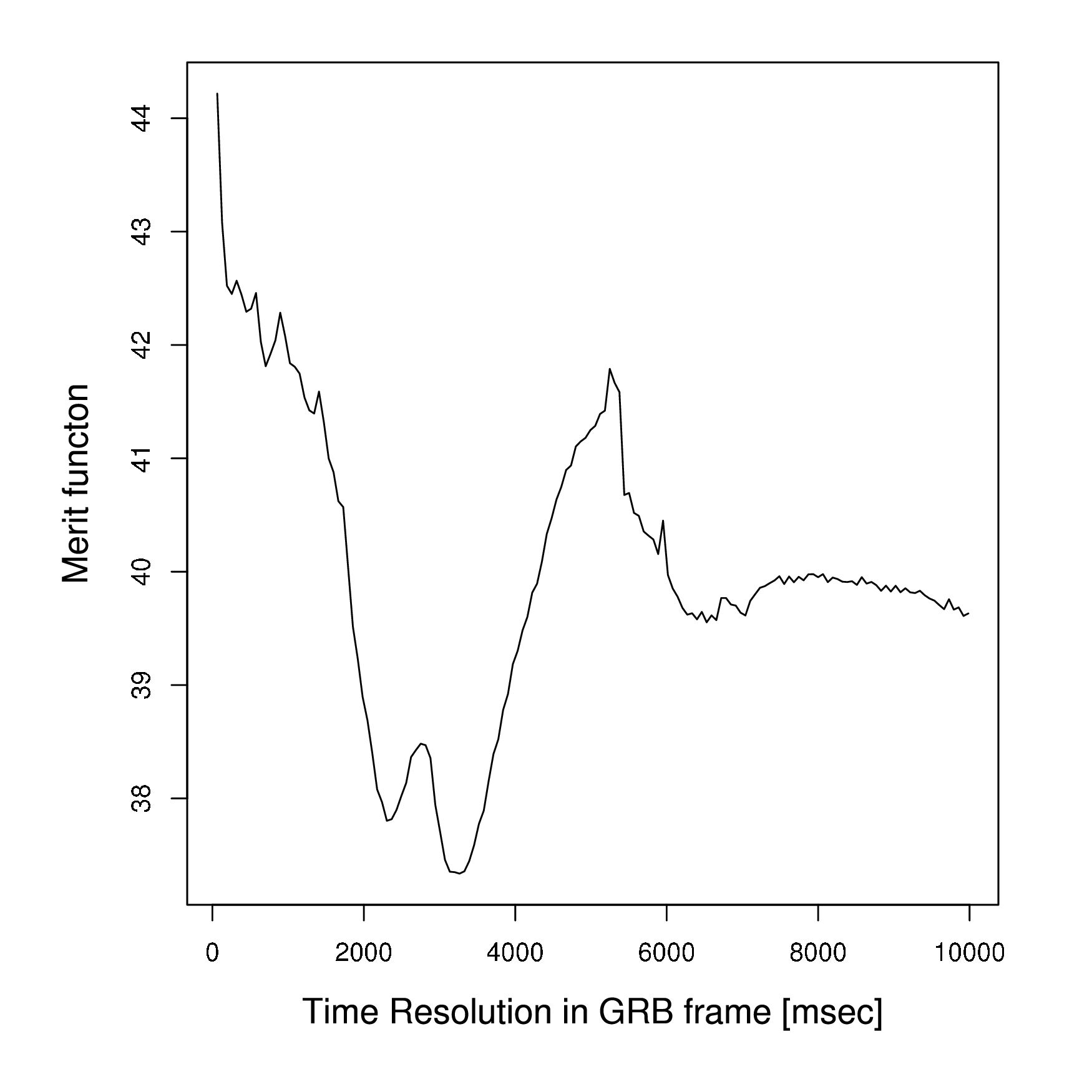}
 \FigureFile(80mm,50mm){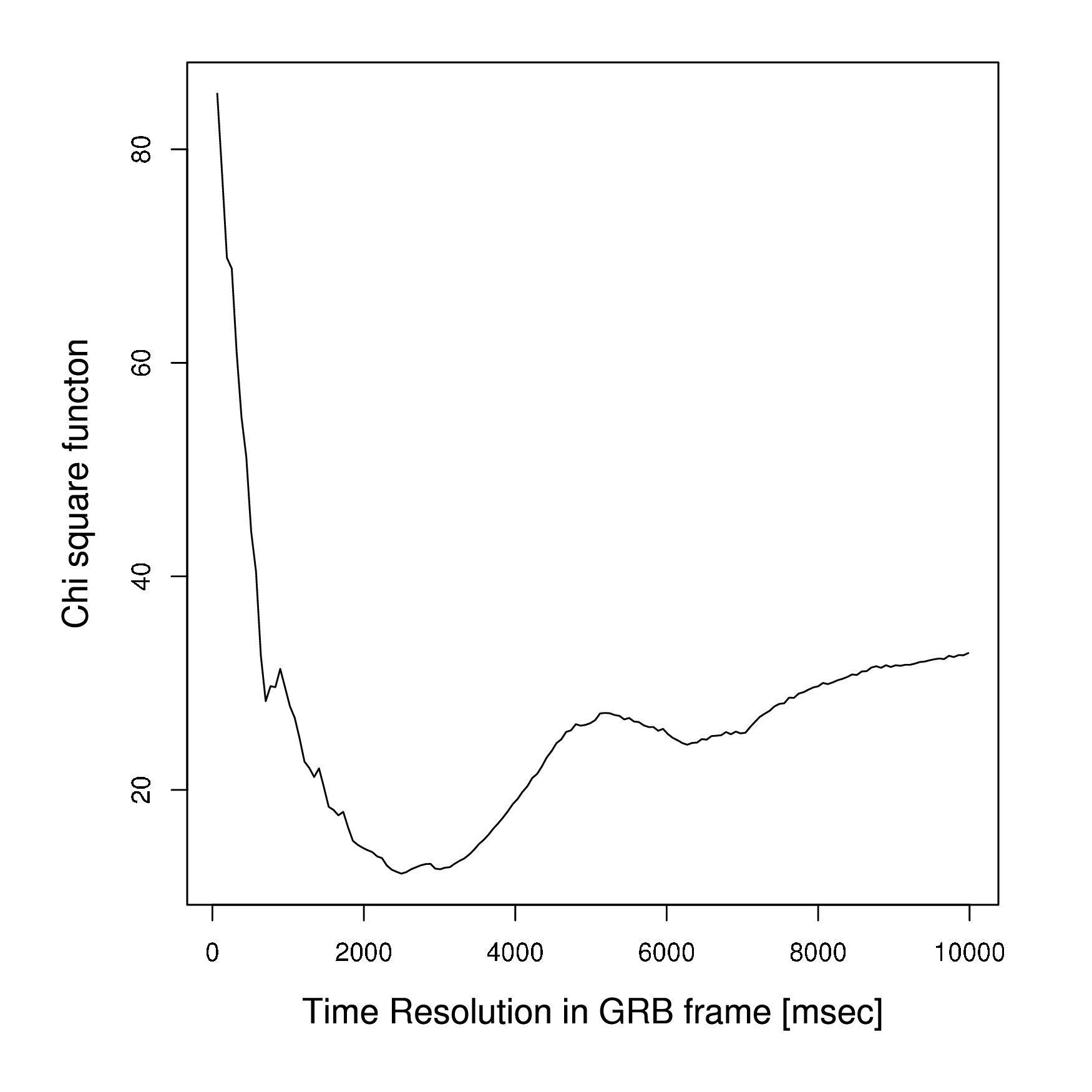}\\
\end{tabular}
\end{center}
\caption{The minimum Lorentzian merit function (left) and chi square
function (right) as a function of time resolution of $L_p$.
It should be noted that the Lorentzian merit function is calculated
using all 18 platinum samples, while the chi square function is
calculated after eliminating 6 outliers (080319B, 081222, 
090328, 090926, 091003, 091127).
}
\label{fig:resolution}
\end{figure}

\begin{figure}
\begin{center}
 \FigureFile(80mm,50mm){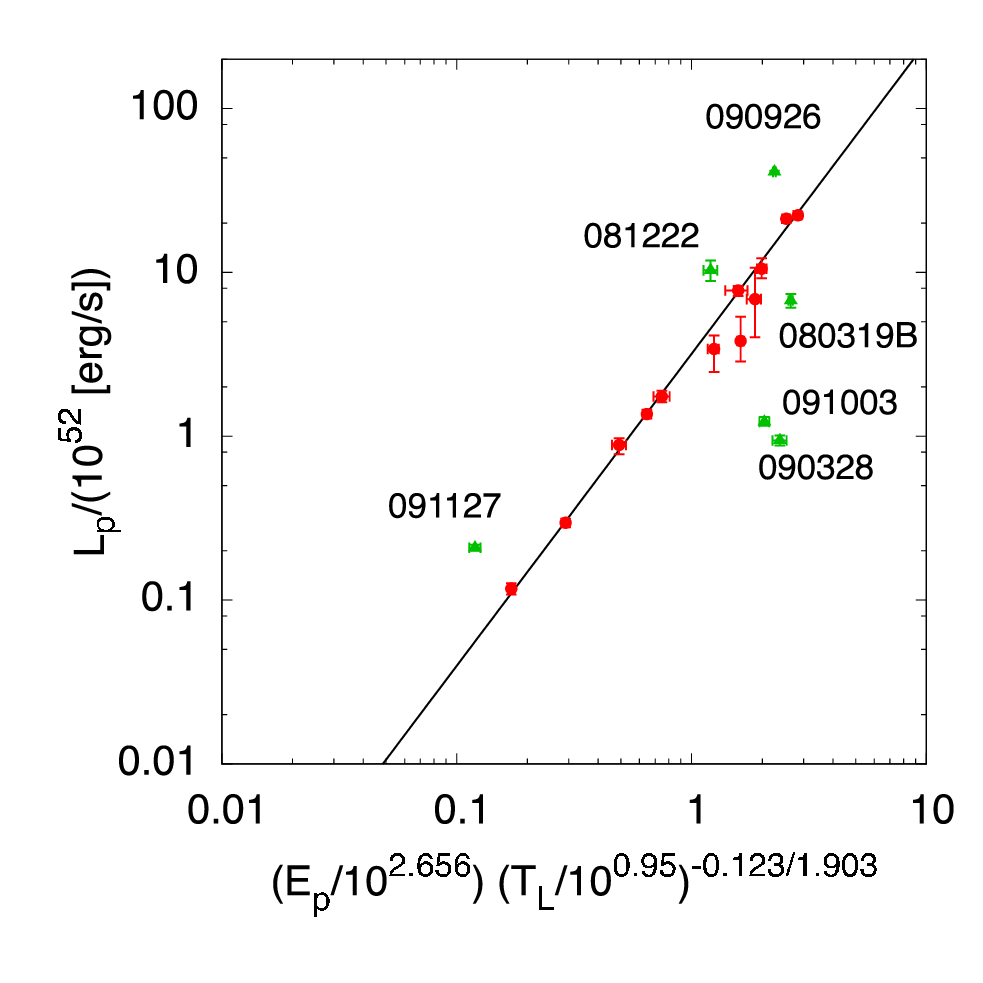}
\end{center}
\caption{The \tsutsui diagram for 18 platinum samples.
The red points indicate  GRBs which is used to derive the relation
and the green points indicate 6 outliers eliminated by our method
described in section~\ref{sec:method}. Solid line indicates
the best-fit model in Eq.~(\ref{eq:tsutsui}). 
}
\label{fig:platinum}
\end{figure}

We use only the platinum data set to derive the \tsutsui relation
as discussed in section~\ref{sec:method}. 
In the left of Fig.~\ref{fig:resolution}, we show the minimum of the Lorentzian
merit function as a function of time resolution of $L_p$.
We found the most favored time resolution is approximately 3 seconds
with $\sigma_{\rm int}=0$, and six outliers (080319B, 081222, 
090328, 090926, 091003, 091127)
are identified. After removing these outliers, we perform chi square
regression with $\sigma_{\rm int}=0$. Then we obtained following result,
\begin{equation}
\label{eq:tsutsui}
L_p = 10^{52.50 \pm 0.017}
      \left( \frac{E_p}{10^{2.656} {\rm keV}} \right)^{1.90 \pm 0.036}
      \left( \frac{T_L}{10^{0.95} {\rm sec}} \right)^{-0.12 \pm 0.053},
\end{equation}
with $\chi^2_{\rm min}/{\rm dof} = 12.16/9$. The most favored time
resolution is $2.5^{+0.6}_{-0.2}$. In Fig.~\ref{fig:platinum},
we show the best-fit \tsutsui relation of (18-6) platinum GRBs.
The red points indicate the GRBs which are used to estimate parameters,
and the green points are the GRBs which are eliminated as outliers.
We should emphasize that, because our method removes outliers
in automatic way, there is no artificial choice of outliers.

If we include the four mid-outliers (080319B, 081222, 090926, 091127), the chi square
value becomes unity when $\sigma_{\rm int}=0.20$. As discussed
in section~\ref{sec:method} and appendix~\ref{sec:MC}, if we assume
the intrinsic dispersion $\sigma_{\rm int} = 0.20$, the chance
probability to obtain the value $\sigma_{\rm int} = 0.0$ using
our method with 15 samples is approximately 1\%
(See top central of figure~\ref{fig:MC2}). 
Thus, we can conclude that these four GRBs surely outliers of
the \tsutsui plane, or at least the true intrinsic dispersion of
the \tsutsui plane is much less than 0.20.

\begin{figure}
\begin{center}
 \FigureFile(80mm,80mm){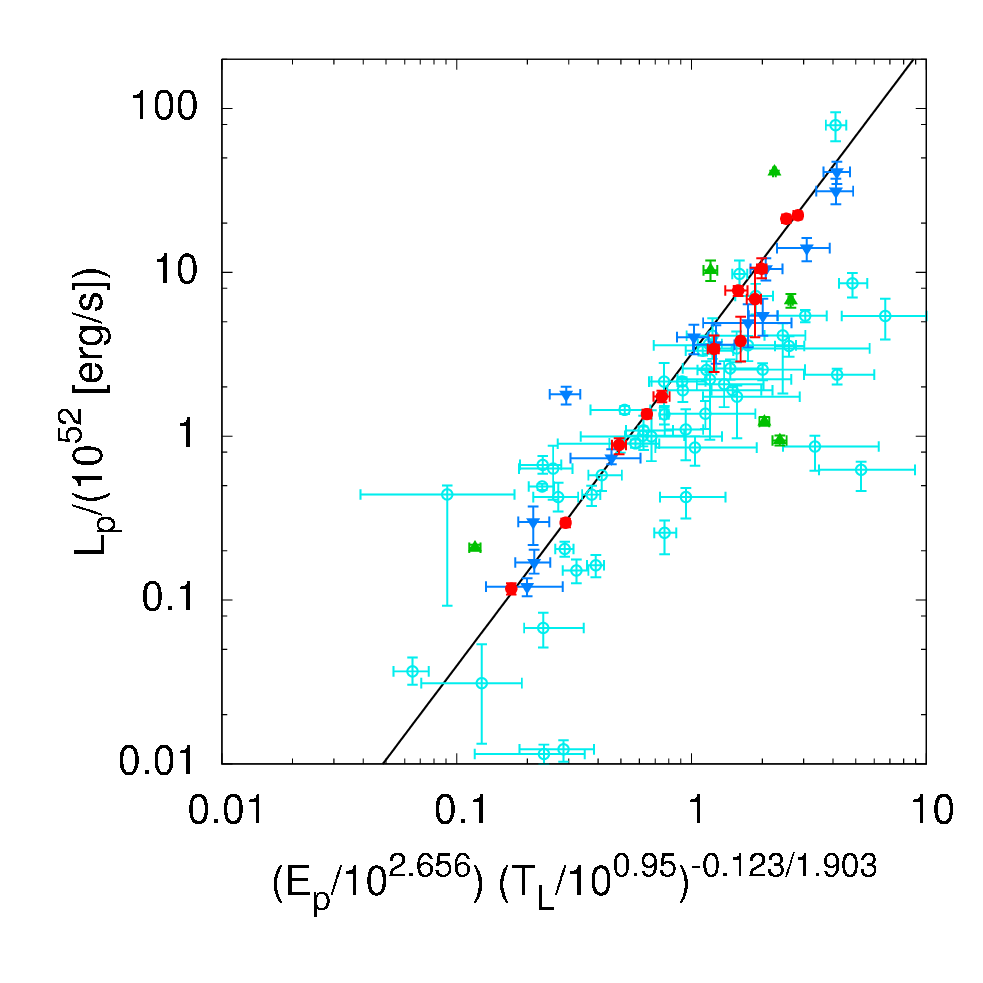}\\
\begin{tabular}{cc}
 \FigureFile(80mm,80mm){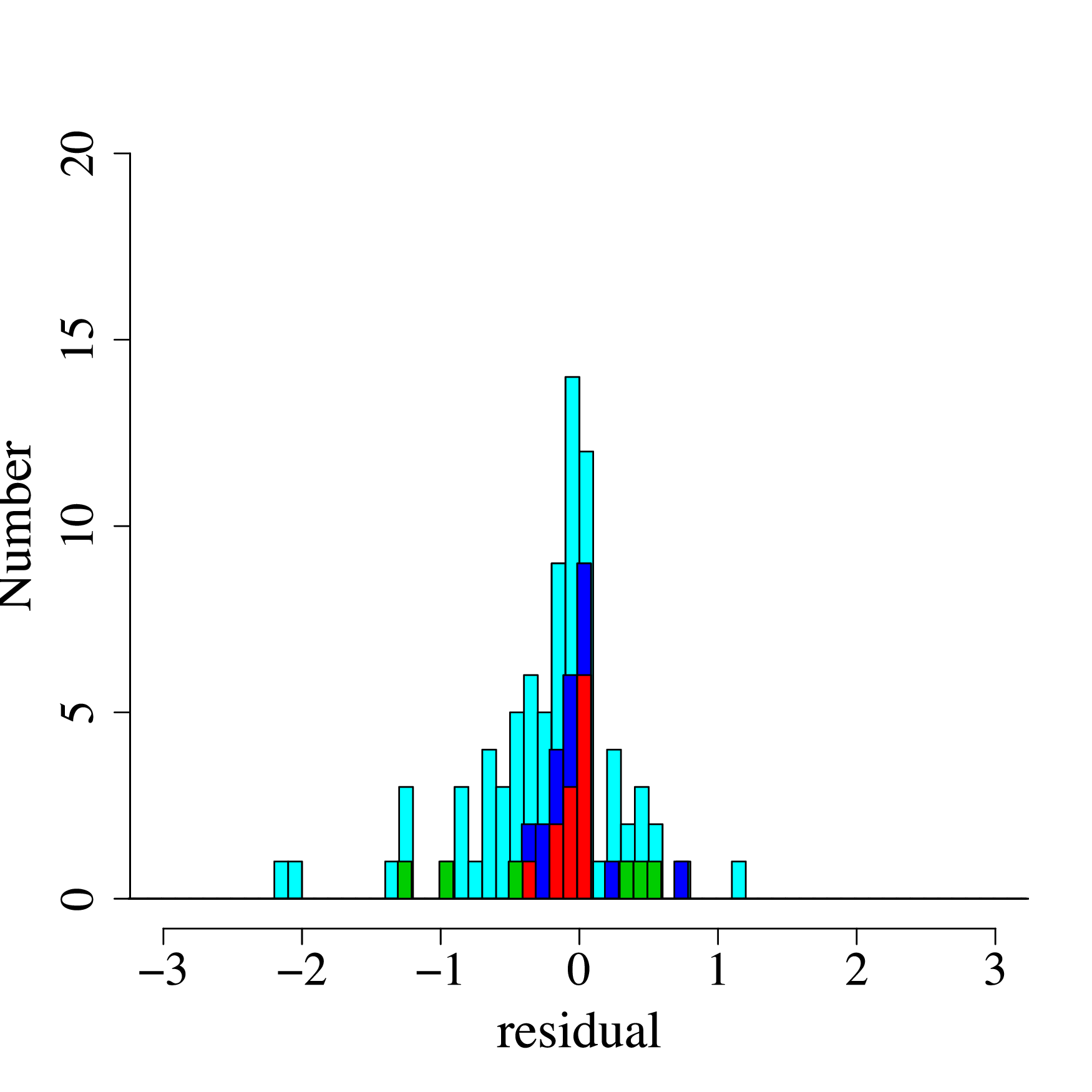}
 \FigureFile(80mm,80mm){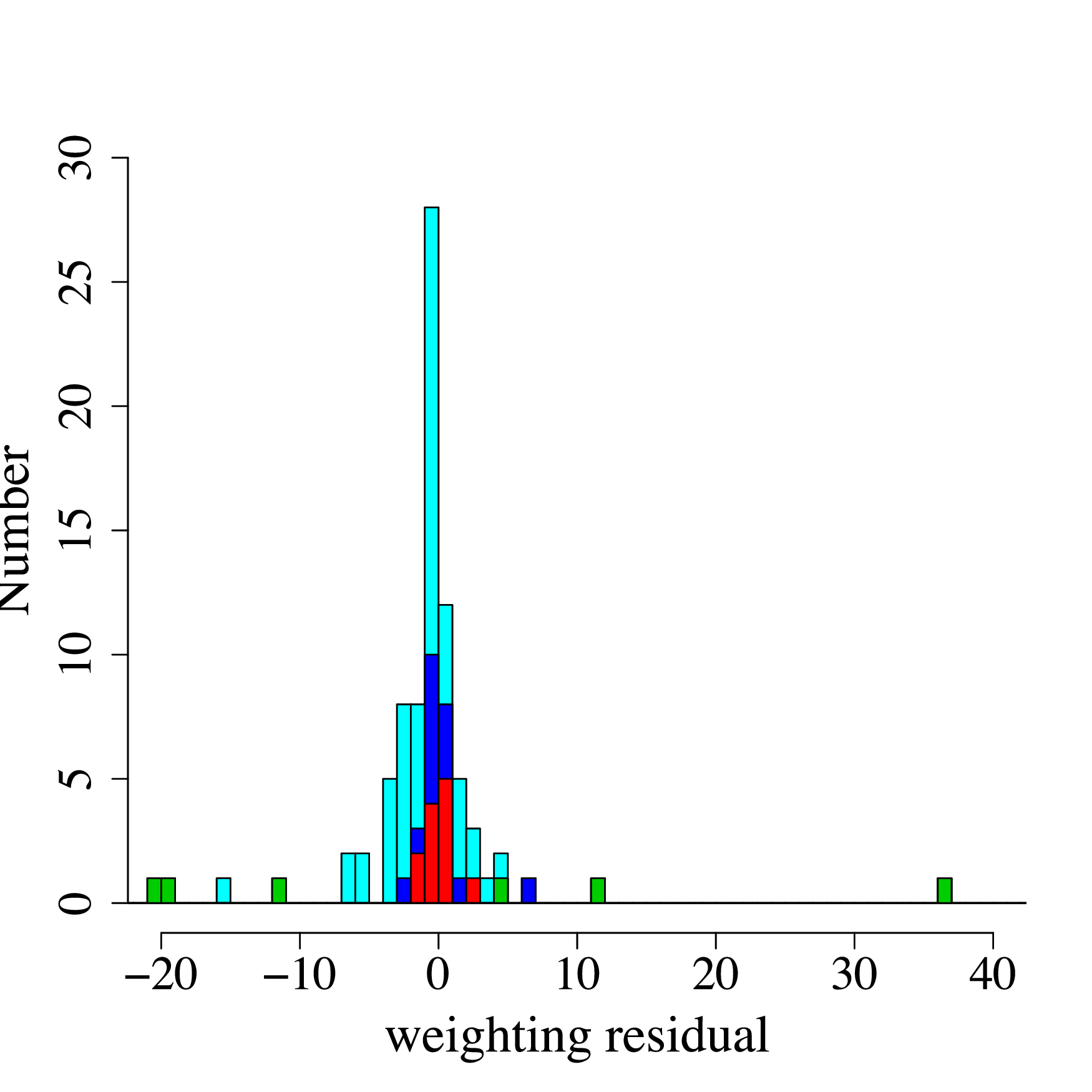}
\end{tabular}
\end{center}
\caption{(Top): Same as figure~\ref{fig:platinum}, but gold data
(blue points) and bronze data (light blue points) are also plotted.
(left bottom): The histogram of the weighting residual from
the best-fit line. The color follows the top figure.
(right bottom): The histogram of the unweighting residual from
the best-fit line. 
}
\label{fig:all}
\end{figure}

In top of figure~\ref{fig:all}, we showed not only the data used
for calibration but also other data (gold and bronze data sets).
Bottom left figure indicates the histogram of weighting residuals
($z_i$) and bottom right indicates the histogram of unweighting
residuals ($\log L_{p,i}-\log L_p(E_{p,i},T_{L,i})$).
As one can see, the gold data are consistent with best-fit model
within statistical error, but the most of bronze data distribute
below the best-fit line of Eq.~(\ref{eq:tsutsui}). It might be
explained by the fact that the peak energies estimated by the CPL model
becomes systematically larger than that estimated by the Band model
\citep{Kaneko:2006p1899,Krimm:2009p1283}.

\section{Summary and Discussion}
\label{sec:discussion}

In this paper, we considered two possible origins of the systematic error
in the \tsutsui relation. The first is concerned with the calculation of
the spectral peak energies ($E_p$) and we removed samples whose $E_p$
was estimated by CPL model. The second is concerned with the definition
of the peak luminosity ($L_p$) which were conventionally estimated by
1-second in observer time. We converted the time resolution to the time
scale in GRB rest frame. Furthermore, because there might be some outliers
like 980425 and some short GRBs, we developed a new method using robust
regression and following outliers rejection. Using our method, we found that
the \tsutsui relation is the tightest around 3-second peak luminosity
with $\sigma_{\rm int} = 0$.

We briefly discuss the properties of six outliers detected
by our method. First, we should note that three of them
(090328, 090926, 091003) are detected  by \fermi/LAT.
Especially GRB 090328 has long-lived GeV emission after
the prompt emission and GRB 090926 has extra high energy component.
It might be possible that \fermi/LAT selectively observe some kind of GRBs 
which are harder than normal GRBs. Second, GRB 080319B is one of
the remarkable GRBs ever observed because of the optical emission
bright enough to be seen by naked-eyes. Although the others
(081222, 091127) seem to be normal GRBs, these might belong to
peculiar GRBs mentioned above.

Of course, because the number of the sample is small, our result must be
confirmed by future experiments. The collaboration of \swift and \fermi
can provide a few GRBs with measured redshift and small $E_p$ error
with $\sigma_{E_p}/E_p<0.1$. Future mission \svom will be able to
determine the redshift and $E_p$ by itself, and can confirm our result
with a large number of data. In future studies, we should study
the classification of GRBs using comprehensive observation of
both prompt and afterglow emission. The \tsutsui plane might become
useful discriminator of GRBs like HR diagram of stars. Once we establish
the relation, GRBs might be a very powerful and unique distance
indicator to probe the high redshift universe.

\section*{Acknowledgments}

This work is supported in part by the Grant-in-Aid from the 
Ministry of Education, Culture, Sports, Science and Technology
(MEXT) of Japan, No.19540283, No.19047004(TN),
No.18684007 (DY) and No.21840028(KT), and by the Grant-in-Aid
for the global COE program {\it The Next Generation of Physics,
Spun from Universality and Emergence} at Kyoto University and
"Quest for Fundamental Principles in the Universe: from Particles
to the Solar System and the Cosmos" at Nagoya University
from MEXT of Japan. RT is supported by a Grant-in-Aid for
the Japan Society for the Promotion of Science (JSPS) Fellows
and is a research fellow of JSPS.


\begin{thebibliography}{32}
\expandafter\ifx\csname natexlab\endcsname\relax\def\natexlab#1{#1}\fi

\bibitem[{Amati(2006)}]{Amati:2006p1577}
Amati, L. 2006, Monthly Notices of the Royal Astronomical Society, 372, 233

\bibitem[{Amati {et~al.}(2009)Amati, Frontera, \& Guidorzi}]{Amati:2009p761}
Amati, L., Frontera, F., \& Guidorzi, C. 2009, Astronomy and Astrophysics, 508,
  173

\bibitem[{Amati {et~al.}(2002)Amati, Frontera, Tavani, in't Zand, Antonelli,
  Costa, Feroci, Guidorzi, Heise, Masetti, Montanari, Nicastro, Palazzi, Pian,
  Piro, \& Soffitta}]{Amati:2002p454}
Amati, L., {et~al.} 2002, Astronomy and Astrophysics, 390, 81

\bibitem[{Band {et~al.}(1993)Band, Matteson, Ford, Schaefer, Palmer, Teegarden,
  Cline, Briggs, Paciesas, Pendleton, Fishman, Kouveliotou, Meegan, Wilson, \&
  Lestrade}]{Band:1993p1901}
Band, D., {et~al.} 1993, Astrophysical Journal, 413, 281

\bibitem[{Barthelmy(1997)}]{gcn}
Barthelmy, S. 1997, GCN Circulars Archive, http://gcn.gsfc.nasa.gov/

\bibitem[{Benjamini \& Hochberg(1995)}]{Benjamini:1995p2956}
Benjamini, Y., \& Hochberg, Y. 1995, Journal of the Royal Statistical Society.
  Series B {\ldots}, 57, 289

\bibitem[Zhang et al.(2010)]{Zhang:2010} Zhang, B.-B., et al.\ 
2010, arXiv:1009.3338 

\bibitem[{Butler {et~al.}(2007)Butler, Kocevski, Bloom, \&
  Curtis}]{Butler:2007p1782}
Butler, N.~R., Kocevski, D., Bloom, J.~S., \& Curtis, J.~L. 2007, The
  Astrophysical Journal, 671, 656

\bibitem[{Collazzi \& Schaefer(2008)}]{Collazzi:2008p1106}
Collazzi, A.~C., \& Schaefer, B.~E. 2008, The Astrophysical Journal, 688, 456

\bibitem[{Fenimore \& Ramirez-Ruiz(2000)}]{Fenimore:2000p1151}
Fenimore, E.~E., \& Ramirez-Ruiz, E. 2000, eprint arXiv, 4176

\bibitem[{Firmani {et~al.}(2006)Firmani, Ghisellini, Avila-Reese, \&
  Ghirlanda}]{Firmani:2006p302}
Firmani, C., Ghisellini, G., Avila-Reese, V., \& Ghirlanda, G. 2006, Monthly
  Notices of the Royal Astronomical Society, 370, 185

\bibitem[{Ghirlanda {et~al.}(2006)Ghirlanda, Ghisellini, \&
  Firmani}]{Ghirlanda:2006p318}
Ghirlanda, G., Ghisellini, G., \& Firmani, C. 2006, New Journal of Physics, 8,
  123

\bibitem[{Ghirlanda {et~al.}(2004)Ghirlanda, Ghisellini, \&
  Lazzati}]{Ghirlanda:2004p266}
Ghirlanda, G., Ghisellini, G., \& Lazzati, D. 2004, The Astrophysical Journal,
  616, 331

\bibitem[{Ghirlanda {et~al.}(2008)Ghirlanda, Nava, Ghisellini, Firmani, \&
  Cabrera}]{Ghirlanda:2008p432}
Ghirlanda, G., Nava, L., Ghisellini, G., Firmani, C., \& Cabrera, J.~I. 2008,
  Monthly Notices of the Royal Astronomical Society, 387, 319

\bibitem[{Hampel {et~al.}(2005)Hampel, Ronchetti, Rousseeuw, \&
  Stahel}]{Hampel2005robust}
Hampel, F., Ronchetti, E., Rousseeuw, P., \& Stahel, W. 2005, Robust
  statistics: the approach based on influence functions, John Wiley \& Sons

\bibitem[{Kaneko {et~al.}(2006)Kaneko, Preece, Briggs, Paciesas, Meegan, \&
  Band}]{Kaneko:2006p1899}
Kaneko, Y., Preece, R., Briggs, M., Paciesas, W., Meegan, C., \& Band, D. 2006,
  The Astrophysical Journal Supplement Series, 166, 298

\bibitem[{Kowalski {et~al.}(2008)Kowalski, Rubin, Aldering, Agostinho, Amadon,
  Amanullah, Balland, Barbary, Blanc, \& Challis}]{Kowalski:2008p1580}
Kowalski, M., {et~al.} 2008, The Astrophysical Journal, 686, 749

\bibitem[{Krimm {et~al.}(2009)Krimm, Yamaoka, Sugita, Ohno, Sakamoto,
  Barthelmy, Gehrels, Hara, Norris, Ohmori, Onda, Sato, Tanaka, Tashiro, \&
  Yamauchi}]{Krimm:2009p1283}
Krimm, H.~A., {et~al.} 2009, The Astrophysical Journal, 704, 1405

\bibitem[{Levenberg(1944)}] {Levenberg:1944}
Levenberg, K., 1994, The Quarterly of Applied Mathematics, 2 ,164

\bibitem[{Liang \& Zhang(2005)}]{Liang:2005p2354}
Liang, E., \& Zhang, B. 2005, The Astrophysical Journal, 633, 611

\bibitem[{Marquart(1963)}]{Marquat:1968}
Marquardt, D.,  1963, SIAM J. Appl. Math. 11, 431 

\bibitem[{Motulsky \& Brown(2006)}]{Motulsky:2006p1549}
Motulsky, H., \& Brown, R. 2006, BMC bioinformatics, 7 ,123

\bibitem[{Nava {et~al.}(2008)Nava, Ghirlanda, Ghisellini, \&
  Firmani}]{Nava:2008p433}
Nava, L., Ghirlanda, G., Ghisellini, G., \& Firmani, C. 2008, Monthly Notices
  of the Royal Astronomical Society, 391, 639

\bibitem[{Norris {et~al.}(2000)Norris, Marani, \& Bonnell}]{Norris:2000p753}
Norris, J.~P., Marani, G.~F., \& Bonnell, J.~T. 2000, The Astrophysical
  Journal, 534, 248

\bibitem[{Pendleton {et~al.}(1997)Pendleton, Paciesas, Briggs, Preece, Mallozzi,
  Meegan, Horack, Fishman, Band, Matteson, Skelton, Hakkila, Ford, Kouveliotou,
  \& Koshut}]{Pendleton:1997p2118}
Pendleton, G., {et~al.} 1997, The Astrophysical Journal, 489, 175

\bibitem[{Preece {et~al.}(2000)Preece, Briggs, Mallozzi, Pendleton, Paciesas,
  \& Band}]{Preece:2000p2353}
Preece, R.~D., Briggs, M.~S., Mallozzi, R.~S., Pendleton, G.~N., Paciesas,
  W.~S., \& Band, D.~L. 2000, The Astrophysical Journal Supplement Series, 126,
  19

\bibitem[{Press {et~al.}(2007)}]{Press2007numerical}
Press, W. et al., 2007, Numerical recipes: the art of scientific computing, Cambridge University Press

\bibitem[{Quimby {et~al.}(2004)Quimby, Mcmahon, \& Murphy}]{Quimby:2004p2271}
Quimby, R., Mcmahon, E., \& Murphy, J. 2004, Gamma-Ray Bursts: 30 Years of
  Discovery, 727, 529

\bibitem[{Reichart {et~al.}(2001)Reichart, Lamb, Fenimore, Ramirez-Ruiz, Cline,
  \& Hurley}]{Reichart:2001p1424}
Reichart, D.~E., Lamb, D.~Q., Fenimore, E.~E., Ramirez-Ruiz, E., Cline, T.~L.,
  \& Hurley, K. 2001, The Astrophysical Journal, 552, 57

\bibitem[{Rossi {et~al.}(2008)Rossi, Guidorzi, Amati, Frontera, Romano,
  Campana, Chincarini, Montanari, Moretti, \& Tagliaferri}]{Rossi:2008p765}
Rossi, F., {et~al.} 2008, Monthly Notices of the Royal Astronomical Society,
  388, 1284

\bibitem[{Sakamoto {et~al.}(2004)Sakamoto, Lamb, Graziani, Donaghy, Suzuki,
  Ricker, Atteia, Kawai, Yoshida, Shirasaki, Tamagawa, Torii, Matsuoka,
  Fenimore, Galassi, Tavenner, Doty, Vanderspek, Crew, Villasenor, Butler,
  Prigozhin, Jernigan, Barraud, Boer, Dezalay, Olive, Hurley, Levine, Monnelly,
  Martel, Morgan, Woosley, Cline, Braga, Manchanda, Pizzichini, Takagishi, \&
  Yamauchi}]{Sakamoto:2004p1731}
Sakamoto, T., {et~al.} 2004, The Astrophysical Journal, 602, 875

\bibitem[{Shahmoradi \& Nemiroff(2010)}]{Shahmoradi:2010p2946}
Shahmoradi, A., \& Nemiroff, R. 10, Monthly Notices of the Royal Astronomical
  Society, 407, 2075

\bibitem[{Tsutsui {et~al.}(2009)Tsutsui, Nakamura, Yonetoku, Murakami, Kodama,
  \& Takahashi}]{Tsutsui:2009p158}
Tsutsui, R., Nakamura, T., Yonetoku, D., Murakami, T., Kodama, Y., \&
  Takahashi, K. 2009, Journal of Cosmology and Astroparticle Physics, 08, 015

\bibitem[{Tsutsui {et~al.}(2008)Tsutsui, Nakamura, Yonetoku, Murakami, Tanabe,
  \& Kodama}]{Tsutsui:2008p162}
Tsutsui, R., Nakamura, T., Yonetoku, D., Murakami, T., Tanabe, S., \& Kodama,
  Y. 2008, Monthly Notices of the Royal Astronomical Society: Letters, 386, L33

\bibitem[{Yonetoku {et~al.}(2004)Yonetoku, Murakami, Nakamura, Yamazaki, Inoue,
  \& Ioka}]{Yonetoku:2004p196}
Yonetoku, D., Murakami, T., Nakamura, T., Yamazaki, R., Inoue, A.~K., \& Ioka,
  K. 2004, The Astrophysical Journal, 609, 935

\bibitem[{Yonetoku {et~al.}(2010)Yonetoku, Murakami, Tsutsui, Nakamura,
  Morihara, \& Takahashi}]{Yonetoku:2010p2942}
Yonetoku, D., Murakami, T., Tsutsui, R., Nakamura, T., Morihara, Y., \&
  Takahashi, K. 2010, Publications of the Astronomical Society of Japan, 62, in press

\end{thebibliography}

\appendix
\begin{longtable}{lccc}					
\caption{Intrinsic property of GRBs for 18 platinum data set$^{\flat}$}
\label{tab:data1}
\hline\hline
GRB &$E_p$(keV)&$L_p$(erg~s$^{-1}$)	$^{\dagger}$&$T_L$ (sec)\\
\hline
\endhead
\hline
\multicolumn{4}{l}{$\flat$ The data with $E_{p}$ fitted with the Band function and $\sigma_{E_{p}}/E_{p}<0.1$}\\
\multicolumn{4}{l}{$\dagger$ 2.496 second peak luminosity in GRB rest frame}\\
\multicolumn{4}{l}{$\sharp$ Outliers detected by our method}\\
\endfoot

971214	& $	807.1 	^{+	48.6 	}_{	-63.2	}$ & $ (	6.87 	^{+	3.80 	}_{	-2.86 	})\times 10^{	52	}$ & $	4.47	^{+	2.50	}_{	-1.89	} $\\
990123	& $	1333.8 	^{+	49.9 	}_{	-56.9	}$ & $ (	2.23 	^{+	0.09 	}_{	-0.10 	})\times 10^{	53	}$ & $	15.94	^{+	0.73	}_{	-0.77	} $\\
990506	& $	737.6 	^{+	69.2 	}_{	-87.9	}$ & $ (	7.75 	^{+	0.53 	}_{	-0.57 	})\times 10^{	52	}$ & $	14.33	^{+	1.07	}_{	-1.15	} $\\
990510	& $	538.2 	^{+	25.1 	}_{	-32.2	}$ & $ (	3.41 	^{+	0.71 	}_{	-0.94 	})\times 10^{	52	}$ & $	4.23	^{+	0.93	}_{	-1.21	} $\\
990705	& $	348.3 	^{+	27.6 	}_{	-27.6	}$ & $ (	1.75 	^{+	0.15 	}_{	-0.14 	})\times 10^{	52	}$ & $	13.90	^{+	2.62	}_{	-2.43	} $\\
991216	& $	1083.7 	^{+	37.4 	}_{	-41.2	}$ & $ (	2.12 	^{+	0.10 	}_{	-0.13 	})\times 10^{	53	}$ & $	3.73	^{+	0.20	}_{	-0.26	} $\\
000210	& $	754.8 	^{+	25.9 	}_{	-25.9	}$ & $ (	3.81 	^{+	1.55 	}_{	-0.96 	})\times 10^{	52	}$ & $	13.95	^{+	8.02	}_{	-4.94	} $\\
030329	& $	79.3 	^{+	2.7 	}_{	-2.5	}$ & $ (	1.17 	^{+	0.09 	}_{	-0.09 	})\times 10^{	51	}$ & $	12.96	^{+	1.14	}_{	-1.18	} $\\
050525	& $	130.4 	^{+	3.7 	}_{	-3.7	}$ & $ (	2.97 	^{+	0.17 	}_{	-0.18 	})\times 10^{	51	}$ & $	7.62	^{+	0.56	}_{	-0.63	} $\\
061007	& $	902.1 	^{+	43.0 	}_{	-40.7	}$ & $ (	1.05 	^{+	0.17 	}_{	-0.13 	})\times 10^{	53	}$ & $	9.08	^{+	1.57	}_{	-1.21	} $\\
080319B$^{\sharp}$	& $	1261.0 	^{+	25.2 	}_{	-27.1	}$ & $ (	6.73 	^{+	0.65 	}_{	-0.65 	})\times 10^{	52	}$ & $	19.46	^{+	1.94	}_{	-1.93	} $\\
081222$^{\sharp}$	& $	505.2 	^{+	33.9 	}_{	-33.9	}$ & $ (	1.03 	^{+	0.15 	}_{	-0.14 	})\times 10^{	53	}$ & $	2.72	^{+	0.46	}_{	-0.43	} $\\
090328$^{\sharp}$	& $	1133.6 	^{+	78.1 	}_{	-78.1	}$ & $ (	9.42 	^{+	0.70 	}_{	-0.64 	})\times 10^{	51	}$ & $	20.07	^{+	1.89	}_{	-1.72	} $\\
090424	& $	273.3 	^{+	4.6 	}_{	-4.6	}$ & $ (	1.36 	^{+	0.09 	}_{	-0.08 	})\times 10^{	52	}$ & $	3.19	^{+	0.24	}_{	-0.22	} $\\
090618	& $	239.5 	^{+	17.1 	}_{	-16.2	}$ & $ (	8.85 	^{+	0.87 	}_{	-1.10 	})\times 10^{	51	}$ & $	27.92	^{+	3.34	}_{	-4.17	} $\\
090926$^{\sharp}$	& $	975.3 	^{+	12.4 	}_{	-12.4	}$ & $ (	4.10 	^{+	0.06 	}_{	-0.07 	})\times 10^{	53	}$ & $	4.39	^{+	0.17	}_{	-0.17	} $\\
091003$^{\sharp}$	& $	922.3 	^{+	44.8 	}_{	-44.8	}$ & $ (	1.22 	^{+	0.09 	}_{	-0.07 	})\times 10^{	52	}$ & $	8.50	^{+	0.86	}_{	-0.62	} $\\
091127$^{\sharp}$	& $	53.6 	^{+	3.0 	}_{	-3.0	}$ & $ (	2.09 	^{+	0.09 	}_{	-0.09 	})\times 10^{	51	}$ & $	7.57	^{+	0.36	}_{	-0.34	} $\\
\end{longtable}	

\begin{longtable}{lccc}					
\caption{Intrinsic property of GRBs for 14 gold data set$^{\flat}$}
\label{tab:data2}
\hline\hline
GRB &$E_p$(keV)&$L_p$(erg~s$^{-1}$)	$^{\dagger}$&$T_L$ (sec)\\
\hline
\endhead
\hline
\multicolumn{4}{l}{$\flat$ The data with $E_{p}$ fitted with the Band function and $\sigma_{E_{p}}/E_{p}\ge0.1$}\\
\multicolumn{4}{l}{$\dagger$ 2.496 second peak luminosity in GRB rest frame}\\
\multicolumn{4}{l}{$*$ Low luminosity GRB}\\
\endfoot
970228	& $	194.9 	^{+	64.4 	}_{	-64.4	}$ & $ (	7.34 	^{+	0.99 	}_{	-0.60 	})\times 10^{	51	}$ & $	3.68	^{+	0.89	}_{	-0.60	} $\\
970508	& $	89.7 	^{+	37.8 	}_{	-29.7	}$ & $ (	1.21 	^{+	0.16 	}_{	-0.15 	})\times 10^{	51	}$ & $	8.11	^{+	1.45	}_{	-1.58	} $\\
980425	& $	55.4 	^{+	11.6 	}_{	-11.6	}$ & $ (	1.00 	^{+	0.20 	}_{	-0.20 	})\times 10^{	47	}$ & $	10.70	^{+	2.80	}_{	-2.74	} $\\
990712	& $	93.0 	^{+	15.7 	}_{	-15.7	}$ & $ (	1.70 	^{+	0.33 	}_{	-0.25 	})\times 10^{	51	}$ & $	4.70	^{+	1.20	}_{	-0.87	} $\\
020405	& $	615.2 	^{+	123.4 	}_{	-123.4	}$ & $ (	3.61 	^{+	1.15 	}_{	-0.84 	})\times 10^{	52	}$ & $	25.33	^{+	11.40	}_{	-8.32	} $\\
021211	& $	91.6 	^{+	15.8 	}_{	-12.5	}$ & $ (	3.00 	^{+	0.73 	}_{	-0.83 	})\times 10^{	51	}$ & $	4.40	^{+	1.35	}_{	-1.54	} $\\
050401	& $	458.3 	^{+	70.2 	}_{	-70.2	}$ & $ (	4.02 	^{+	0.77 	}_{	-0.85 	})\times 10^{	52	}$ & $	7.49	^{+	1.70	}_{	-1.95	} $\\
050603	& $	1313.3 	^{+	332.4 	}_{	-332.4	}$ & $ (	1.41 	^{+	0.21 	}_{	-0.24 	})\times 10^{	53	}$ & $	3.29	^{+	0.68	}_{	-0.76	} $\\
060124	& $	784.4 	^{+	415.3 	}_{	-280.2	}$ & $ (	4.91 	^{+	1.49 	}_{	-1.41 	})\times 10^{	52	}$ & $	8.17	^{+	2.48	}_{	-2.37	} $\\
070125	& $	934.7 	^{+	165.6 	}_{	-129.9	}$ & $ (	1.05 	^{+	0.17 	}_{	-0.16 	})\times 10^{	53	}$ & $	8.49	^{+	1.61	}_{	-1.47	} $\\
080721	& $	1747.0 	^{+	241.3 	}_{	-212.5	}$ & $ (	4.10 	^{+	0.63 	}_{	-0.64 	})\times 10^{	53	}$ & $	2.85	^{+	0.48	}_{	-0.49	} $\\
081121	& $	871.0 	^{+	133.5 	}_{	-112.4	}$ & $ (	5.44 	^{+	1.46 	}_{	-1.34 	})\times 10^{	52	}$ & $	4.43	^{+	1.49	}_{	-1.31	} $\\
090323	& $	1901.1 	^{+	347.3 	}_{	-333.6	}$ & $ (	3.13 	^{+	0.60 	}_{	-0.53 	})\times 10^{	53	}$ & $	11.83	^{+	2.86	}_{	-2.42	} $\\
091020	& $	129.8 	^{+	19.2 	}_{	-19.2	}$ & $ (	1.80 	^{+	0.20 	}_{	-0.24 	})\times 10^{	52	}$ & $	6.54	^{+	1.62	}_{	-1.67	} $\\
\end{longtable}																					
																					
\begin{longtable}{lccc}					
\caption{Intrinsic property of GRBs for 54 bronze data set$^{\flat}$}
\label{tab:data3}
\hline\hline
GRB &$E_p$(keV)&$L_p$(erg~s$^{-1}$)	$^{\dagger}$&$T_L$ (sec)\\
\hline
\endhead
\hline
\multicolumn{4}{l}{$\flat$ The data with $E_{p}$ fitted with the CPL function}\\
\multicolumn{4}{l}{$\dagger$ 2.496 second peak luminosity in GRB rest frame}\\
\endfoot
040924	& $	124.6 	^{+	11.2 	}_{	-11.2	}$ & $ (	2.05 	^{+	0.22 	}_{	-0.22 	})\times 10^{	51	}$ & $	4.22	^{+	0.48	}_{	-0.48	} $\\
050126	& $	107.6 	^{+	52.7 	}_{	-18.3	}$ & $ (	6.75 	^{+	1.62 	}_{	-1.62 	})\times 10^{	50	}$ & $	11.67	^{+	3.02	}_{	-3.02	} $\\
050223	& $	110.0 	^{+	54.1 	}_{	-54.1	}$ & $ (	1.15 	^{+	0.16 	}_{	-0.16 	})\times 10^{	50	}$ & $	14.70	^{+	2.55	}_{	-2.55	} $\\
050315	& $	118.8 	^{+	25.0 	}_{	-33.6	}$ & $ (	6.34 	^{+	2.41 	}_{	-2.25 	})\times 10^{	51	}$ & $	12.13	^{+	5.57	}_{	-5.34	} $\\
050318	& $	114.9 	^{+	24.9 	}_{	-24.9	}$ & $ (	4.25 	^{+	0.95 	}_{	-0.78 	})\times 10^{	51	}$ & $	3.32	^{+	1.01	}_{	-0.84	} $\\
050319	& $	296.8 	^{+	296.8 	}_{	-148.4	}$ & $ (	9.97 	^{+	4.43 	}_{	-2.92 	})\times 10^{	51	}$ & $	5.69	^{+	3.38	}_{	-2.24	} $\\
050416A	& $	27.1 	^{+	4.6 	}_{	-4.6	}$ & $ (	3.67 	^{+	0.79 	}_{	-0.63 	})\times 10^{	50	}$ & $	2.59	^{+	0.92	}_{	-0.72	} $\\
050505	& $	658.8 	^{+	245.1 	}_{	-245.1	}$ & $ (	2.59 	^{+	0.35 	}_{	-0.39 	})\times 10^{	52	}$ & $	8.33	^{+	1.38	}_{	-1.48	} $\\
050803	& $	137.9 	^{+	48.3 	}_{	-48.3	}$ & $ (	1.23 	^{+	0.16 	}_{	-0.20 	})\times 10^{	50	}$ & $	24.90	^{+	3.84	}_{	-4.41	} $\\
050814	& $	340.2 	^{+	47.3 	}_{	-47.3	}$ & $ (	2.15 	^{+	0.65 	}_{	-0.66 	})\times 10^{	52	}$ & $	6.83	^{+	2.34	}_{	-2.39	} $\\
050820A	& $	888.6 	^{+	458.7 	}_{	-238.4	}$ & $ (	2.55 	^{+	0.24 	}_{	-0.35 	})\times 10^{	52	}$ & $	6.23	^{+	0.73	}_{	-1.04	} $\\
050904	& $	3180.6 	^{+	2443.8 	}_{	-1101.5	}$ & $ (	5.42 	^{+	1.48 	}_{	-1.52 	})\times 10^{	52	}$ & $	19.37	^{+	5.78	}_{	-8.27	} $\\
050908	& $	178.1 	^{+	39.1 	}_{	-21.7	}$ & $ (	5.78 	^{+	0.00 	}_{	-1.16 	})\times 10^{	51	}$ & $	3.99	^{+	0.50	}_{	-0.92	} $\\
050922C	& $	629.4 	^{+	204.7 	}_{	-118.3	}$ & $ (	1.91 	^{+	0.08 	}_{	-0.16 	})\times 10^{	52	}$ & $	2.85	^{+	0.18	}_{	-0.34	} $\\
051016B	& $	53.3 	^{+	25.6 	}_{	-23.8	}$ & $ (	3.11 	^{+	2.27 	}_{	-1.78 	})\times 10^{	50	}$ & $	2.42	^{+	2.05	}_{	-1.62	} $\\
051022	& $	550.8 	^{+	55.8 	}_{	-46.8	}$ & $ (	2.54 	^{+	0.33 	}_{	-0.33 	})\times 10^{	52	}$ & $	19.31	^{+	2.52	}_{	-2.52	} $\\
051109A	& $	466.8 	^{+	388.1 	}_{	-150.6	}$ & $ (	8.52 	^{+	1.49 	}_{	-1.91 	})\times 10^{	51	}$ & $	8.38	^{+	1.80	}_{	-2.14	} $\\
060115	& $	280.9 	^{+	140.4 	}_{	-45.3	}$ & $ (	1.01 	^{+	0.14 	}_{	-0.14 	})\times 10^{	52	}$ & $	8.85	^{+	1.44	}_{	-1.44	} $\\
060206	& $	380.6 	^{+	98.4 	}_{	-98.4	}$ & $ (	2.16 	^{+	0.13 	}_{	-0.13 	})\times 10^{	52	}$ & $	2.54	^{+	0.20	}_{	-0.20	} $\\
060210	& $	731.6 	^{+	1964.0 	}_{	-171.9	}$ & $ (	3.43 	^{+	0.92 	}_{	-0.61 	})\times 10^{	52	}$ & $	16.08	^{+	6.87	}_{	-2.99	} $\\
060218	& $	5.1 	^{+	0.3 	}_{	-0.3	}$ & $ (	3.00 	^{+	1.00 	}_{	-1.00 	})\times 10^{	46	}$ & $	289.35	^{+	106.80	}_{	-114.95	} $\\
060223A	& $	384.1 	^{+	541.0 	}_{	-54.1	}$ & $ (	1.91 	^{+	0.25 	}_{	-0.29 	})\times 10^{	52	}$ & $	2.65	^{+	1.09	}_{	-0.45	} $\\
060510B	& $	560.5 	^{+	354.0 	}_{	-177.0	}$ & $ (	1.37 	^{+	0.27 	}_{	-0.27 	})\times 10^{	52	}$ & $	29.27	^{+	5.92	}_{	-5.87	} $\\
060522	& $	427.7 	^{+	79.4 	}_{	-79.4	}$ & $ (	1.10 	^{+	0.37 	}_{	-0.38 	})\times 10^{	52	}$ & $	8.83	^{+	3.05	}_{	-3.19	} $\\
060526	& $	105.5 	^{+	21.1 	}_{	-21.1	}$ & $ (	6.68 	^{+	0.88 	}_{	-0.79 	})\times 10^{	51	}$ & $	9.25	^{+	2.04	}_{	-1.84	} $\\
060604	& $	147.2 	^{+	18.4 	}_{	-18.4	}$ & $ (	1.51 	^{+	0.25 	}_{	-0.25 	})\times 10^{	51	}$ & $	9.77	^{+	3.04	}_{	-3.04	} $\\
060707	& $	278.8 	^{+	92.9 	}_{	-44.3	}$ & $ (	1.09 	^{+	0.25 	}_{	-0.26 	})\times 10^{	52	}$ & $	7.06	^{+	1.74	}_{	-1.82	} $\\
060714	& $	233.8 	^{+	107.6 	}_{	-107.6	}$ & $ (	9.00 	^{+	0.96 	}_{	-1.06 	})\times 10^{	51	}$ & $	15.33	^{+	3.01	}_{	-2.07	} $\\
060814	& $	437.7 	^{+	207.8 	}_{	-98.8	}$ & $ (	4.25 	^{+	0.58 	}_{	-1.10 	})\times 10^{	51	}$ & $	12.26	^{+	1.75	}_{	-4.80	} $\\
060908	& $	517.9 	^{+	631.1 	}_{	-140.6	}$ & $ (	2.22 	^{+	0.25 	}_{	-1.27 	})\times 10^{	52	}$ & $	4.40	^{+	0.92	}_{	-2.86	} $\\
060927	& $	475.2 	^{+	165.0 	}_{	-72.6	}$ & $ (	3.33 	^{+	0.21 	}_{	-0.21 	})\times 10^{	52	}$ & $	3.46	^{+	0.30	}_{	-0.30	} $\\
070508	& $	342.2 	^{+	14.6 	}_{	-14.6	}$ & $ (	1.36 	^{+	0.17 	}_{	-0.18 	})\times 10^{	52	}$ & $	7.08	^{+	0.89	}_{	-1.03	} $\\
070521	& $	344.8 	^{+	41.9 	}_{	-32.6	}$ & $ (	2.57 	^{+	0.49 	}_{	-0.67 	})\times 10^{	51	}$ & $	7.90	^{+	1.52	}_{	-2.46	} $\\
070714B	& $	2150.4 	^{+	1497.6 	}_{	-729.6	}$ & $ (	6.24 	^{+	0.75 	}_{	-1.61 	})\times 10^{	51	}$ & $	1.75	^{+	0.65	}_{	-0.53	} $\\
071003	& $	2080.9 	^{+	322.9 	}_{	-260.4	}$ & $ (	8.58 	^{+	1.34 	}_{	-1.55 	})\times 10^{	52	}$ & $	3.94	^{+	0.65	}_{	-0.87	} $\\
071010B	& $	101.2 	^{+	12.5 	}_{	-12.5	}$ & $ (	4.93 	^{+	0.19 	}_{	-0.23 	})\times 10^{	51	}$ & $	5.27	^{+	0.45	}_{	0.38	} $\\
071020	& $	1012.7 	^{+	251.6 	}_{	-166.7	}$ & $ (	4.12 	^{+	0.76 	}_{	-2.30 	})\times 10^{	52	}$ & $	2.22	^{+	0.43	}_{	-1.85	} $\\
071117	& $	648.0 	^{+	550.1 	}_{	-184.1	}$ & $ (	1.74 	^{+	0.30 	}_{	-0.77 	})\times 10^{	52	}$ & $	2.31	^{+	0.41	}_{	-1.39	} $\\
080411	& $	525.8 	^{+	71.1 	}_{	-54.8	}$ & $ (	3.65 	^{+	0.46 	}_{	-0.46 	})\times 10^{	52	}$ & $	6.33	^{+	0.85	}_{	-0.84	} $\\
080413A	& $	583.6 	^{+	274.6 	}_{	-137.3	}$ & $ (	2.08 	^{+	0.80 	}_{	-0.57 	})\times 10^{	52	}$ & $	3.18	^{+	1.26	}_{	-1.33	} $\\
080603B	& $	262.0 	^{+	59.0 	}_{	-59.0	}$ & $ (	9.03 	^{+	0.52 	}_{	-0.52 	})\times 10^{	51	}$ & $	9.45	^{+	0.67	}_{	-0.67	} $\\
080605	& $	665.2 	^{+	52.8 	}_{	-44.9	}$ & $ (	9.78 	^{+	2.03 	}_{	-2.03 	})\times 10^{	52	}$ & $	2.37	^{+	0.50	}_{	-0.50	} $\\
080607	& $	1691.1 	^{+	185.7 	}_{	-153.4	}$ & $ (	7.92 	^{+	1.59 	}_{	-1.59 	})\times 10^{	53	}$ & $	2.07	^{+	0.43	}_{	-0.43	} $\\
080810	& $	1363.7 	^{+	320.2 	}_{	-320.2	}$ & $ (	5.43 	^{+	0.47 	}_{	-0.47 	})\times 10^{	52	}$ & $	7.46	^{+	0.84	}_{	-0.84	} $\\
080913	& $	716.4 	^{+	431.7 	}_{	-431.7	}$ & $ (	3.58 	^{+	0.51 	}_{	-0.71 	})\times 10^{	52	}$ & $	2.08	^{+	0.37	}_{	-0.47	} $\\
080916A	& $	184.1 	^{+	15.2 	}_{	-15.2	}$ & $ (	1.63 	^{+	0.25 	}_{	-0.25 	})\times 10^{	51	}$ & $	16.13	^{+	5.92	}_{	-5.92	} $\\
090102	& $	1148.7 	^{+	185.9 	}_{	-147.7	}$ & $ (	3.55 	^{+	0.52 	}_{	-0.50 	})\times 10^{	52	}$ & $	6.05	^{+	1.04	}_{	-1.02	} $\\
090418	& $	1590.9 	^{+	1382.2 	}_{	-427.7	}$ & $ (	8.64 	^{+	1.41 	}_{	-2.48 	})\times 10^{	51	}$ & $	18.52	^{+	3.72	}_{	-5.74	} $\\
090423	& $	754.4 	^{+	138.0 	}_{	-138.0	}$ & $ (	7.18 	^{+	1.34 	}_{	-1.65 	})\times 10^{	52	}$ & $	1.41	^{+	0.47	}_{	-0.50	} $\\
090715B	& $	536.0 	^{+	224.0 	}_{	-120.0	}$ & $ (	4.10 	^{+	1.14 	}_{	-1.14 	})\times 10^{	52	}$ & $	5.00	^{+	1.60	}_{	-1.51	} $\\
090812	& $	1974.5 	^{+	866.5 	}_{	-548.9	}$ & $ (	2.37 	^{+	0.21 	}_{	-0.30 	})\times 10^{	52	}$ & $	17.74	^{+	2.78	}_{	-3.23	} $\\
090926B	& $	175.4 	^{+	15.7 	}_{	-15.7	}$ & $ (	4.40 	^{+	0.58 	}_{	-0.66 	})\times 10^{	51	}$ & $	14.52	^{+	1.97	}_{	-2.22	} $\\
091018	& $	37.8 	^{+	35.5 	}_{	-21.7	}$ & $ (	4.42 	^{+	0.58 	}_{	-3.49 	})\times 10^{	51	}$ & $	2.35	^{+	0.59	}_{	-1.89	} $\\
091029	& $	230.4 	^{+	65.7 	}_{	-65.7	}$ & $ (	1.45 	^{+	0.08 	}_{	-0.08 	})\times 10^{	52	}$ & $	6.57	^{+	0.70	}_{	-0.50	} $\\
\end{longtable}						
\section{Monte Carlo simulation}
\label{sec:MC}

Here we show the validity and limitation of our method using
Monte Carlo simulations. For simplicity, we consider a relation
between 2 quantities, X and Y, in this section.

\subsection{A case with two populations with different normalizations}

Here we consider two populations whose relations are, respectively,
\begin{eqnarray}
\label{eq:MC1}
\log Y_{1} &=& 48 + 2 \log X_1, \\ 
\log Y_{2} &=& 48.3 + 2 \log X_2
\end{eqnarray}
with $\sigma_{\rm int} = 0$ for both of them.

We generate mock data according to the following equations,
\begin{eqnarray}
X_{i_1} &=& U(100,1500) \\
\sigma_{X_{i_1}} / X_{i_1}
&=& \sigma_{Y_{i_1}} / Y_{i_1}
 =  U(\sigma_{\rm min},\sigma_{\rm max})
~~~~~~~~~~~~~~~~~ (1 \leqq i_1 \leqq N_1) \\
Y_{i_1}
&=& 48 + 2 \log X_{i_1}
    + G(0,\sqrt{\sigma_{\log Y_{i_1}}^2 + B^2 \sigma_{\log X_{i_1}}^2
                + \sigma_{\rm int}^2}) \\
X_{i_2} &=& U(100,1500) \\
\sigma_{X_{i_2}} / X_{i_2}
&=& \sigma_{Y_{i_2}} / Y_{i_2}
 =  U(\sigma_{\rm min},\sigma_{\rm max})
~~~~~~~~~~~~~~~~~ (1 \leqq i_2 \leqq N_2) \\
Y_{i_2}
&=& 48.3 + 2 \log X_{i_2}
    + G(0,\sqrt{\sigma_{\log Y_{i_2}}^2 + B^2 \sigma_{\log X_{i_2}}^2
                + \sigma_{\rm int}^2})
\end{eqnarray}
where $U({\rm min}, {\rm max})$ represents a random number from
a uniform distribution between min and max, and $G(m, \sigma_{\rm SD})$
represents a random number from a Gaussian distribution which has
mean value $m$ and standard deviation $\sigma_{\rm SD}$.
We generate $N_1$ and $N_2$ ($N_1 > N_2$) samples for each population
with $100 \sigma_{X_i} \%$ and $100 \sigma_{Y_i} \%$ observational errors
in $X$ and $Y$, respectively. Then, following the method described in
\S-\ref{sec:method}, we obtain a tentative set of best-fit model
parameters and the intrinsic dispersion, and identify outliers. 
We assume the fitting function of the form
$\log Y = A + B \log (X/{\bar X}) $. 

\begin{figure}
 \begin{center}
 \begin{tabular}{ccc}
  \FigureFile(55mm,60mm){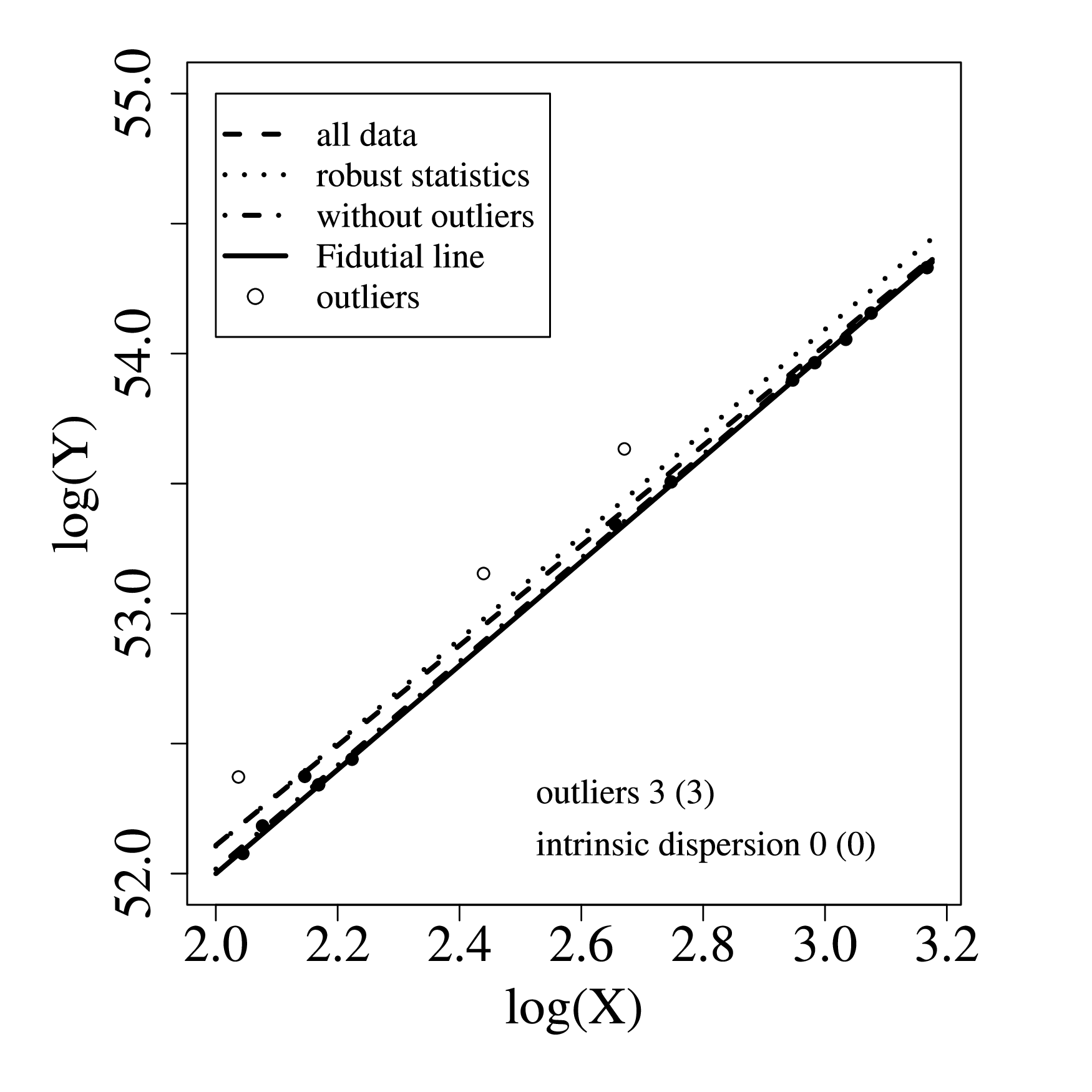}
  \FigureFile(55mm,60mm){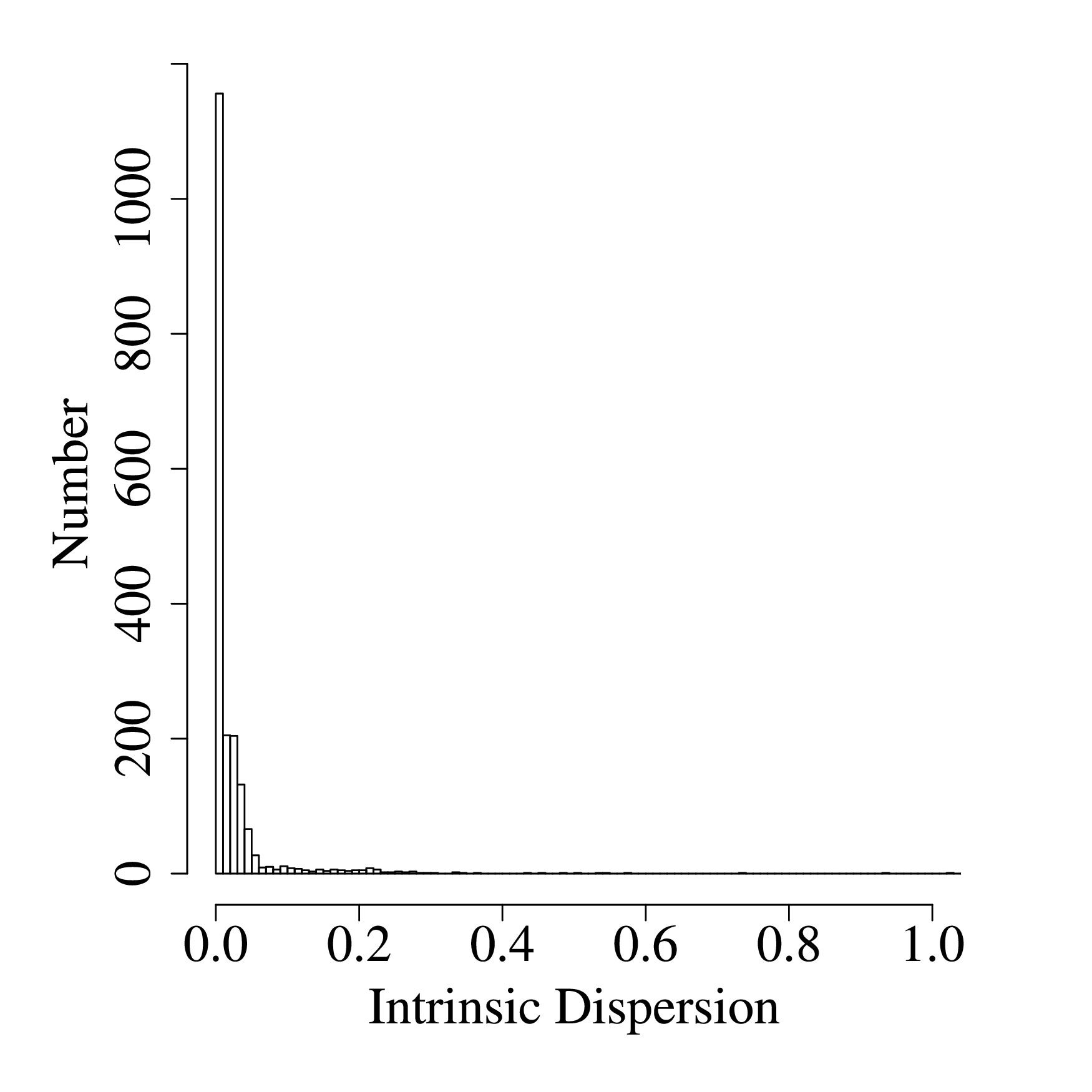}
  \FigureFile(55mm,60mm){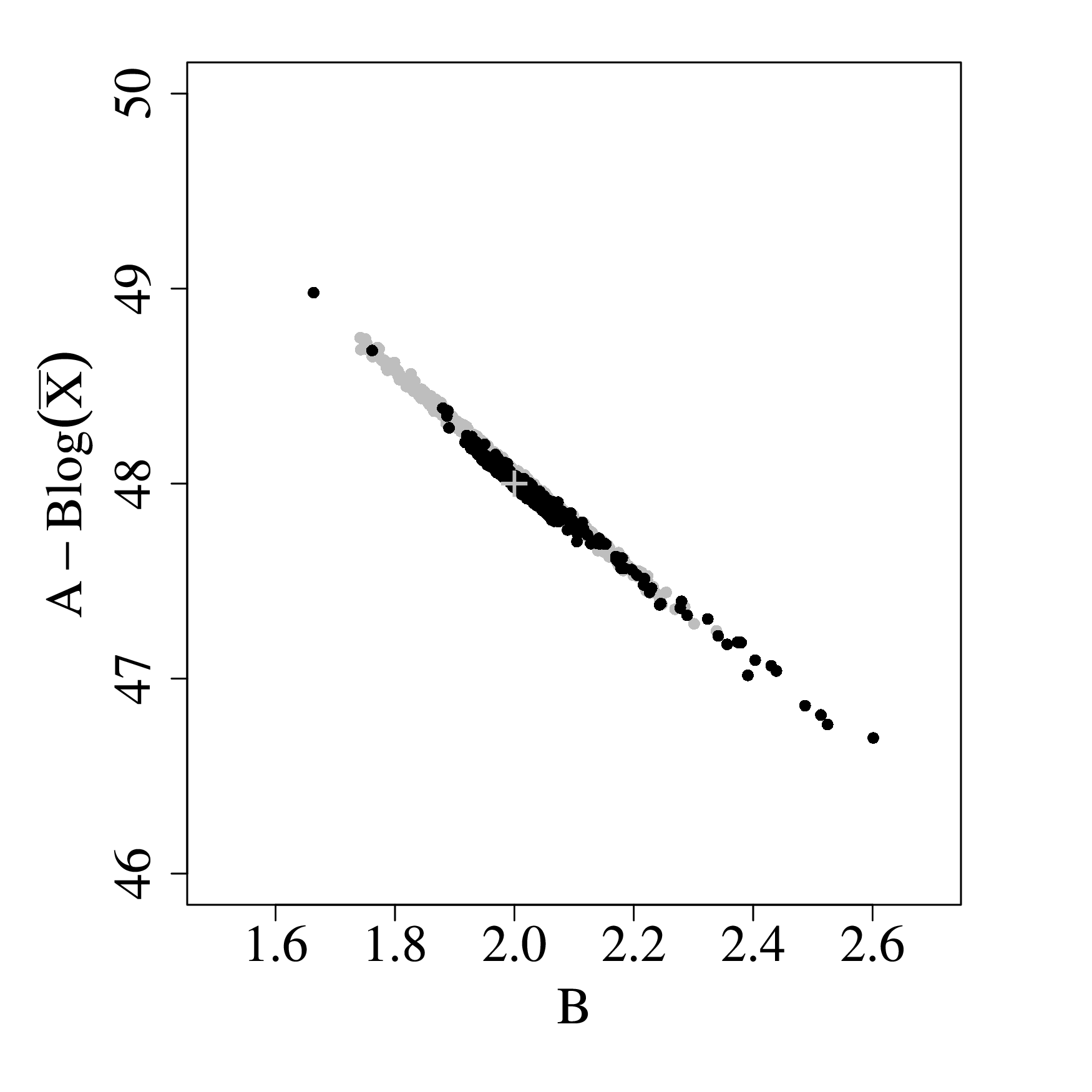}\\
  \FigureFile(55mm,60mm){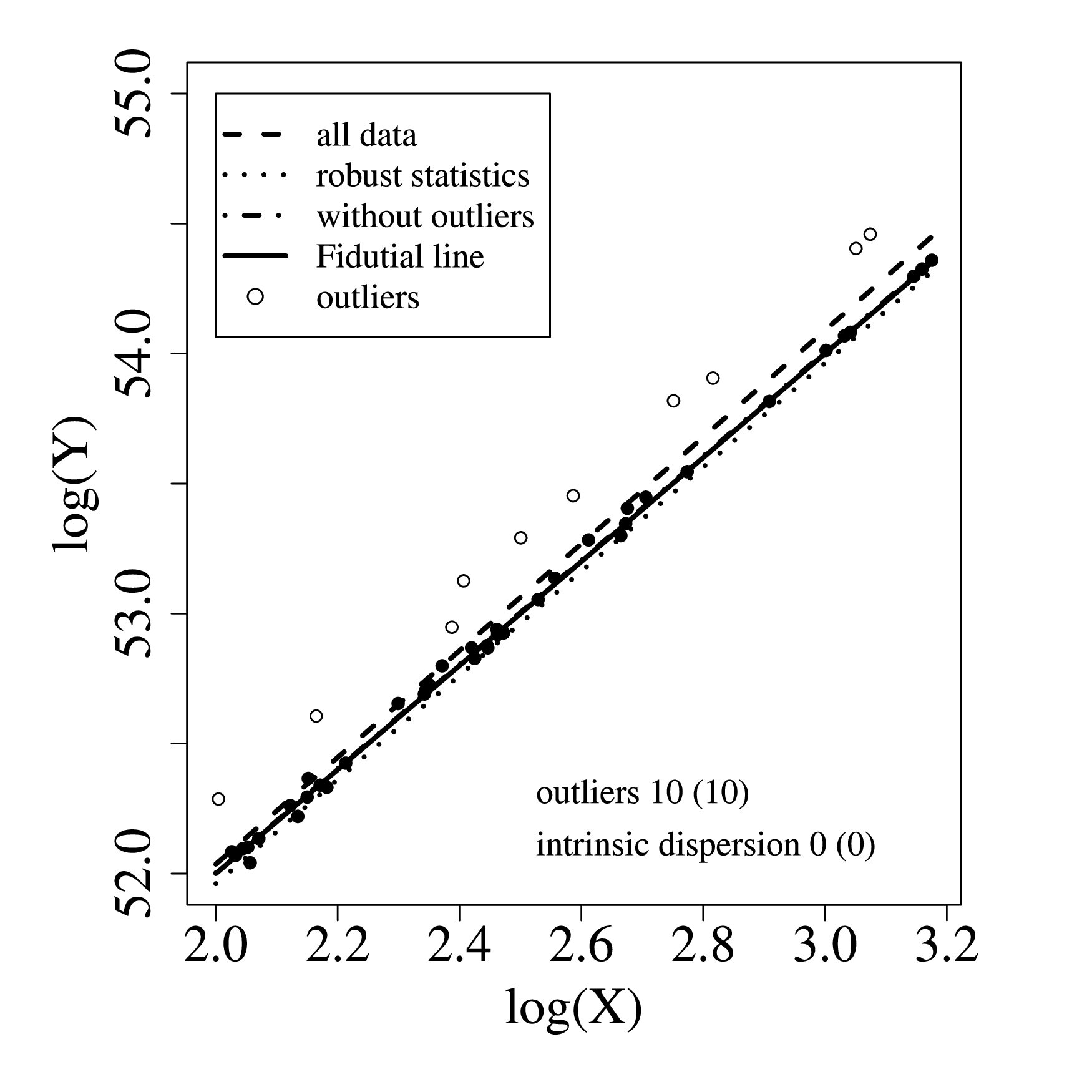}
  \FigureFile(55mm,60mm){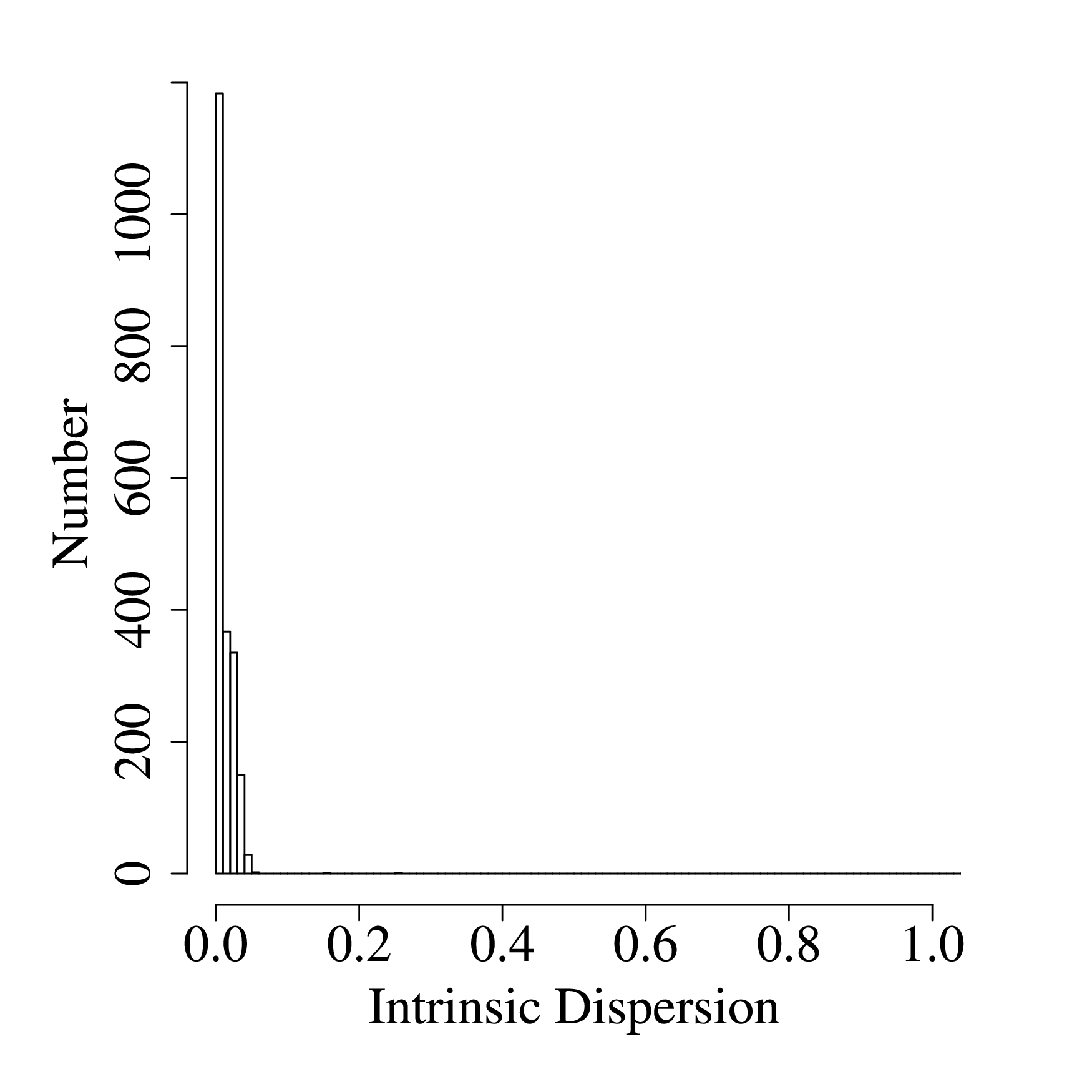}
  \FigureFile(55mm,60mm){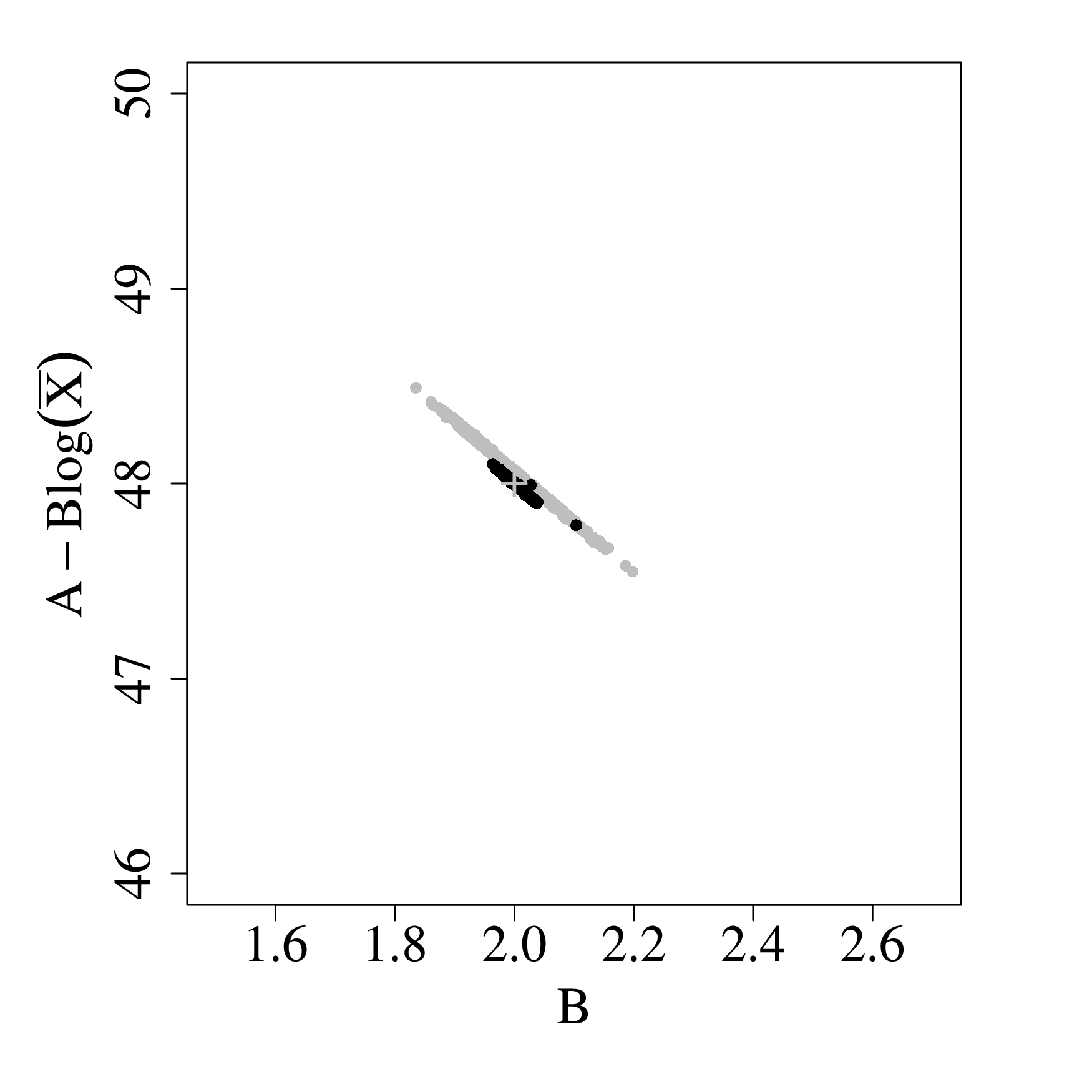}\\
  \FigureFile(55mm,60mm){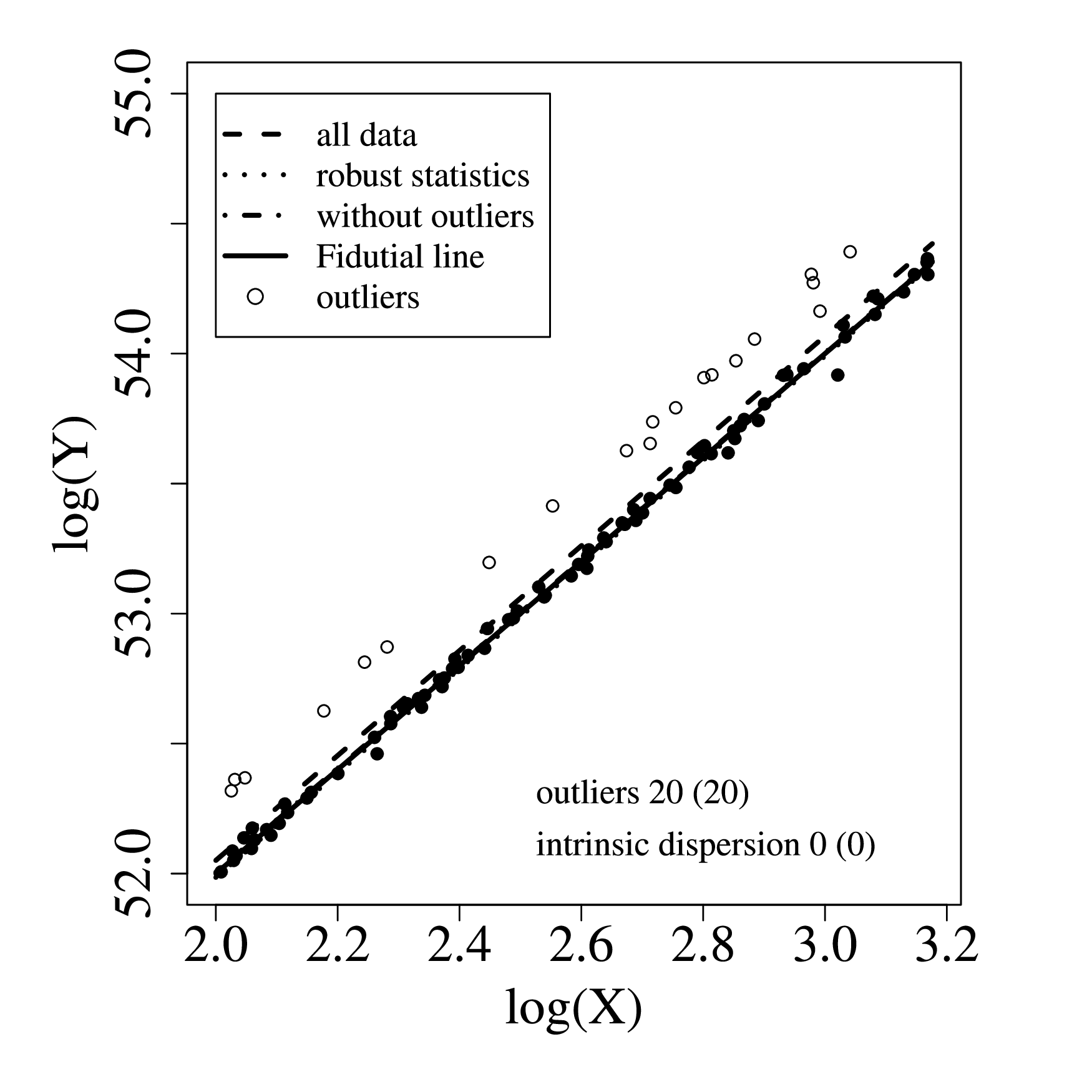}
  \FigureFile(55mm,60mm){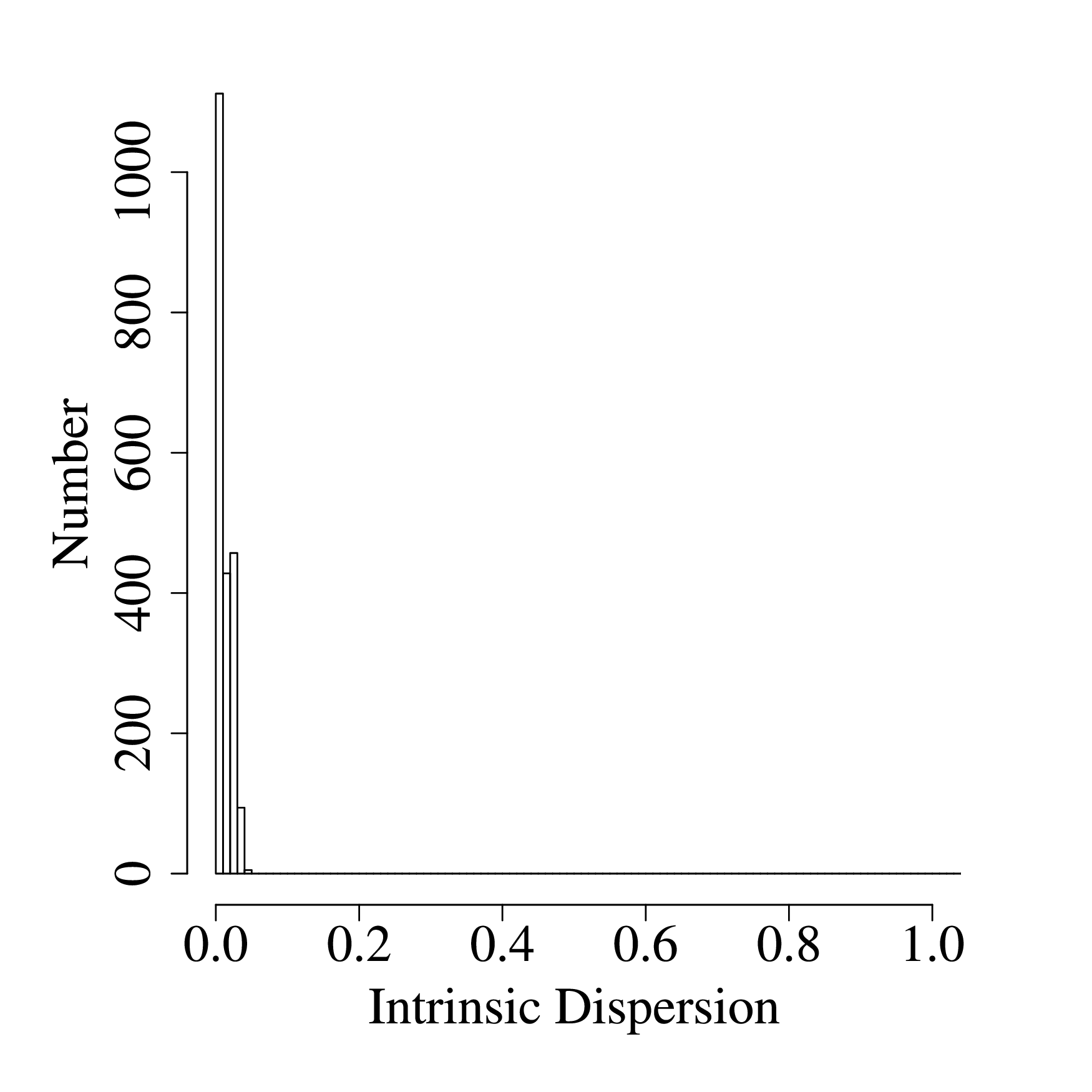}
  \FigureFile(55mm,60mm){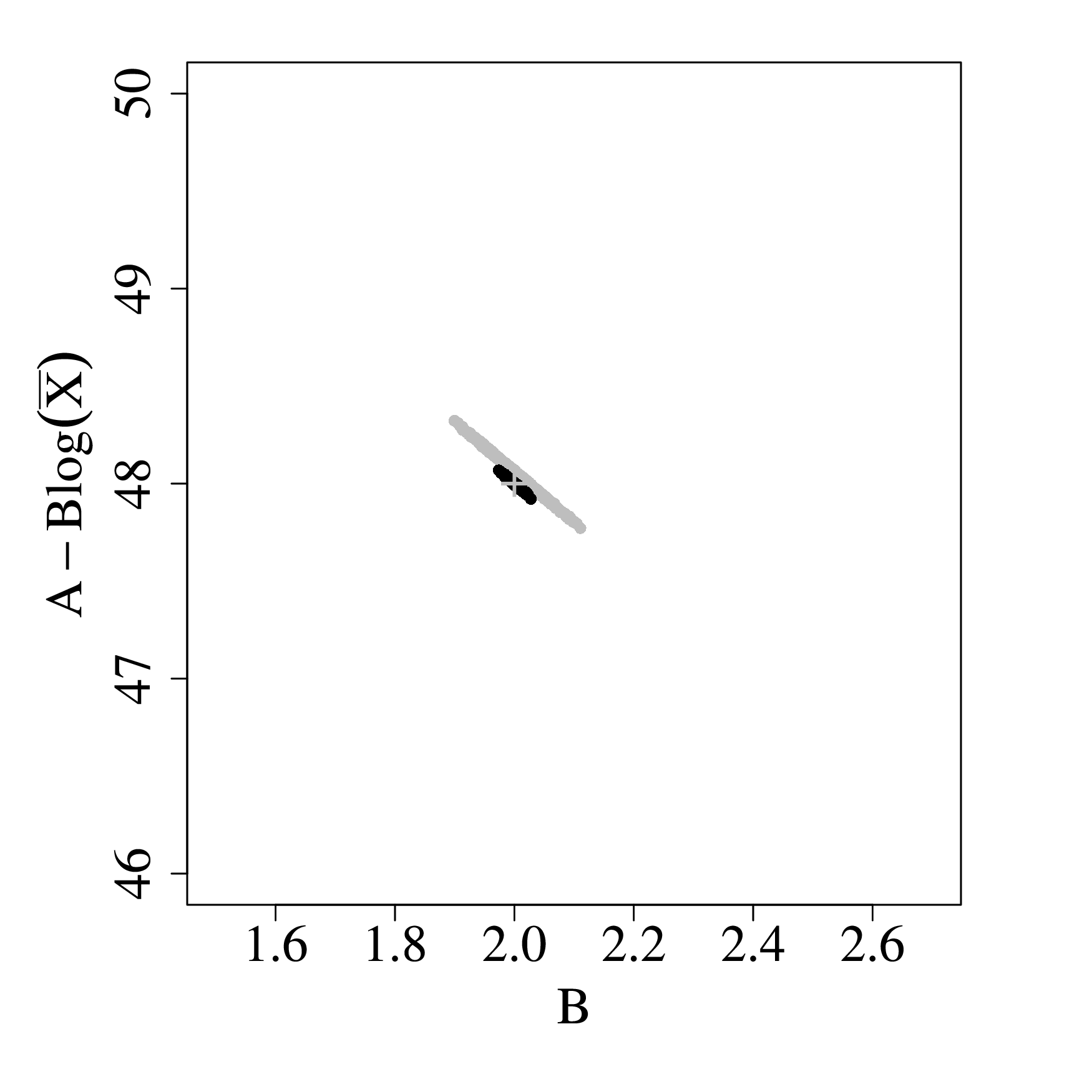}\\
 \end{tabular}
 \end{center}
\caption{Results of Monte Carlo simulations with two populations
whose normalization factors are slightly different. The number
of samples is ($N_1,N_2$) = (12,3) (top), (40,10) (middle) and
(80,20) (bottom), respectively. The left figures show the result
of one realization and each line represents the best-fit line
for ordinary regression (dashed line), robust regression
(dotted line), and our method (dash-dotted line). The assumed
relation Eq~(\ref{eq:MC1}) is also indicated by the thick line.
All lines except the ordinary regression are overlapping.
The white points indicate the outliers which are detected
by our method. The central figures show the histogram of
the estimated $\sigma_{\rm int}$. As the number of sample increases,
the value of $\sigma_{\rm int}$ converges to the true value.
We show distribution of the value of the estimated parameters
in the right figures. The parameters estimated after outlier
elimination are indicated by black points, while the ones obtained
by the normal regression with all samples are indicated by gray points.
See also table~\ref{tab:MC1}.}
\label{fig:MC1}
\end{figure}

First, we consider a case with relatively small observational
errors, $({\rm min}, {\rm max}) = (0.01,0.05)$.
In figure~\ref{fig:MC1}, we show the result of 1000 simulations
for ($N_1,N_2$) = (12,3) (top), (40,10) (middle) and (80,20) (bottom),
respectively. The left figures show the result of one realization
and each line represents the best-fit line for ordinary regression
(dashed line), robust regression (dotted line), and our method
(dash-dotted line). The assumed relation Eq~(\ref{eq:MC1}) is also
indicated by the thick line. All lines except the ordinary regression
are overlapping. The white points indicate the outliers which are
detected by our method. The central figures show the histogram of
the estimated $\sigma_{\rm int}$. As the number of sample increases,
the value of $\sigma_{\rm int}$ converges to the true value.
We show distribution of the value of the estimated parameters
in the right figures. The parameters estimated after outlier
elimination are indicated by black points, while the ones obtained
by the normal regression with all samples are indicated by gray points.
Apparently our method is more effective than ordinary regression.
We summarized the parameters and results in table~\ref{tab:MC1}.
Figure~\ref{fig:MC1} and table~\ref{tab:MC1} show that our method
gives more reasonable results than ordinary regression in a case
with two populations whose normalizations are slightly different
like Cepheid variables.

\begin{table}
 \caption{Results of Monte Carlo simulations with two populations.}
 \label{tab:MC1}
 \begin{center}
  \begin{tabular}{cccc}
   \hline
   $N_{\rm total}$ & $[\sigma_{\rm min},\sigma_{\rm max}]$ &
   $\bar \sigma_{\rm int}$ (fiducial) &
   $\bar N_{\rm out}$ (fiducial)\\ 
   \hline

   15 & [0.01,0.05] & 0.02 (0) & 2.95 (3)\\ 
   50 & [0.01,0.05] & 0.01 (0) & 10.09 (10)\\ 
   100 & [0.01,0.05] & 0.01 (0) & 20.1 (20)\\ 
   \hline
  \end{tabular}
 \end{center}
\end{table}

\begin{figure}
 \begin{center}
 \begin{tabular}{ccc}
  \FigureFile(55mm,60mm){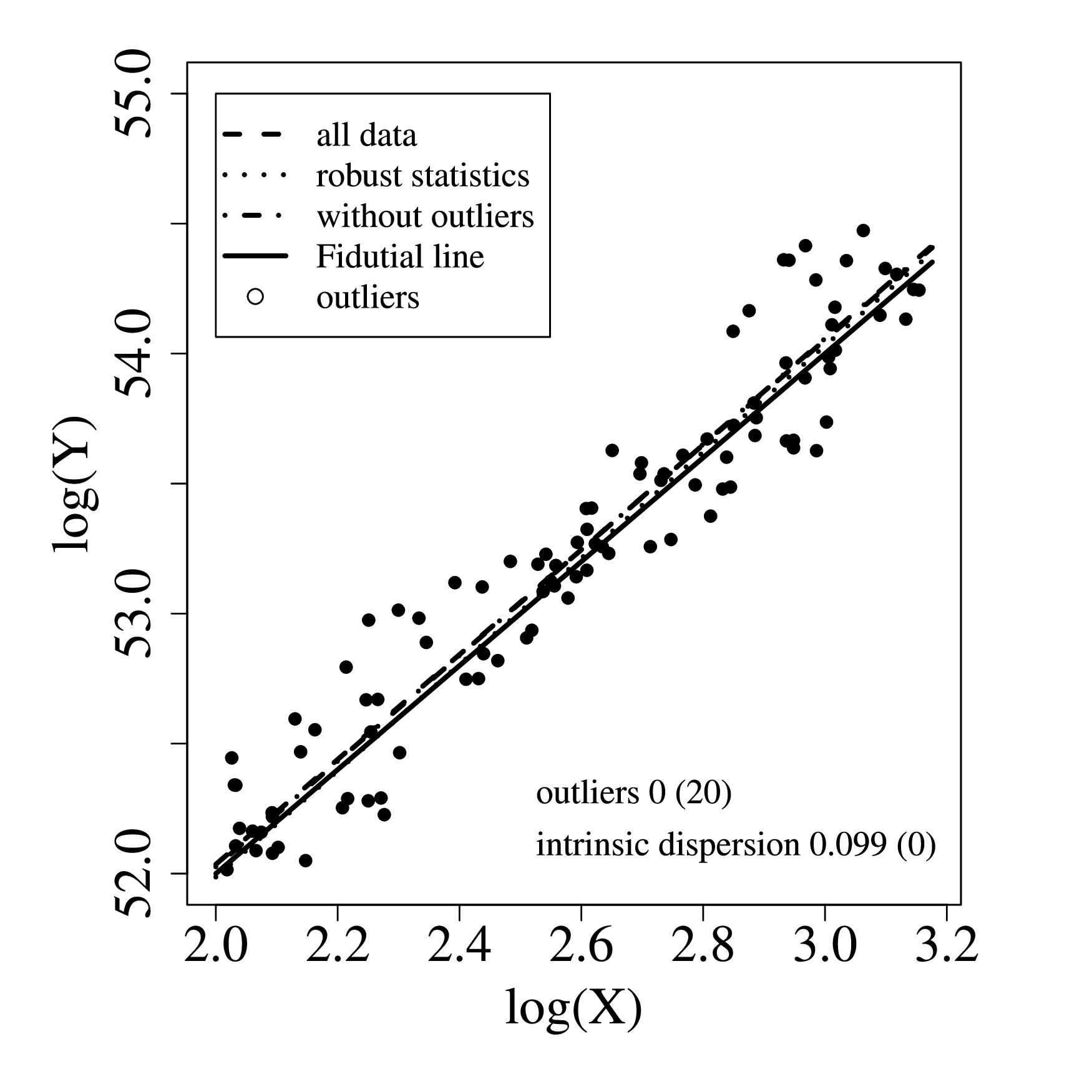}
  \FigureFile(55mm,60mm){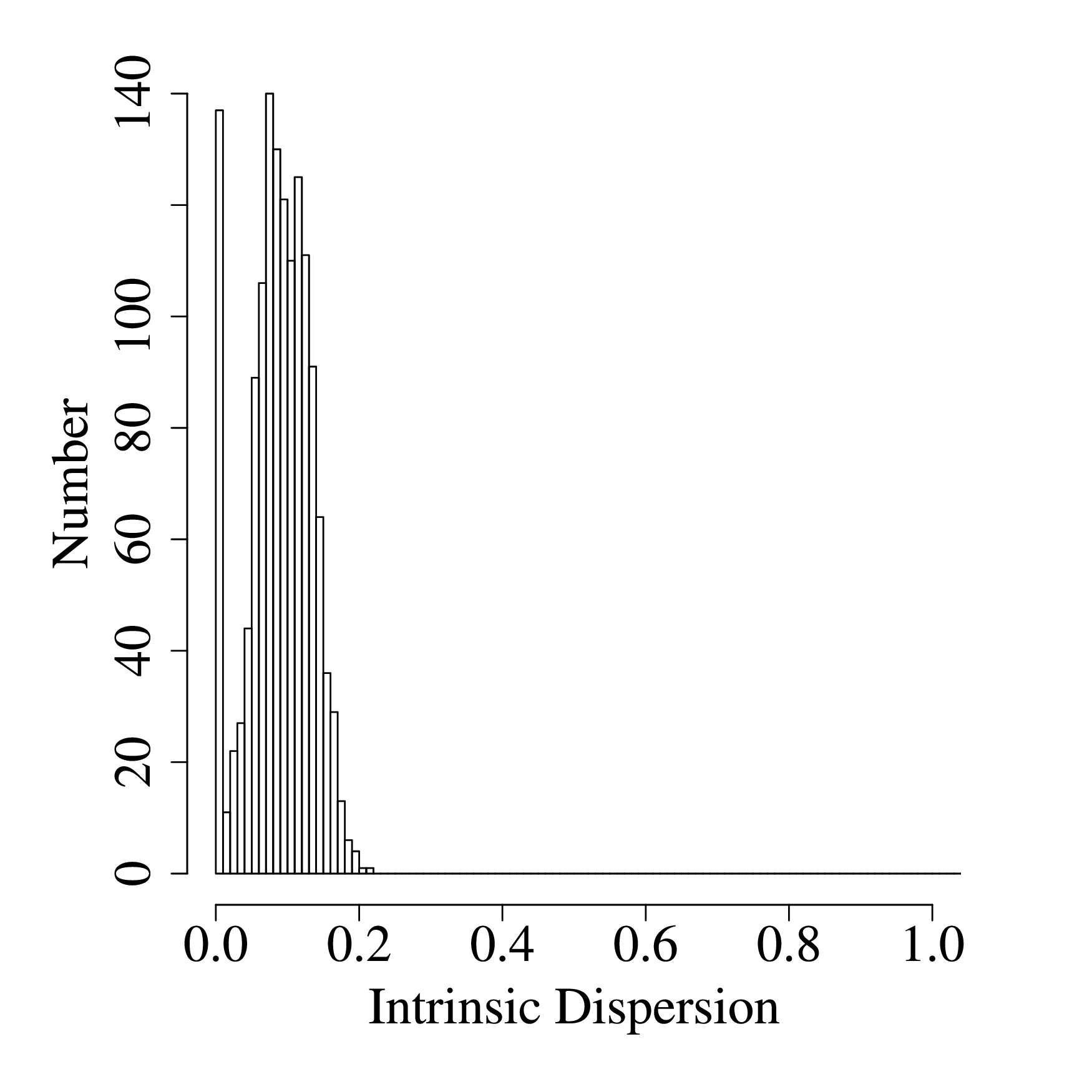}
  \FigureFile(55mm,60mm){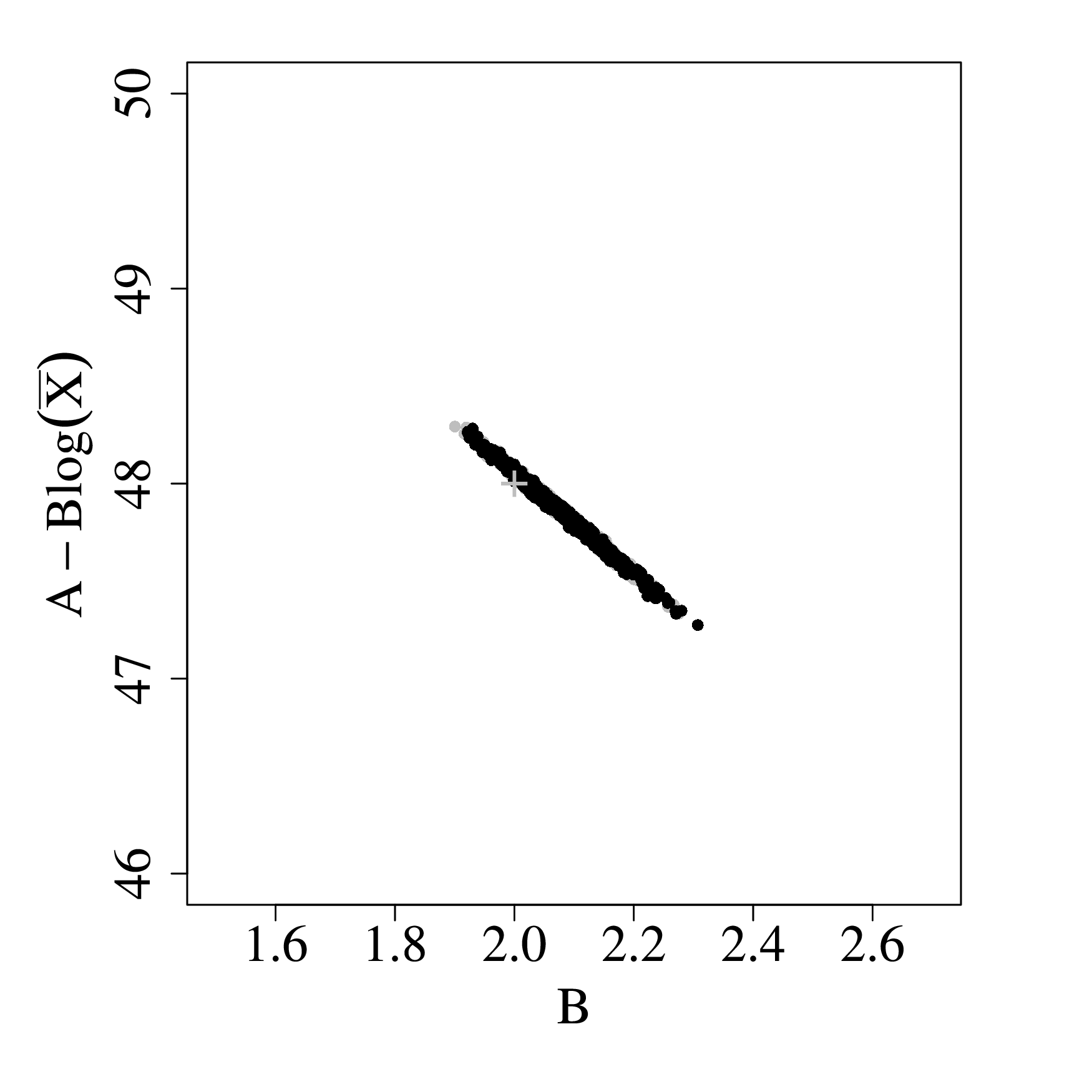}\\
  \FigureFile(55mm,60mm){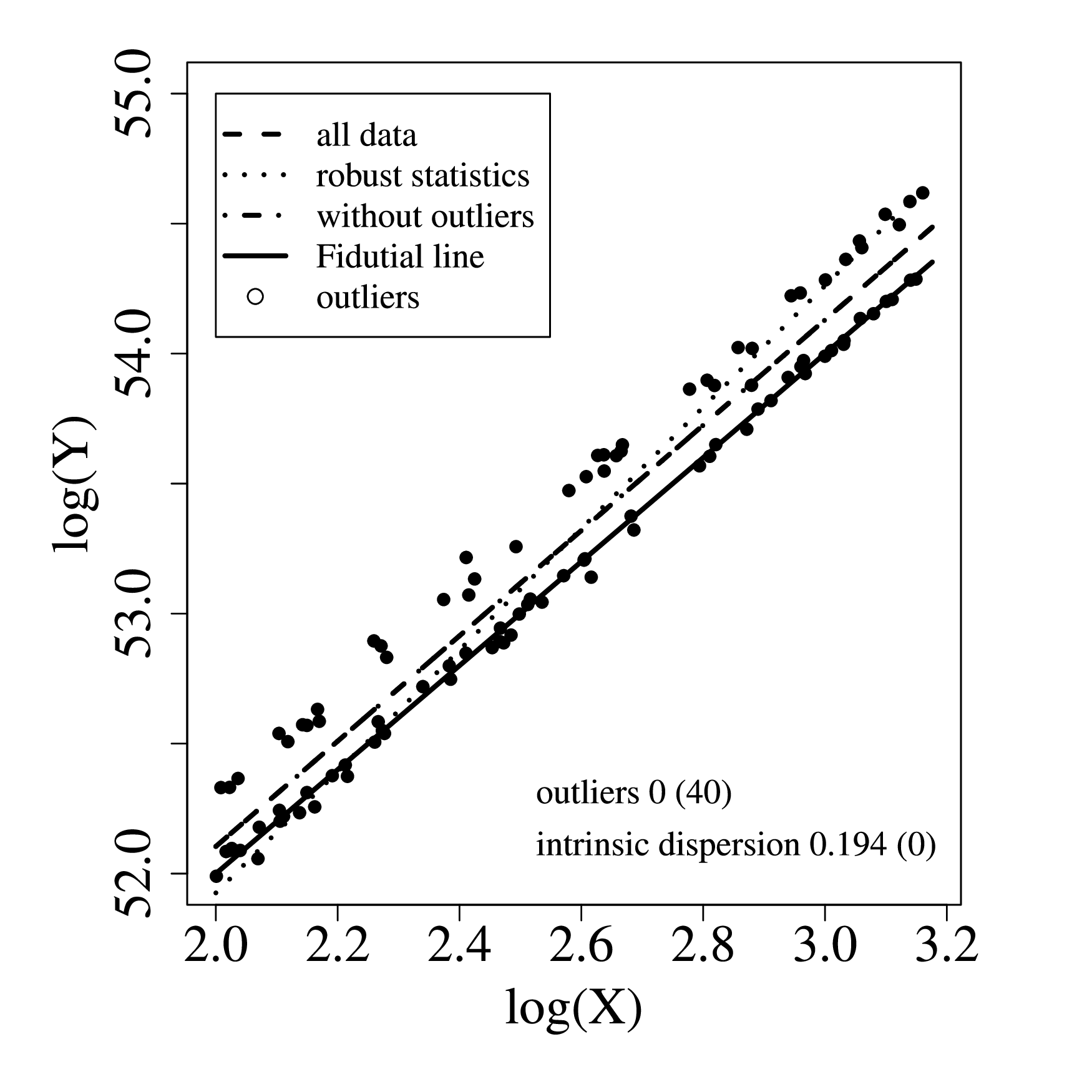}
  \FigureFile(55mm,60mm){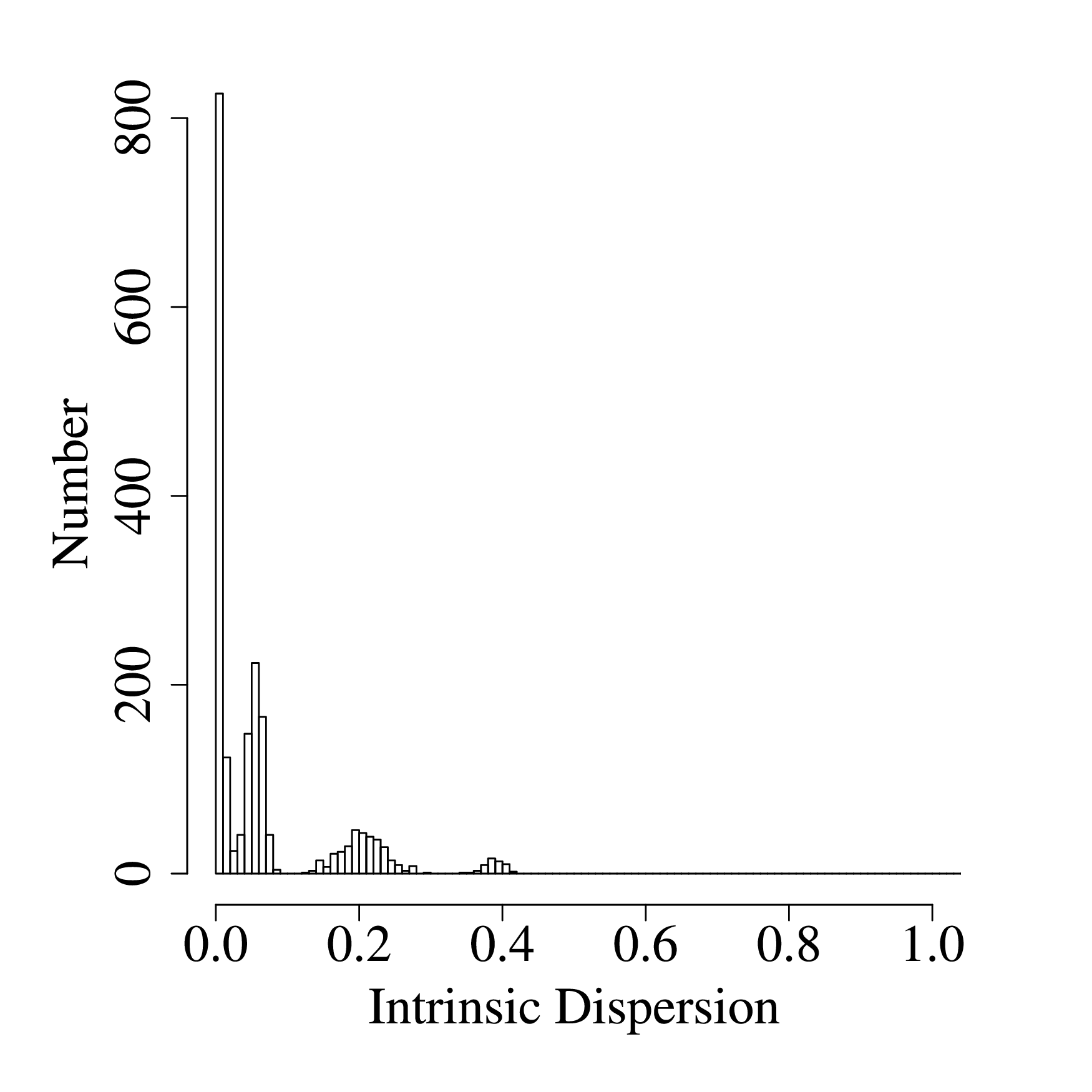}
  \FigureFile(55mm,60mm){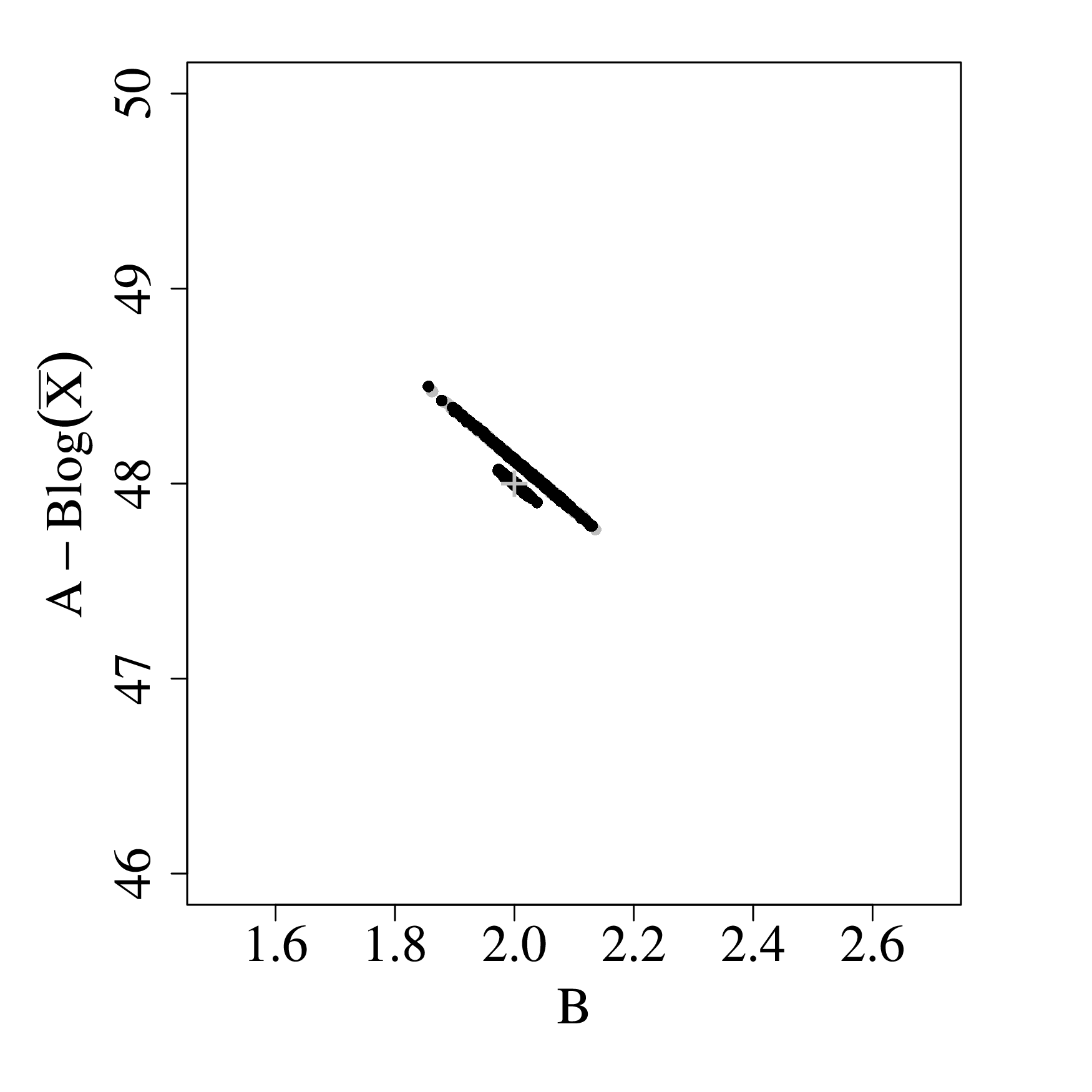}\\
 \end{tabular}
 \end{center}
\caption{Results of Monte Carlo simulations with two populations
whose normalization factors are slightly different. The assumed
two correlations are the same as in figure~\ref{fig:MC1},
but we consider two cases where observational uncertainties are
larger (top) and where the fraction of outliers is larger (bottom).
The meaning of points and lines are the same as in figure~\ref{fig:MC1}.
See also table~\ref{tab:MC4}.}
\label{fig:MC4}
\end{figure}

Next, we consider a case with relatively large observational
errors, $({\rm min}, {\rm max}) = (0.1,0.2)$, to show the limitation
of our method. Top left of figure~\ref{fig:MC4} shows that
there are no points eliminated as outlier, although the assumed
number of outliers is $20$. Thus, if observational uncertainties
are larger than the difference in the normalization of
the two relations, it is difficult to detect outliers.
This is why we use only samples with small observational uncertainties.
Likewise, if there are too many outliers (Bottom), $40\%$ rather
than $20\%$, it is also difficult to detect outliers correctly.
The result is summarized in table~\ref{tab:MC4}.

\begin{table}
 \caption{Results of Monte Carlo simulations with two populations
 with large observational uncertainties or large fraction of outliers.}
 \label{tab:MC4}
 \begin{center}
  \begin{tabular}{cccc}
   \hline
   $N_{\rm total}$ & $[\sigma_{\rm min},\sigma_{\rm max}]$ &
   $\bar \sigma_{\rm int}$ (fiducial) &
   $\bar N_{\rm out}$ (fiducial)\\ 
   \hline
    100 & [0.1,0.2] &0.09 (0) & 0.62 (20)\\ 
   100 & [0.01,0.05] &0.01 (0) & 26.8 (40)\\ 
   \hline
  \end{tabular}
 \end{center}
\end{table}

\subsection{A case with intrinsic dispersion}

Next we consider a single population with the correlation defined as,
\begin{equation}
\label{eq:MC2}
\log Y= 48 + 2 \log X.
\end{equation}
Here we assume that this relation has the intrinsic dispersion
of $\sigma_{\rm int} = 0.20$, which is much larger than observational
errors ($\leq 0.05$).

We generate mock data according to the following equations,
\begin{eqnarray}
X_i &=& U(100,1500) \\
\sigma_{X_i} / X_i
&=& \sigma_{Y_i} / Y_i
 =  U(\sigma_{\rm min}, \sigma_{\rm max}) \\
Y_i &=& 48 + 2 \log X_i
        + G(0,\sqrt{\sigma_{\log Y_i}^2 + B^2 \sigma_{\log X_i}^2
                    + \sigma_{\rm int}^2}).
\end{eqnarray}

\begin{figure}
 \begin{center}
 \begin{tabular}{ccc}
  \FigureFile(55mm,60mm){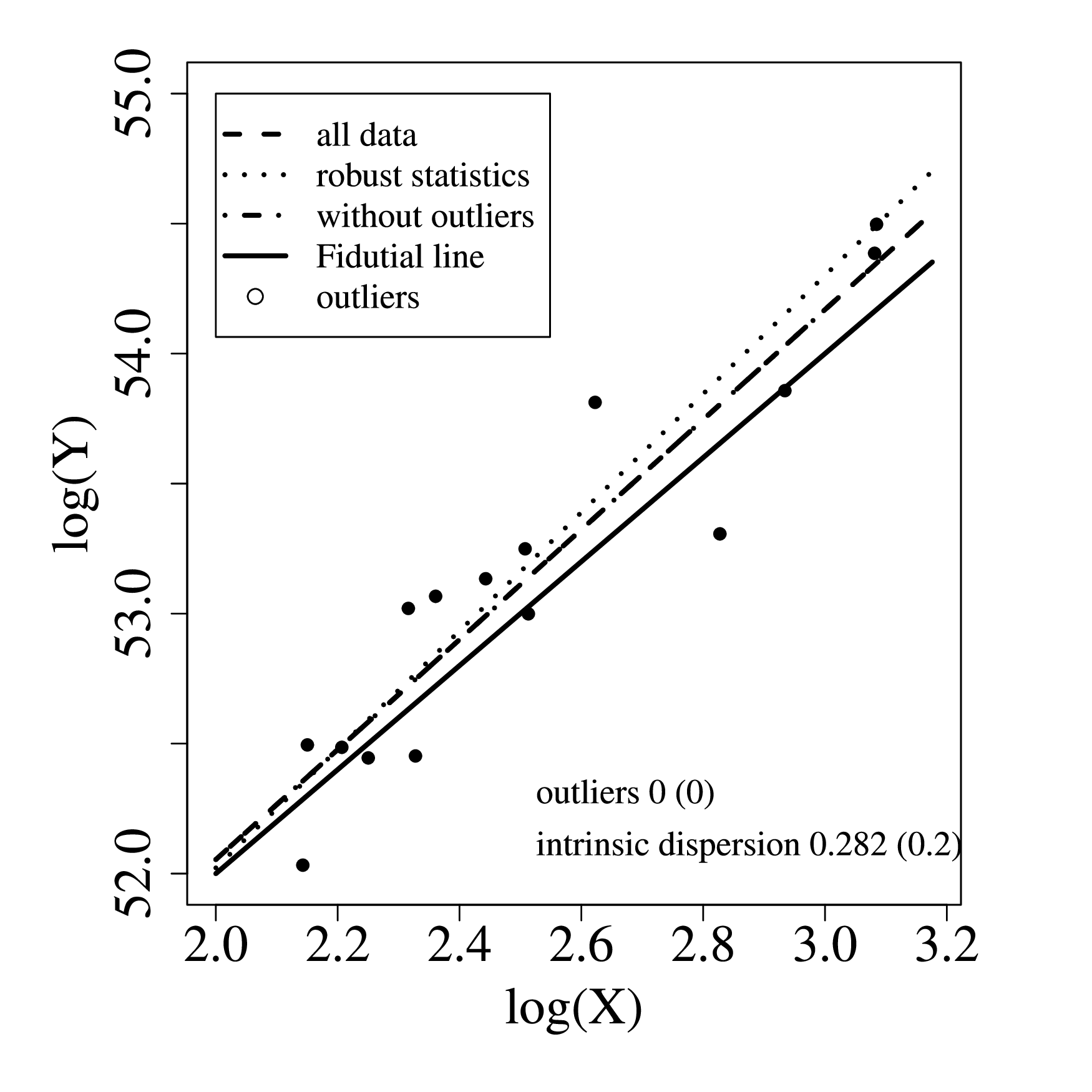}
  \FigureFile(55mm,60mm){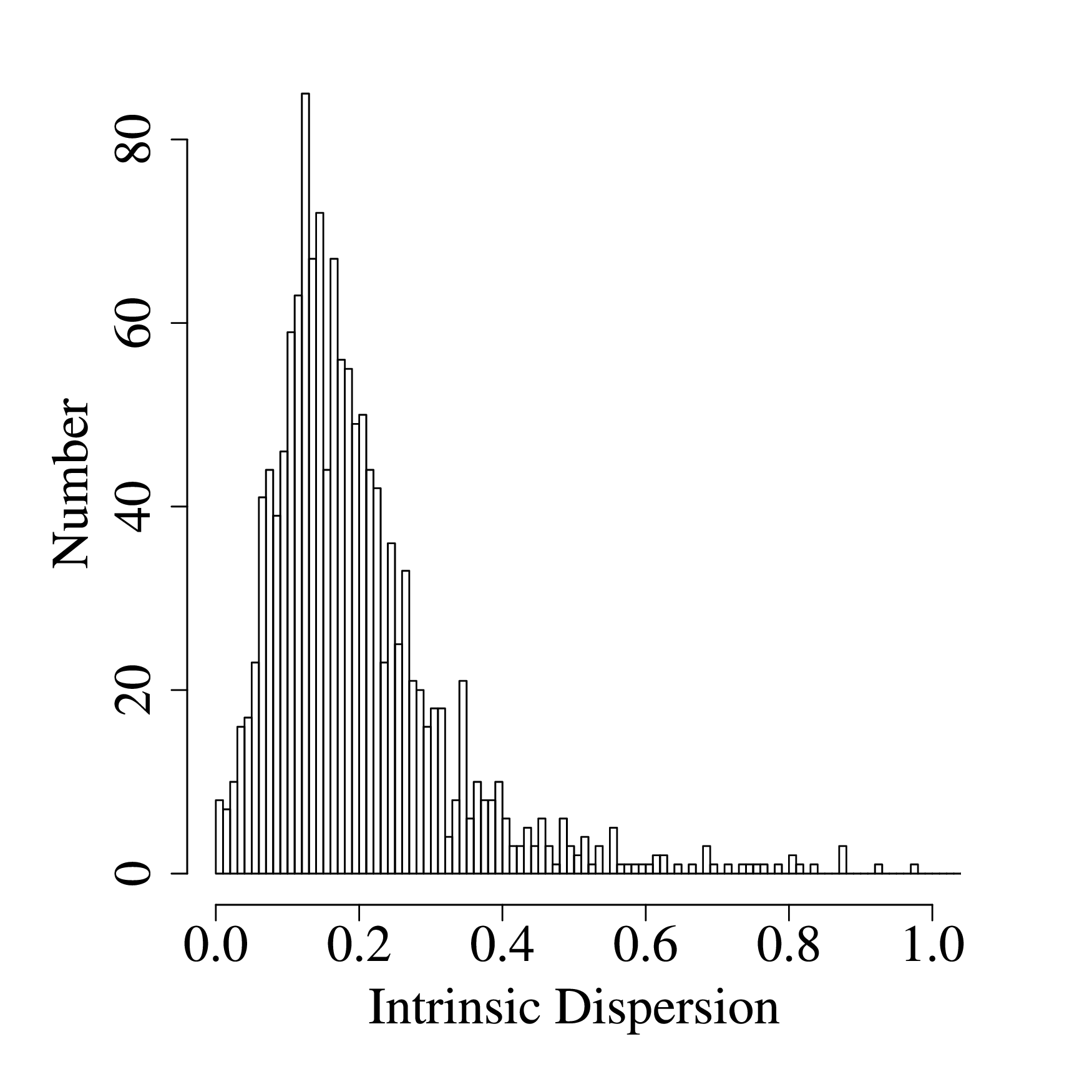}
  \FigureFile(55mm,60mm){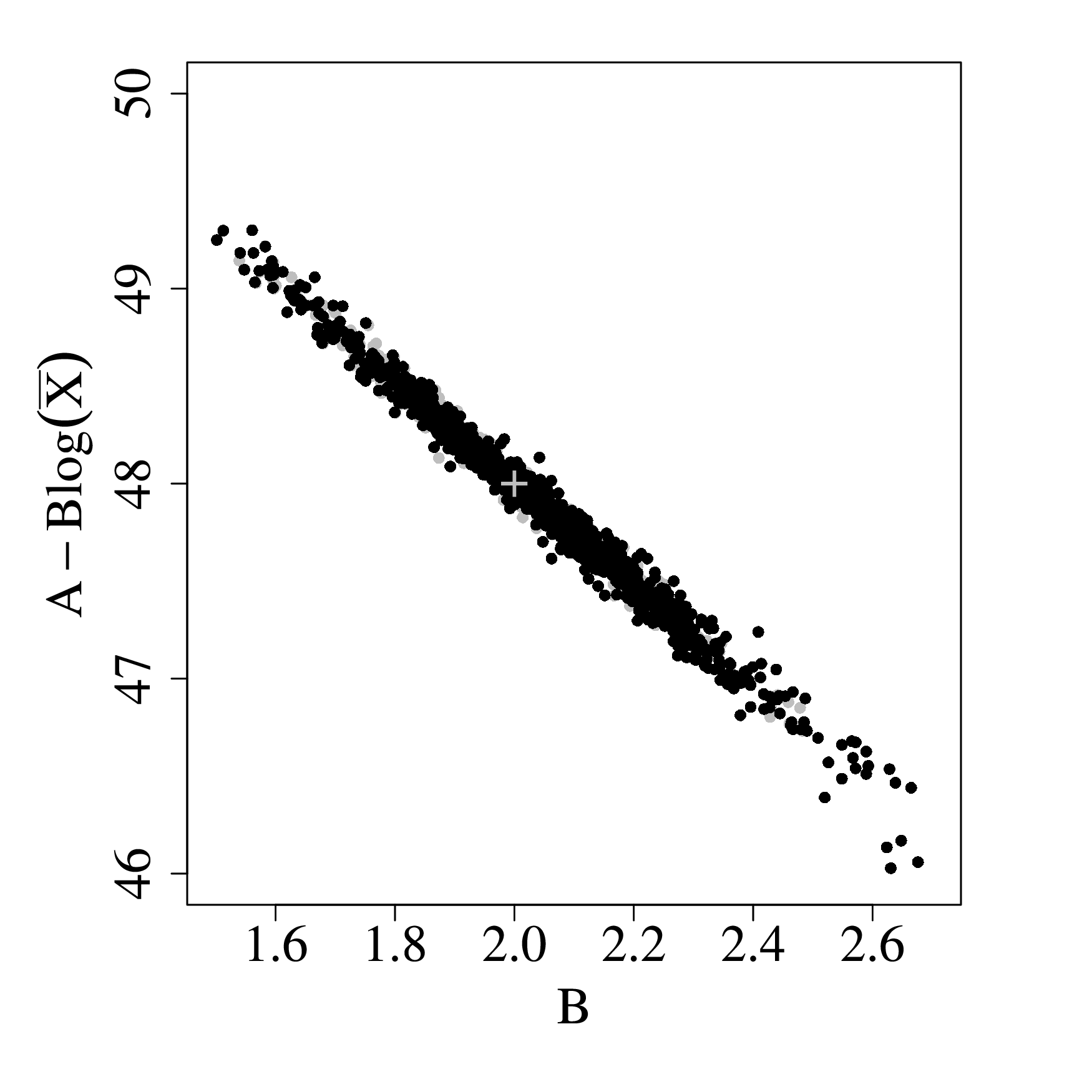}\\
  \FigureFile(55mm,60mm){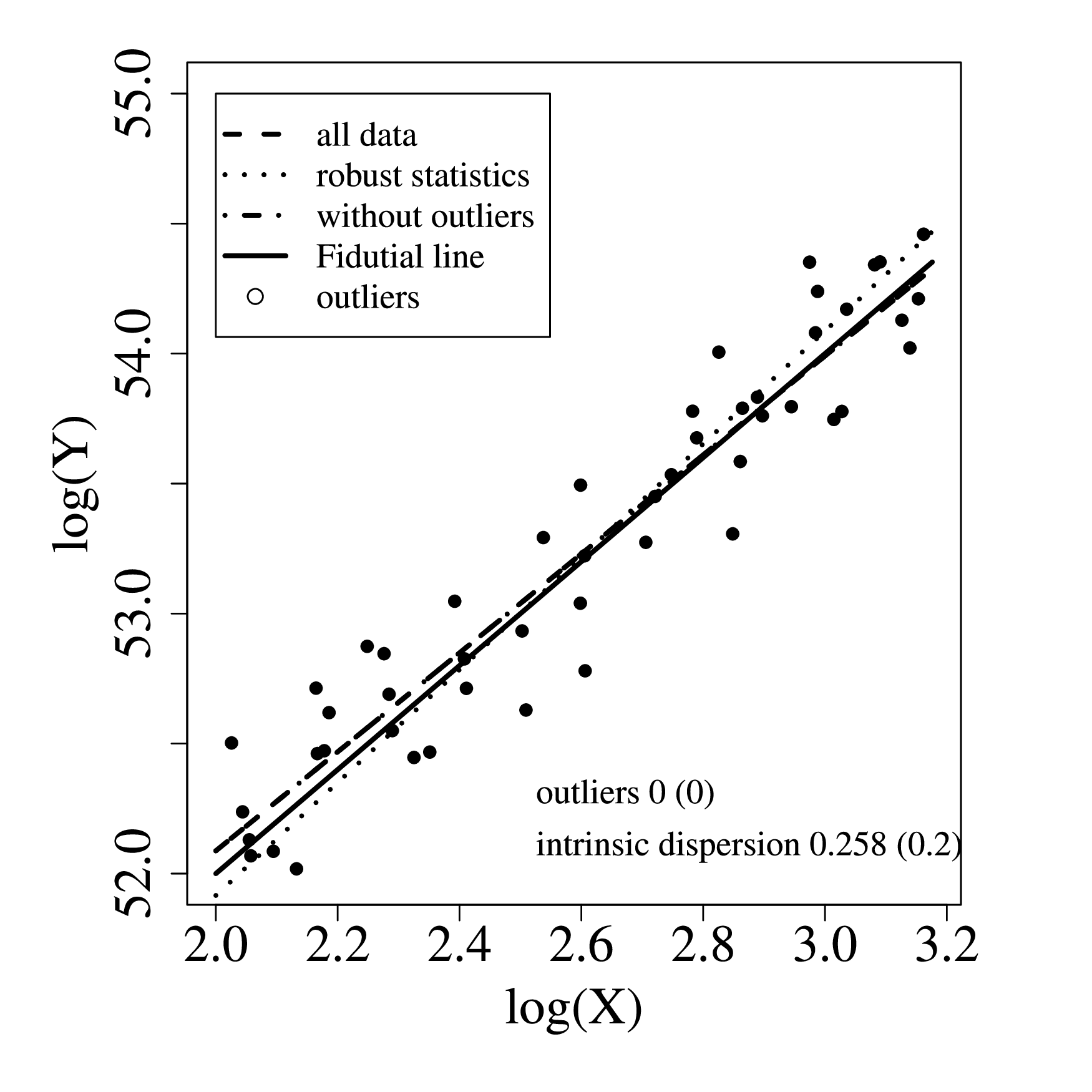}
  \FigureFile(55mm,60mm){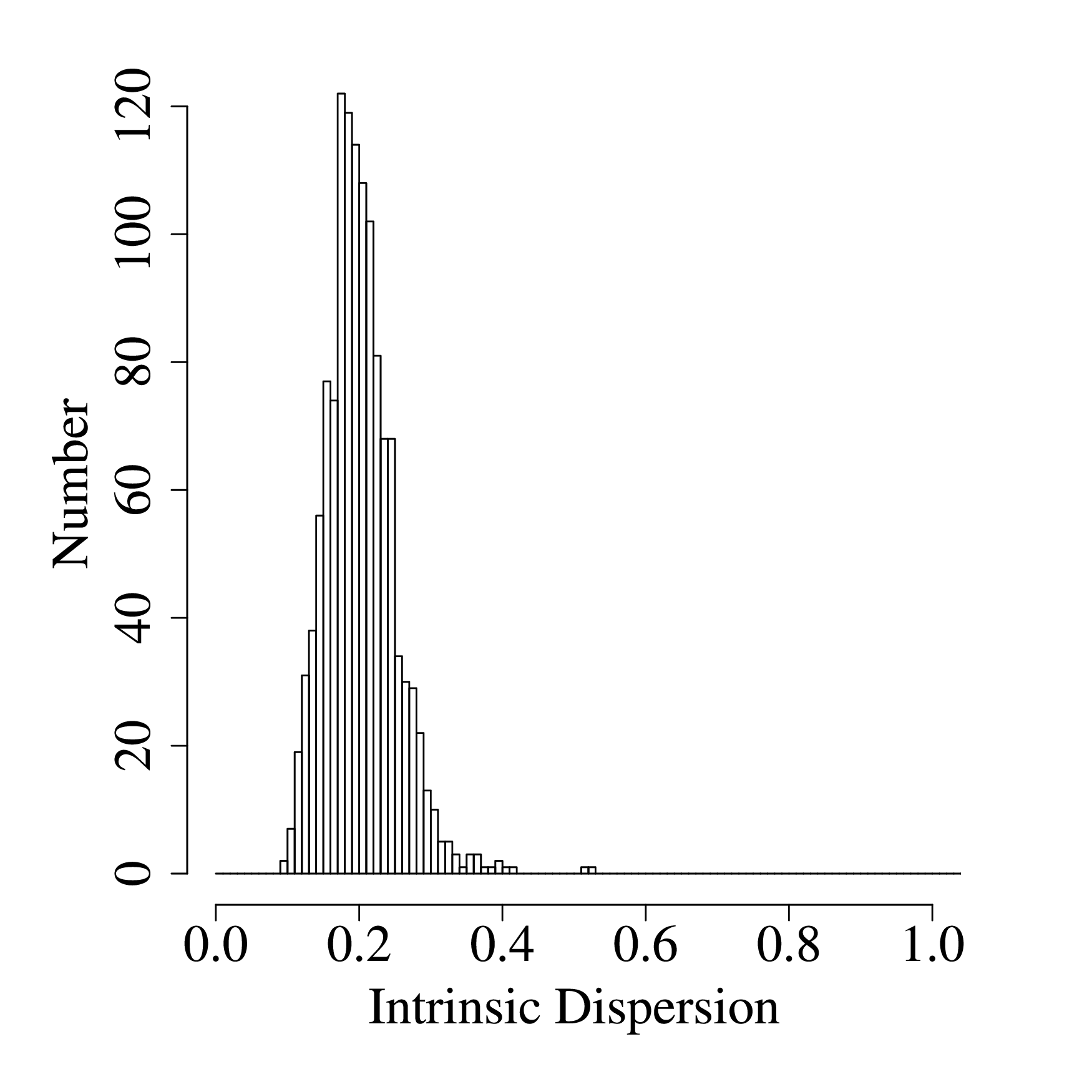}
  \FigureFile(55mm,60mm){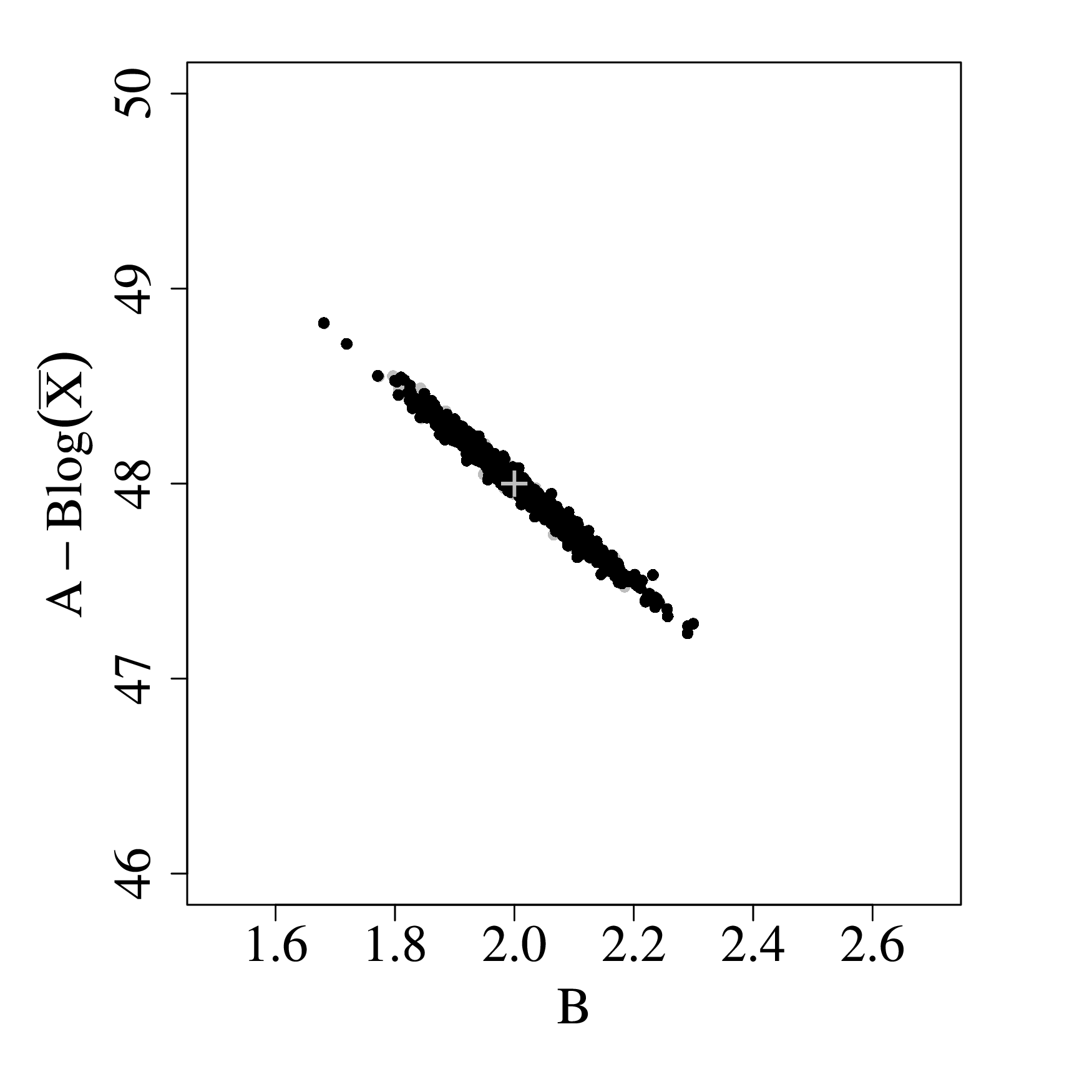}\\
  \FigureFile(55mm,60mm){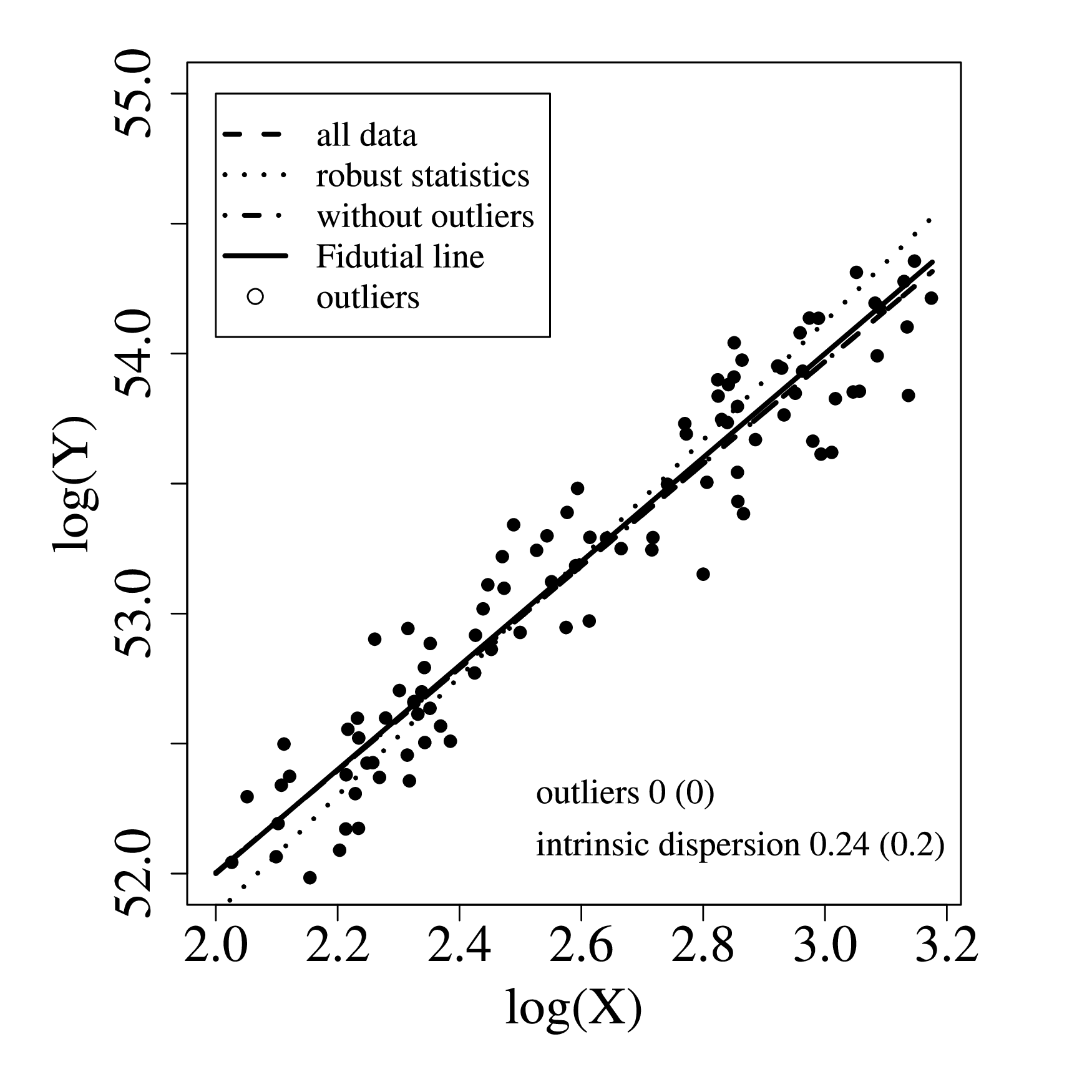}
  \FigureFile(55mm,60mm){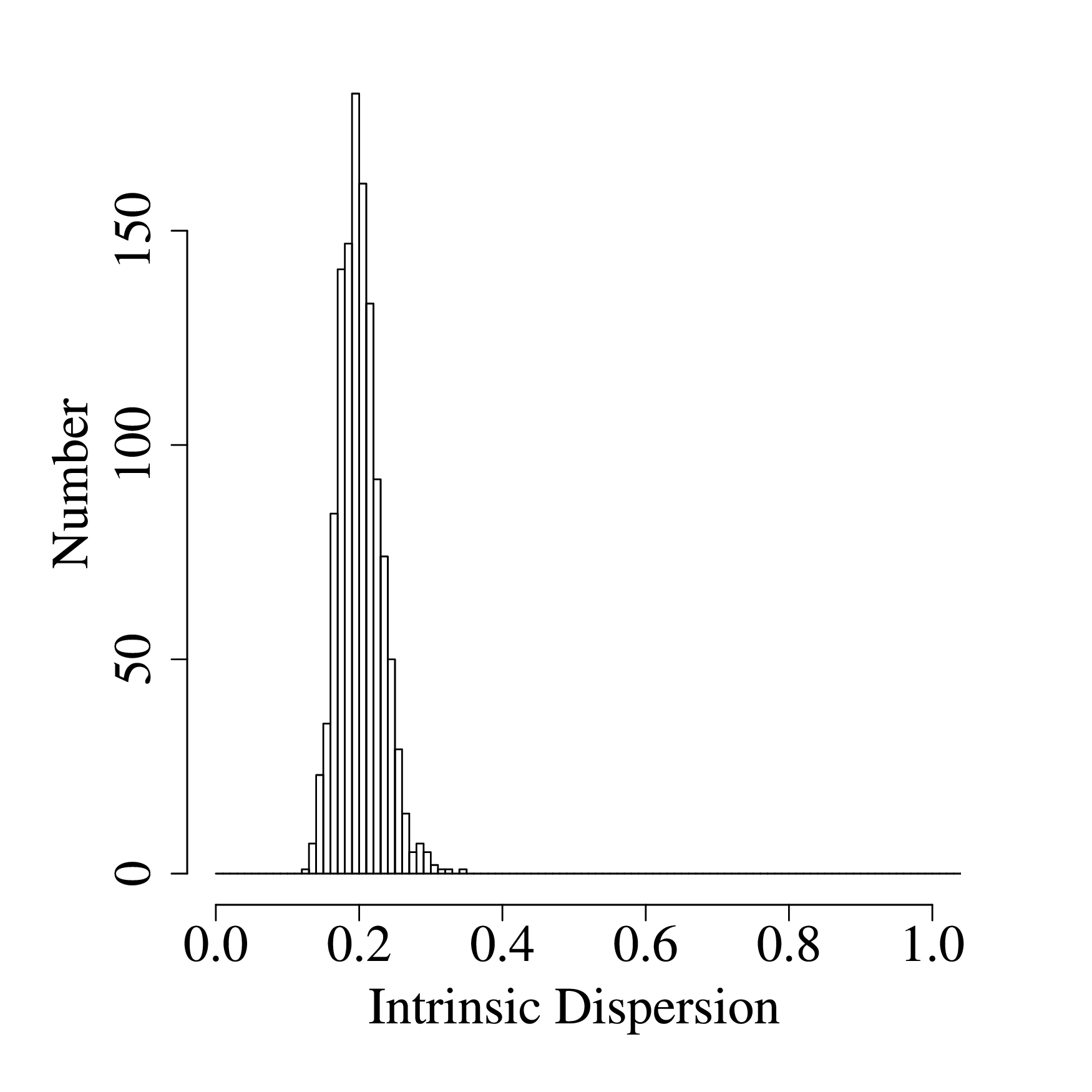}
  \FigureFile(55mm,60mm){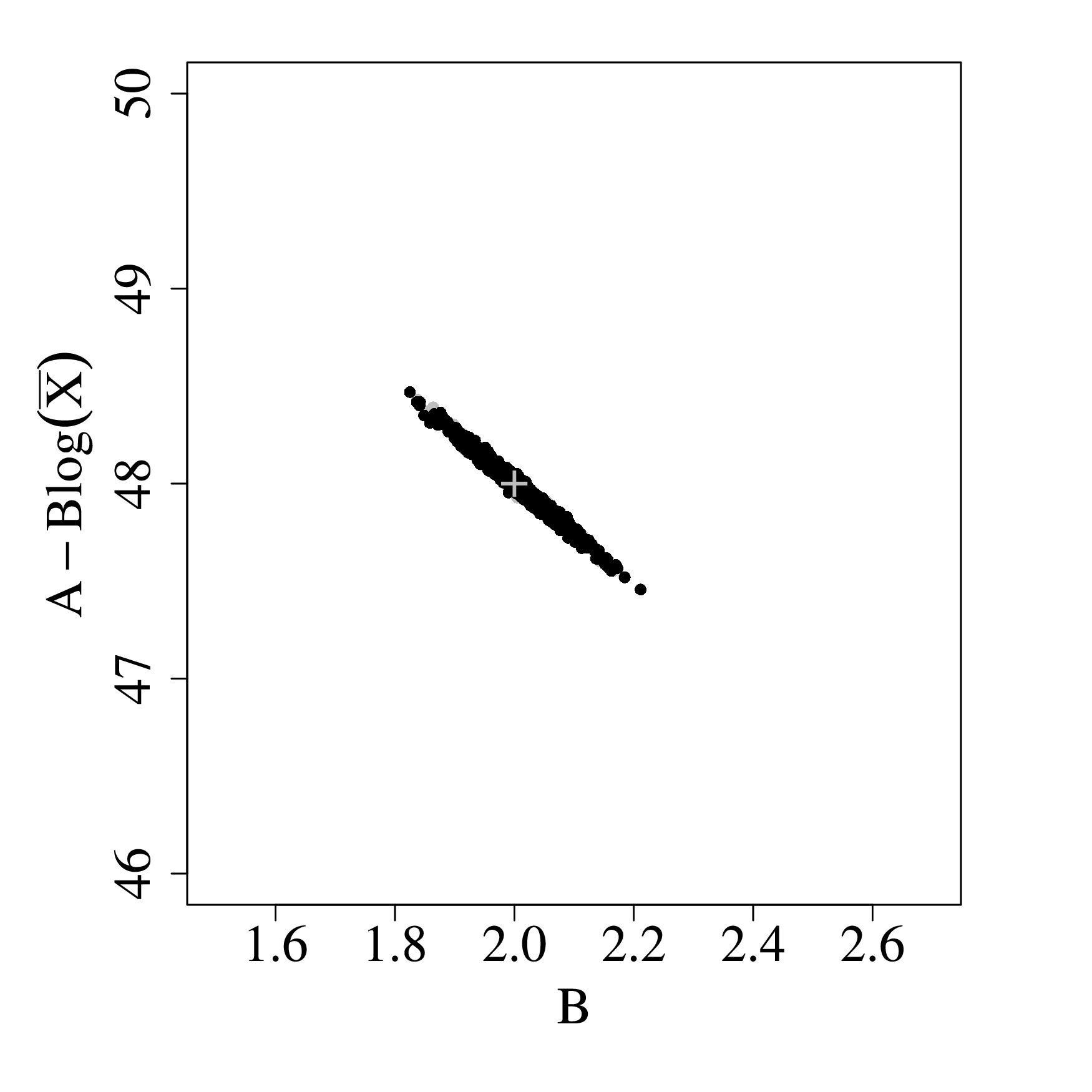}\\
 \end{tabular}
\end{center}
\caption{The results of Monte Carlo simulation with
the intrinsic dispersion $\sigma_{\rm int}=0.2$ for 
$N =$ 15 (top), 50 (middle), 100 (bottom), respectively.
The meaning of points and lines are the same as
in figure~\ref{fig:MC1}. For these simulations, there are
little points eliminated as outliers. As the number of
samples increases, the value of $\sigma_{\rm int}$
converge to the fiducial value. See also table~\ref{tab:MC2}.}
\label{fig:MC2}
\end{figure}

In figure~\ref{fig:MC2}, we show the result for
$N =$ 15 (top), 50 (middle), 100 (bottom), respectively.
The meaning of points and lines are the same as
in figure~\ref{fig:MC1}. For these simulations, there are
little points eliminated as outliers. As the number of
sample increases, the value of $\sigma_{\rm int}$ converges
to the fiducial value. It is indicated that our method return
reasonable values even if the correlation has intrinsic
dispersion which follows Gaussian distribution. We summarized
the parameters and result in table~\ref{tab:MC2}.

\begin{table}
 \caption{Monte Carlo simulation with intrinsic dispersion.}
 \label{tab:MC2}
 \begin{center}
  \begin{tabular}{cccc}
   \hline
   $N_{\rm total}$ & $[\sigma_{\rm min},\sigma_{\rm max}]$ &
   $\bar \sigma_{\rm int}$ (fiducial) &
   $\bar N_{\rm out}$ (fiducial)\\
   \hline
   15 & [0.01,0.05] &0.20 (0.20) & 1.18 (0)\\
   50 & [0.01,0.05] &0.20 (0.20) & 0.51 (0)\\
   100 & [0.01,0.05] & 0.20 (0.20) & 0.28 (0)\\
   \hline
  \end{tabular}
 \end{center}
\end{table}

\subsection{A case with two populations with different intrinsic dispersions}

Next we consider two populations with the same correlation,
\begin{equation}
\label{eq:MC3}
\log Y_j = 48 + 2 \log X_j ~~~~~~~~~~~ (j=1,2),
\end{equation}
but with different intrinsic dispersion. For $j=1$, we assume
$\sigma_{\rm int} = 0$ and, for $j=2$, we assume the scatter
around the relation is uniform between $-1$ and $1$.

We generate mock data according to the following equations,
\begin{eqnarray}
X_{i_1} &=& U(100,1500) \\
\sigma_{X_{i_1}} / X_{i_1}
&=& \sigma_{Y_{i_1}} / Y_{i_1}
 =  U(\sigma_{\rm min},\sigma_{\rm max})
~~~~~~~~~~~~~~~~~ (1\leqq i_1 \leqq N_1) \\
Y_{i_1}
&=& 48 + 2 \log X_{i_1}
    + G(0,\sqrt{\sigma_{\log Y_{i_1}}^2 + B^2 \sigma_{\log X_{i_1}}^2
                + \sigma_{\rm int}^2}) \\
X_{i_2} &=& U(100,1500) \\
\sigma_{X_{i_2}} / X_{i_2}
&=& \sigma_{Y_{i_2}} / Y_{i_2}
 =  U(\sigma_{\rm min},\sigma_{\rm max})
~~~~~~~~~~~~~~~~~ (1 \leqq i_2 \leqq N_2) \\
Y_{i_2} &=& 48 + 2 \log X_{i_2} + U(-1,1)
\end{eqnarray}

\begin{figure}
 \begin{center}
 \begin{tabular}{cc}
  \FigureFile(55mm,60mm){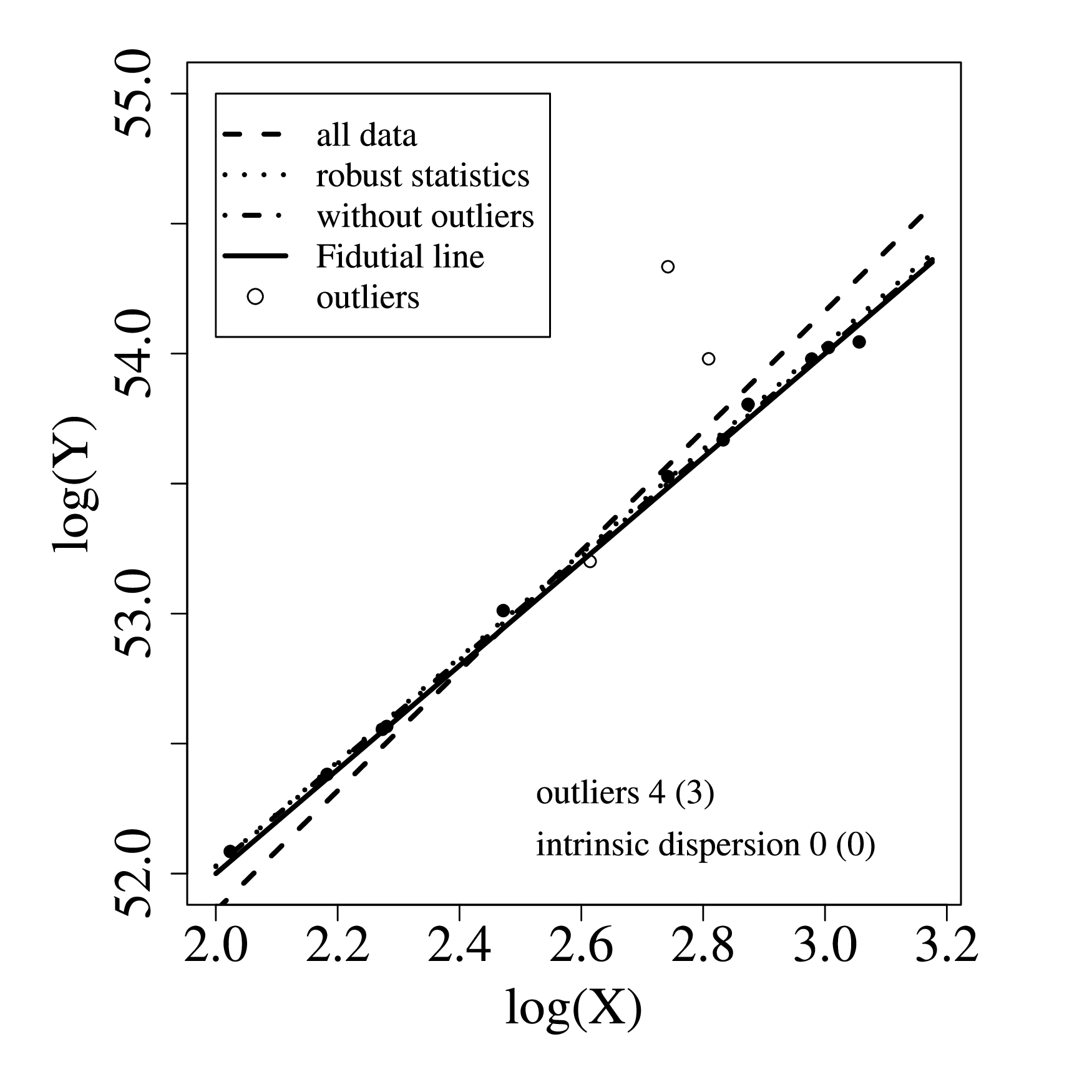}
  \FigureFile(55mm,60mm){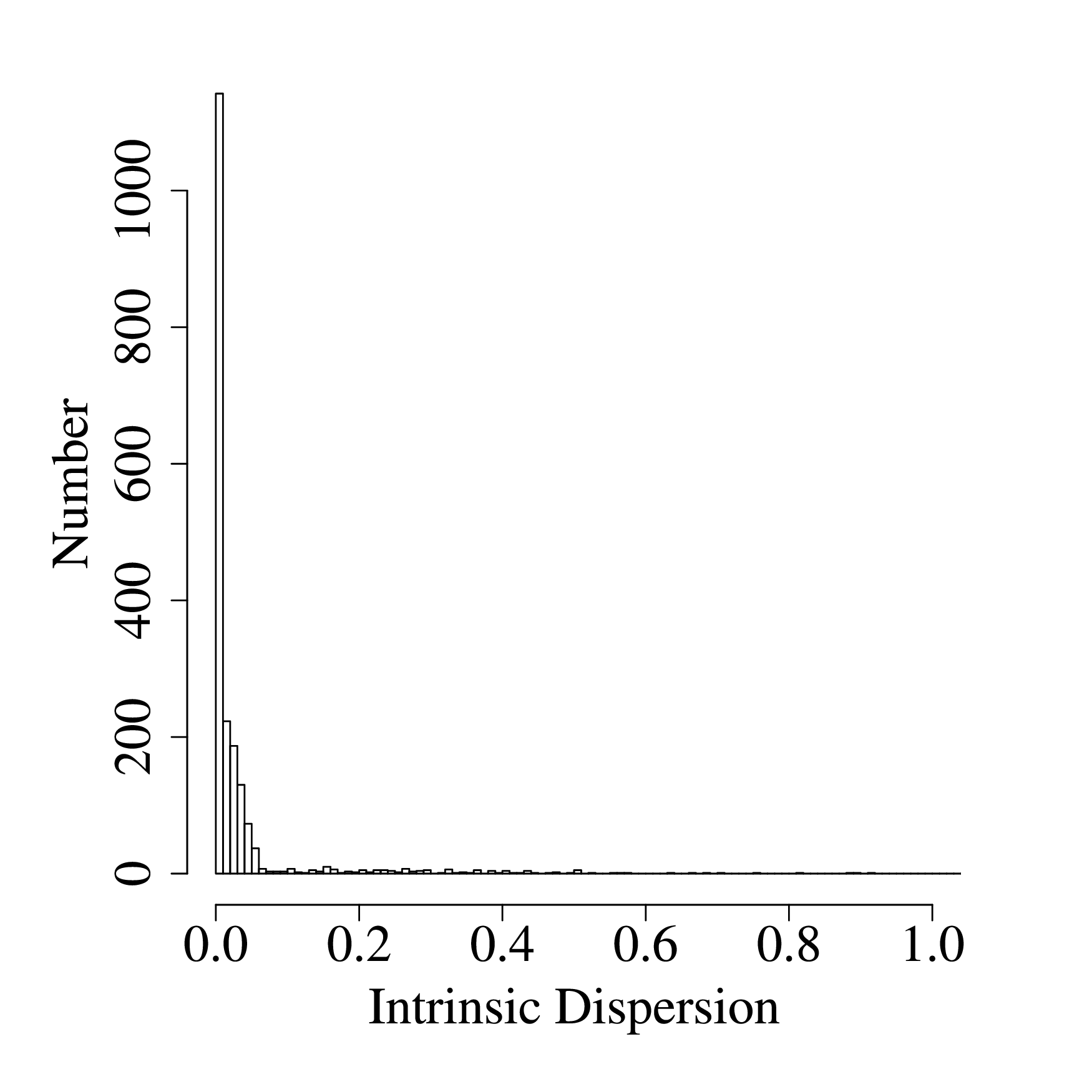}
  \FigureFile(55mm,60mm){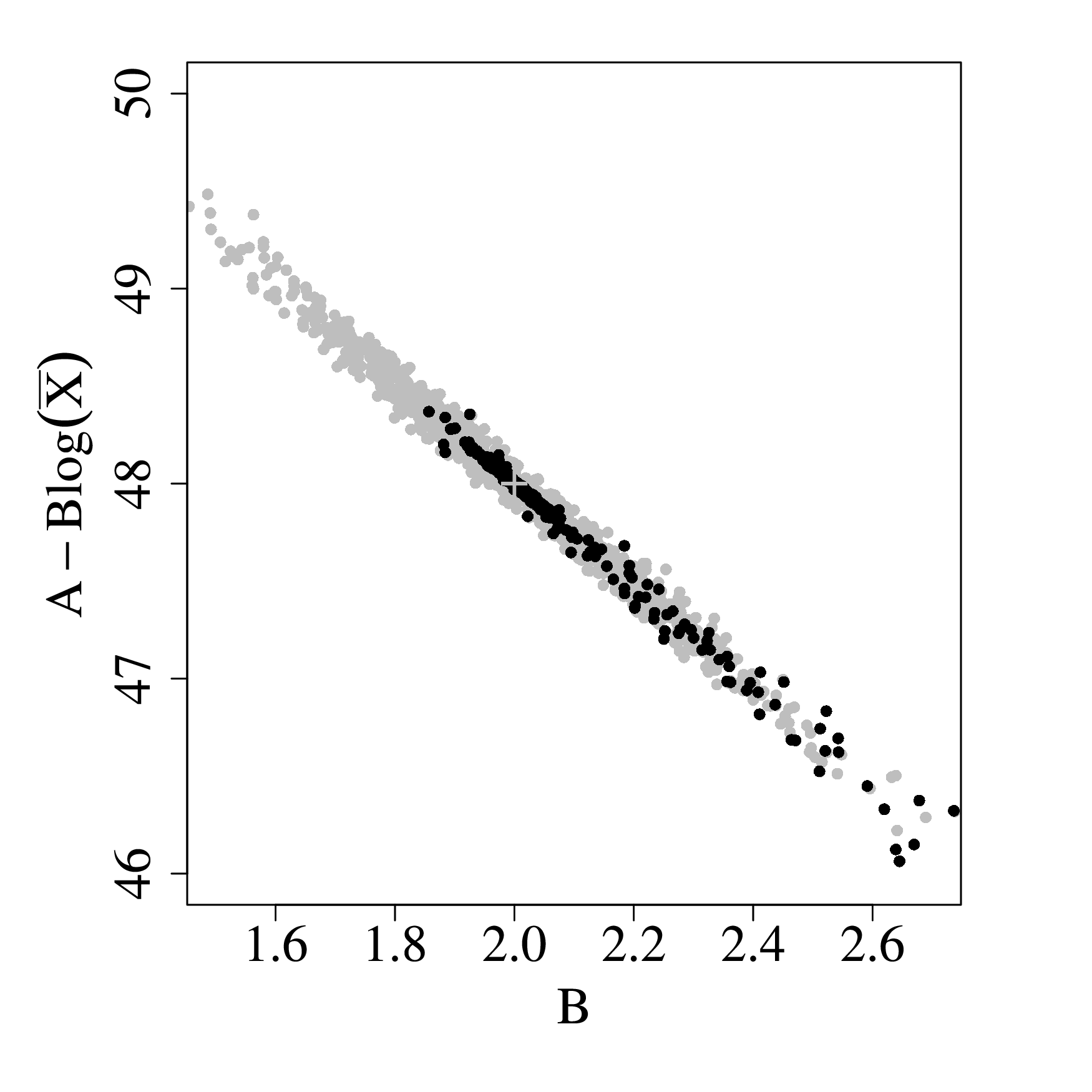}\\
  \FigureFile(55mm,60mm){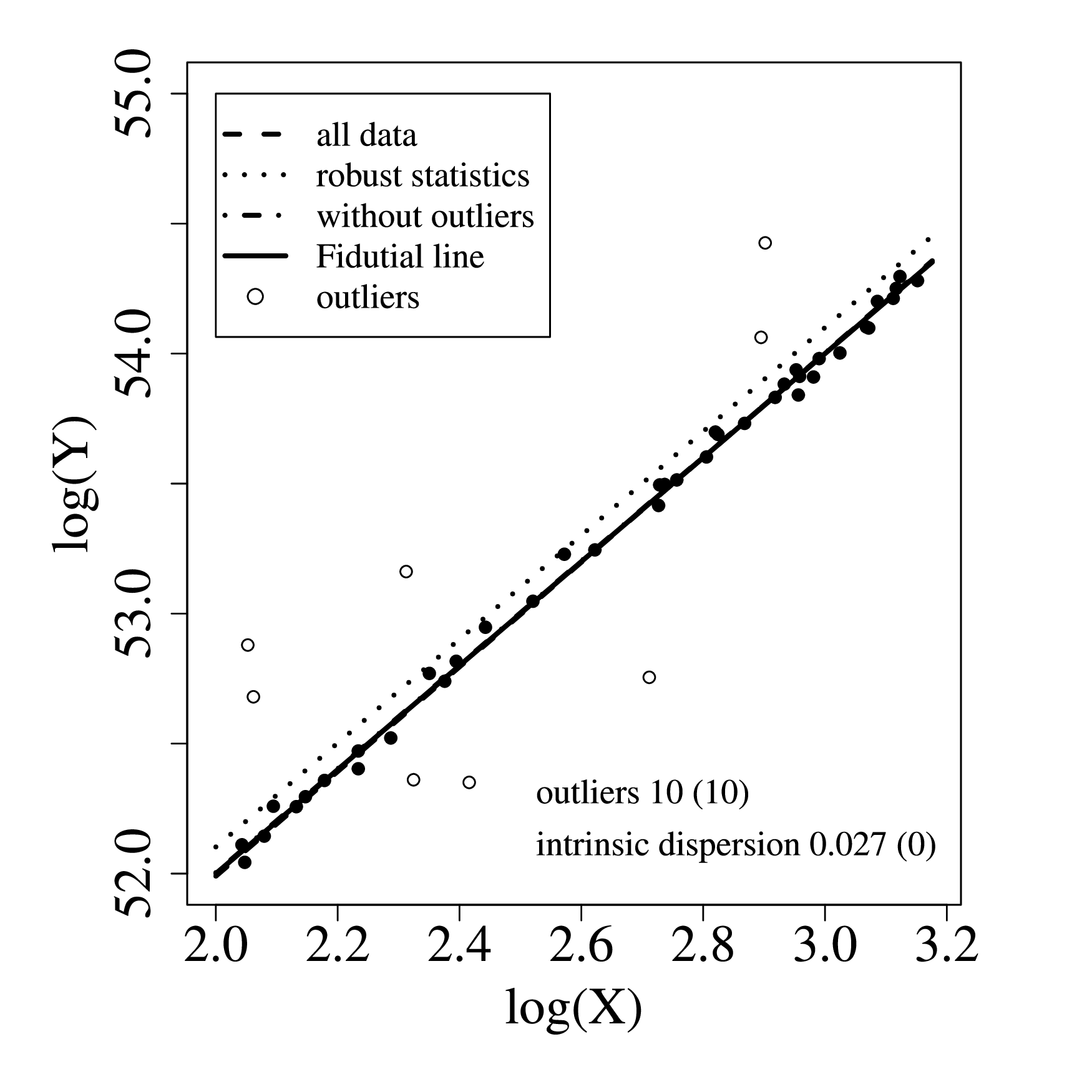}
  \FigureFile(55mm,60mm){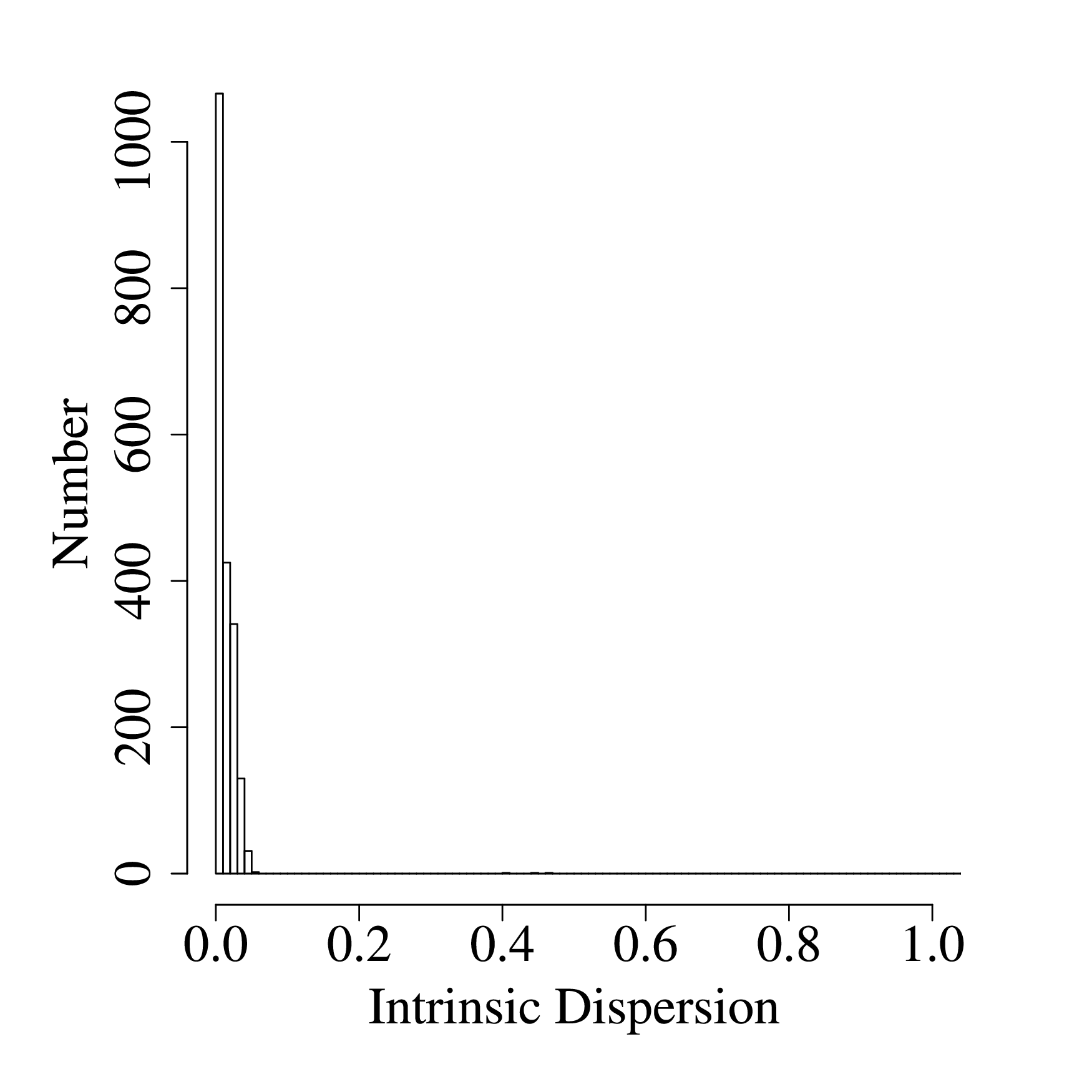}
  \FigureFile(55mm,60mm){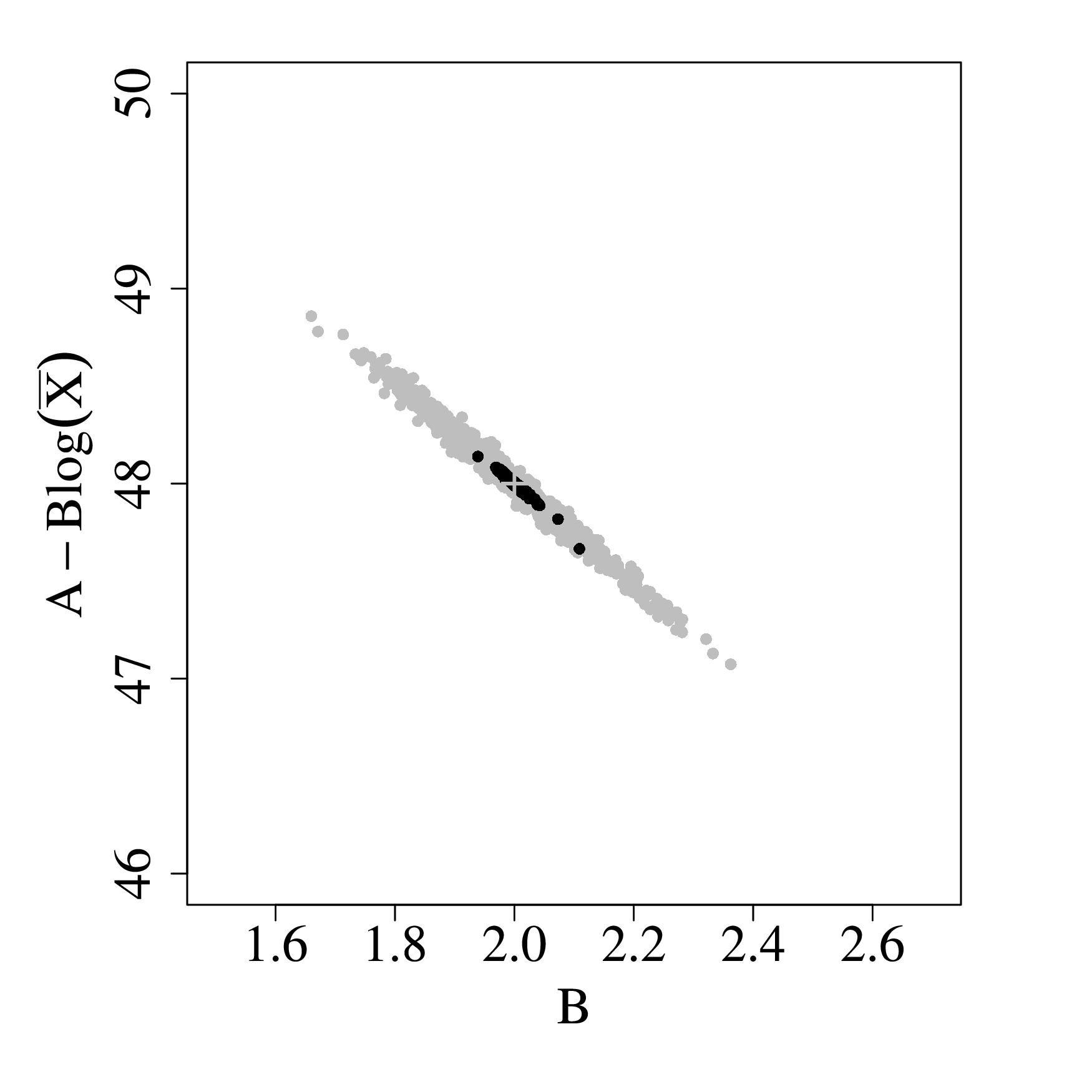}\\
  \FigureFile(55mm,60mm){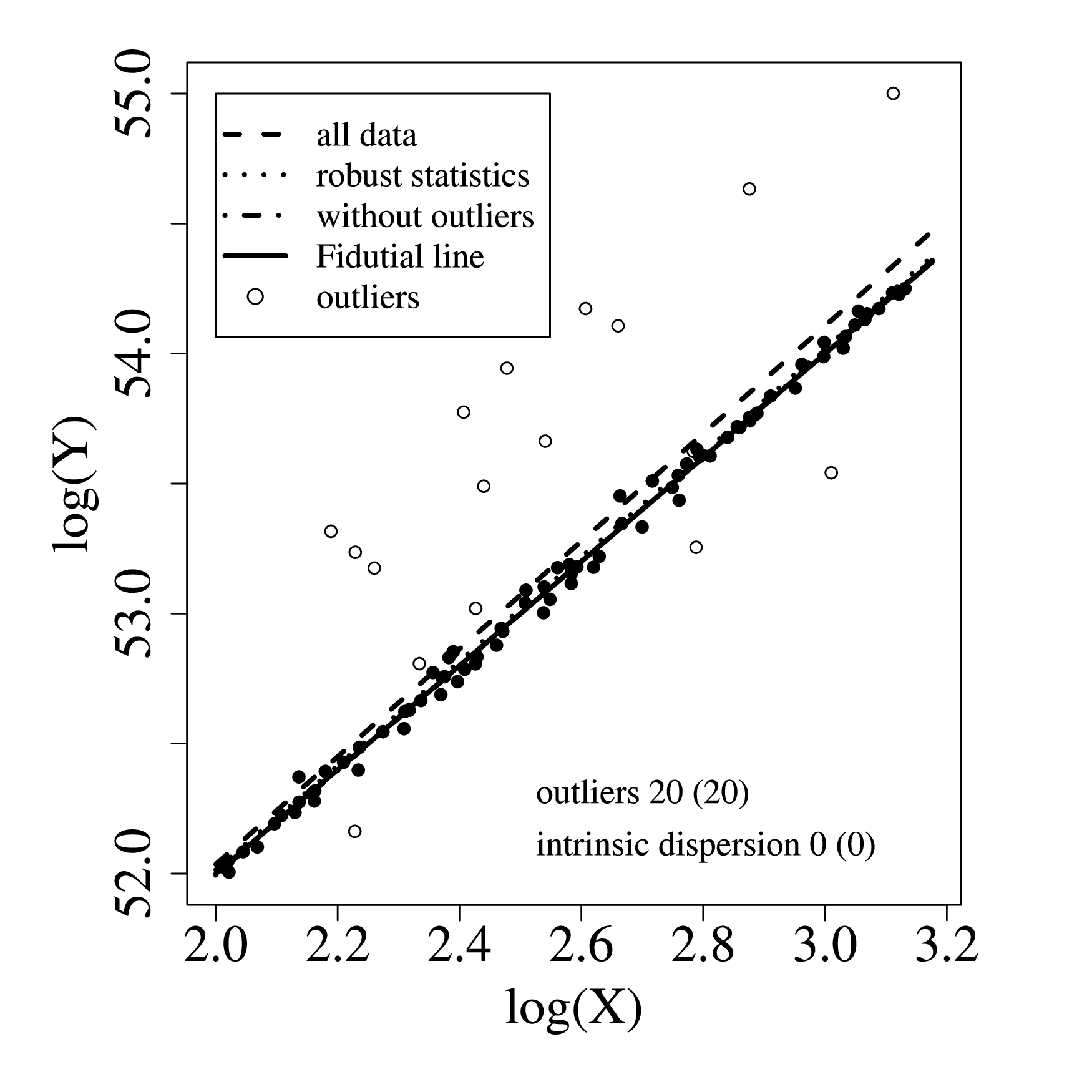}
  \FigureFile(55mm,60mm){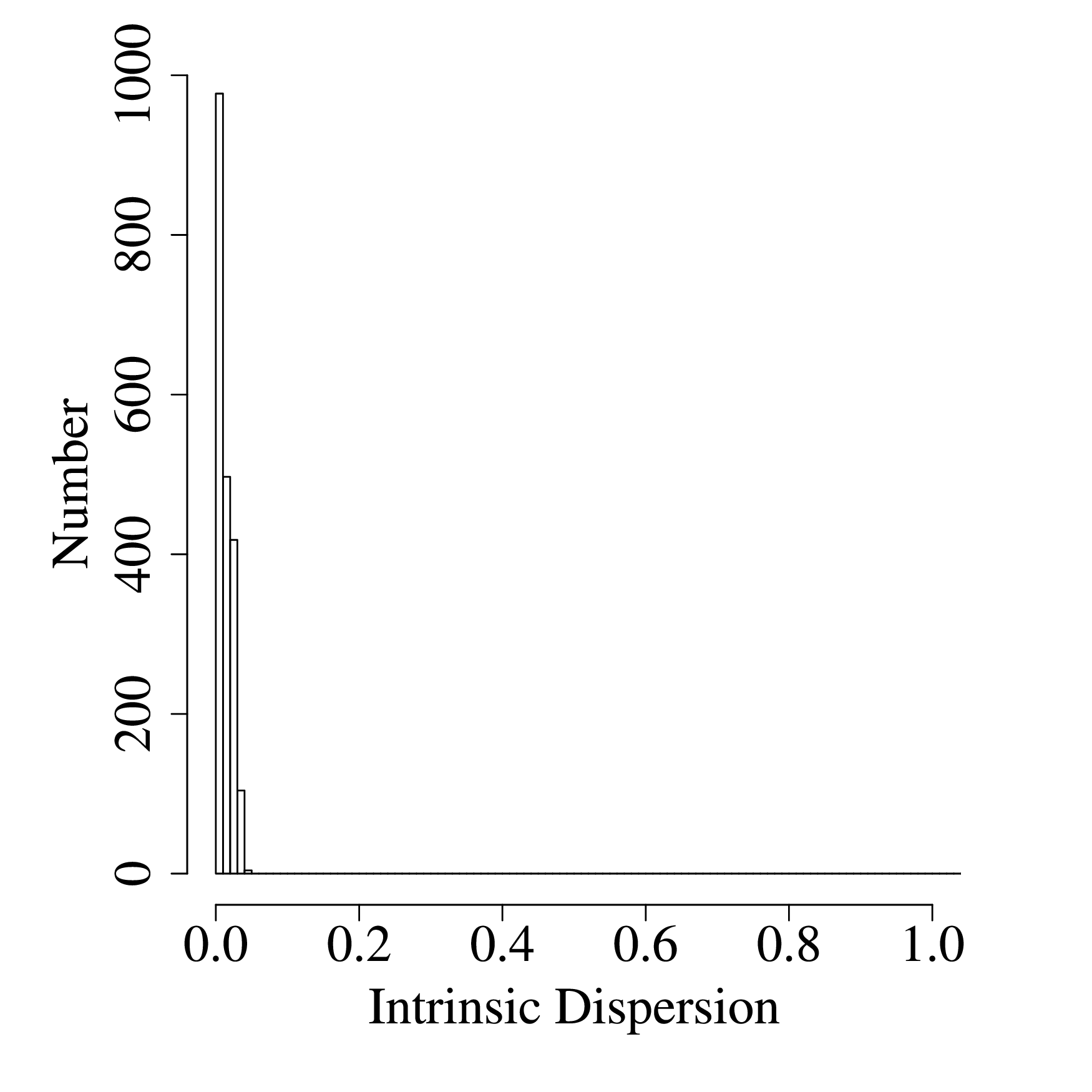}
  \FigureFile(55mm,60mm){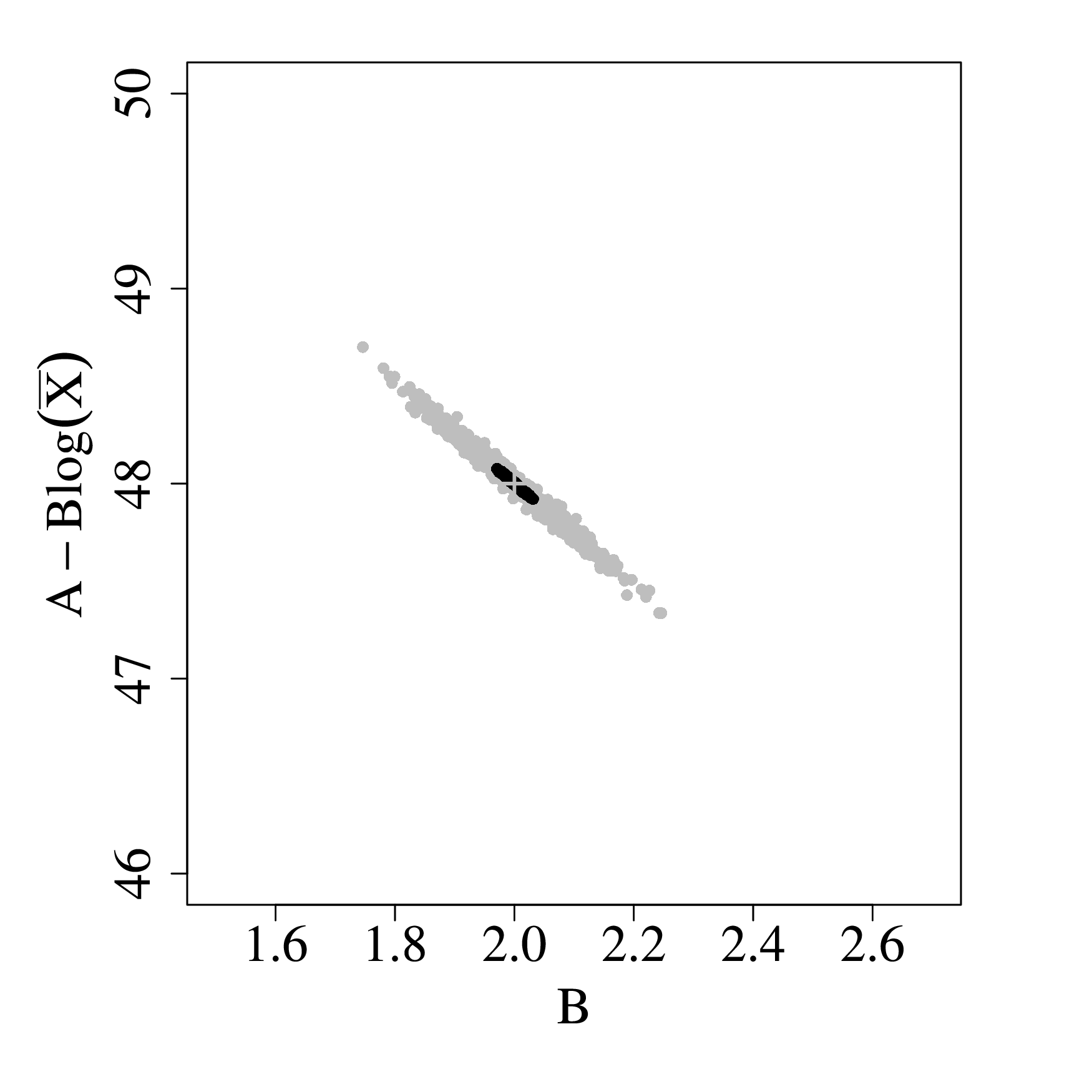}\\
 \end{tabular}
 \end{center}
\caption{The results of Monte Carlo simulation with uniform
distribution outliers for ($N_1,N_2$) = (12,3) (top),
(40,10) (middle), (80,20) (bottom), respectively.
The meaning of points and lines are the same as in
figure~\ref{fig:MC1}. As the number of sample increases,
the value of $\sigma_{\rm int}$ converges to the fiducial value.
See also table~\ref{tab:MC3}.}
\label{fig:MC3}
\end{figure}

In figure~\ref{fig:MC3}, we show the result for $N =$ 15 (top),
50 (middle), 100 (bottom), respectively. The meaning of points
and lines are the same as in figure~\ref{fig:MC1}.
We summarized the parameters and result in table~\ref{tab:MC3}.

These result indicate that our method return reasonable
values even if there are two types of correlation whose
intrinsic dispersions are different.

\begin{table}
 \caption{Results of Monte Carlo simulations with uniform distribution
          outliers.}
 \label{tab:MC3}
 \begin{center}
  \begin{tabular}{cccc}
   \hline
   $N_{\rm total}$ & $[\sigma_{\rm min},\sigma_{\rm max}]$ &
   $\bar \sigma_{\rm int}$ (fiducial) &
   $\bar N_{\rm out}$ (fiducial)\\
   \hline
   15 & [0.01,0.05] &0.032 (0) & 2.7 (2) \\
   50 & [0.01,0.05] &0.011 (0) & 9.15 (10) \\
   100 & [0.01,0.05] &0.011 (0) & 18.13 (20)\\ 
   \hline
 \end{tabular}
\end{center}
\end{table}

\end{document}